\documentclass[aps,pre,superscriptaddress,groupedaddress]{revtex4}
\usepackage{hyperref}
\usepackage{color}
\usepackage{graphics,graphicx,epsfig,wrapfig}
\usepackage{epsf,epstopdf}
\usepackage{amssymb,amsfonts,amsmath,bbold}
\usepackage{enumitem}

\usepackage{ifthen}

\newcommand\mydots{\hbox to 0.8em{.\hss.\hss.}}
\newcommand{\beqn}{\begin{eqnarray}}
\newcommand{\eeqn}{\end{eqnarray}}
\newcommand{\beq}{\begin{equation}}
\newcommand{\eeq}{\end{equation}}
\newcommand{\abs}[1]{|#1|}
\newcommand{\avg}[1]{\langle{#1}\rangle}

\newcommand{\AW}[1]{{\color{black}#1}}
\newcommand{\JD}[1]{{\color{black}#1}}
\newcommand{\HT}[1]{{\color{black}#1}}
\newcommand{\Jon}[1]{{\color{black}#1}}

\newcommand{\mytitle}{Precision of readout at the hunchback gene}

\newcommand{\myauthors}{Jonathan Desponds$^{1,2,3}$, Huy Tran$^{1,2,3}$, Teresa Ferraro$^{1,2,3}$, Tanguy Lucas$^{2,3,4}$,  Carmina Perez Romero$^{5}$, Aurelien Guillou$^{2,3,4}$, Cecile Fradin$^{5}$, Mathieu Coppey$^{2,3,4}$, Nathalie Dostatni$^{2,3,4}$ and Aleksandra M. Walczak$^{1,2,3}$}

\begin{document}
\title{\mytitle}
\author{\myauthors}
\affiliation{~\\
\normalsize{$^{1}$ Ecole Normale Superieure, PSL Research University, Paris, France}\\
\normalsize{$^{2}$ UPMC Univ Paris 06, Sorbonne Universit\'es, Paris, France}\\
\normalsize{$^{3}$ UMR3664/UMR168/UMR8549, CNRS, Paris, France}\\
\normalsize{$^{4}$ Institut Curie, PSL Research University, Paris, France}\\
\normalsize{$^{5}$ McMaster University, Canada}\\
}

\date{\today}
\linespread{1}

\begin{abstract}
 % sections/abstract
The simultaneous expression of the {\it hunchback} gene in the numerous nuclei of the developing fly embryo gives us a unique opportunity to study how transcription is regulated in living organisms.  A recently developed MS2-MCP technique for  imaging nascent messenger RNA in living \textit{Drosophila} embryos allows us to quantify the dynamics of the developmental transcription process. The initial measurement of the morphogens by the {\it hunchback} promoter takes place during very short cell cycles, not only giving each nucleus little time for a precise readout, but also resulting in short time traces of transcription.  Additionally, the relationship between the measured signal and the promoter state depends on the molecular design of the reporting probe. We develop an analysis approach based on tailor made autocorrelation functions that overcomes the short trace problems and quantifies the dynamics of transcription initiation.  Based on \AW{live} imaging data, we identify signatures of bursty transcription initiation from the {\it hunchback} promoter. We show that the precision of the expression of the {\it hunchback} gene to measure its position along the anterior-posterior axis is low both at the boundary and in the anterior even at cycle 13, suggesting additional post-\AW{transcriptional} averaging mechanisms to provide the precision observed in fixed embryos.

\end{abstract}

\maketitle

 % sections/maintext2
\section*{Introduction}
During development the different identities of cells are determined by sequentially expressing particular subsets of genes in different parts of the embryo. Proper development relies on the correct spatial-temporal assignment of cell types. In the fly embryo, the initial information about the position along the anterior-posterior (AP) axis is encoded in the exponentially decaying Bicoid gradient. The simultaneous expression of the Bicoid target gene {\it hunchback}  in the multiple nuclei of the developing fly embryo gives us a unique opportunity to study how transcription is regulated and controlled in a living organism \cite{Jaeger2011, Gregor2014}. \AW{Despite many downstream points
where possible mistakes can be corrected \cite{Driever1988, Jaeger2011, Tikhonov2015}, } the initial mRNA readout of the maternal Bicoid gradient by the {\it hunchback} gene is remarkably accurate and reproducible between embryos \cite{Porcher2010, Little2013}: it is highly expressed in the anterior part of the embryo, quickly decreasing in the middle and not expressed in the posterior part. This precision is even more surprising given the very short duration of the cell cycles (6-15 minutes) during which the initial Bicoid readout takes place and the intrinsic molecular noise in transcription regulation \cite{Elowitz2002, Ozbudak2002, raser2004}. 

Even though most of our understanding of transcription regulation in the fly embryo comes from studies of fixed samples, gene expression is a dynamic process. The process involves the assembly of the transcription machinery and depends on the concentrations of the maternal gradients \cite{Crauk2005}. Recent studies based on single-cell temporal measurements of a short lived luciferase reporter gene under the control of a number of promoters in mouse fibroblast cell cultures  \cite{Suter2011, Zoller2015} and experiments in \textit{E. Coli} and yeast populations  \cite{Taniguchi2010, Kandhavelu2012, Muthukrishnan2012, Chong2014} have quantitatively confirmed that mRNAs are generally produced in bursts, which result from periods of activation and inactivation. In early fly development, what are the dynamical properties of transcription initiation that allow for the concentration of the Bicoid gradient and other maternal factors to be measured in these short intervals between mitoses?

\begin{figure}
\begin{center}
\noindent\includegraphics[width=\linewidth]{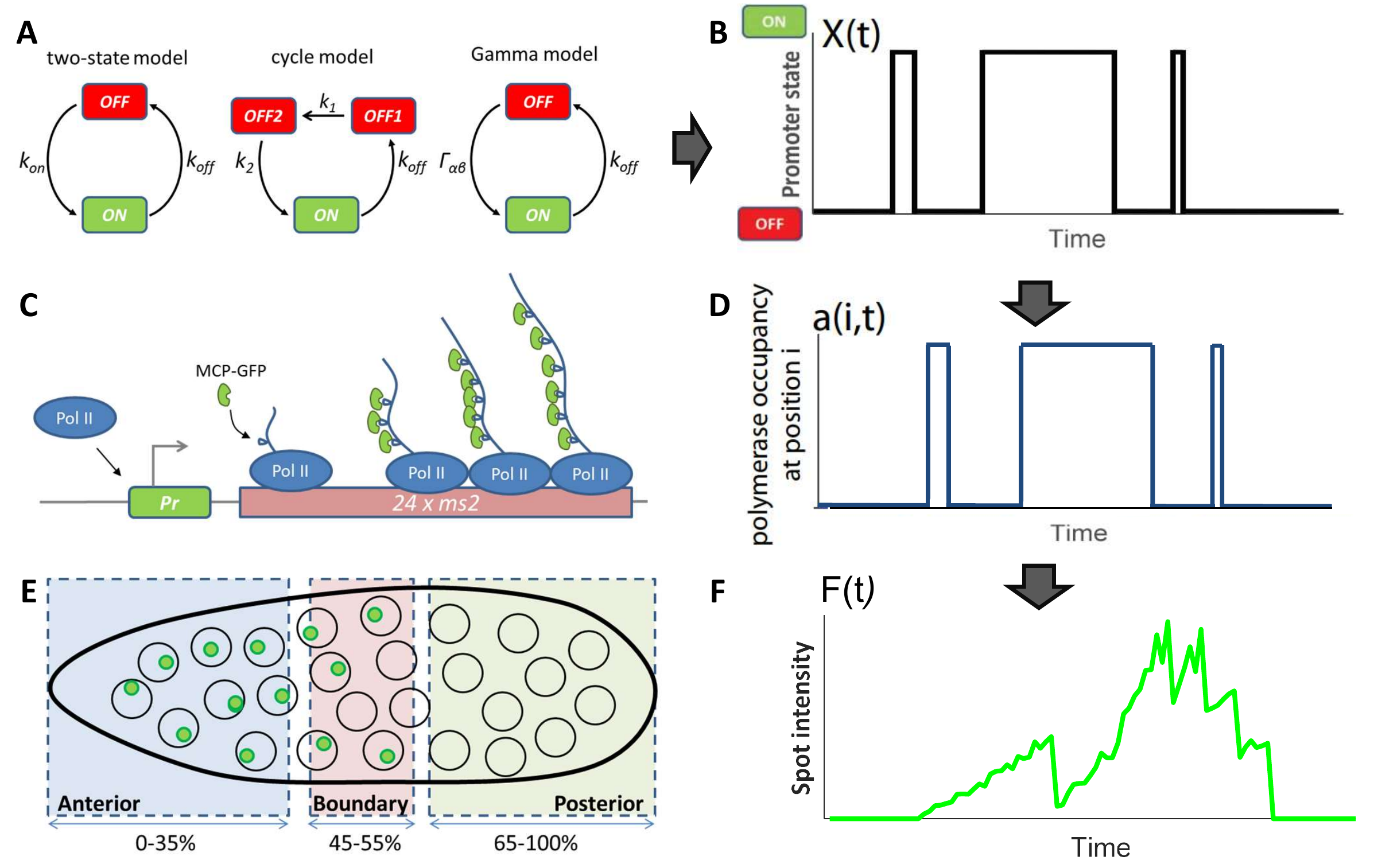}
\caption{
\textbf{Transcription dynamics in the fly embryo}. 
{(A)} The three models of transcription dynamics considered in this paper. From left to right: the two state model, the cycle model and the Gamma model (see Sections B, D and E).
{(B)} Example of the promoter state dynamics (either ON or OFF) as a function of time. We assume that the polymerase is abundant and every time the promoter is ON and is not flanked by the previous polymerase a new polymerase will start transcribing. 
\AW{ The function $X(t)$ in black is non-zero when a polymerase is occupying the transcription initiation site and zero otherwise.}
{(C)} In the ON state, the promoter (\textit{Pr}) is accessible to RNA polymerases (Pol II) that initiate the transcription of the target gene and the \textit{24$\times$} MS2 loops. As the \textit{24$\times$} target mRNA is elongated MCP-GFP fluorescent molecules bind  a detectable fluorescence signal.
{(D)} MCP-GFP molecules labeling several mRNAs co-localize at the transcription loci, which appear as green spots under the confocal microscope. The spot intensities are then extracted over time and classified by each nuclei's position in the \textit{Drosophila} embryo as Anterior, Boundary and Posterior. {\AW{ The spatial resolution of the spots is limited by the Abel limit, which is $\sim200nm$. The ability to identify spots is also limited by the background level of free MCP-GFP. Typical spot sizes are $\sim260nm$, giving an upper bound on the size of the transcription site}.}  
{(E)} \AW{ The gene is divided into $r $ sites of size $150$ base pairs, indexed by $i$.} The presence or absence of a polymerase at site $i$ on the gene as a function of time is given by the promoter occupancy in B and a delay time that depends on the speed of the polymerase.
{(F)} A cartoon representing the type of experimental signal we analyze (see Fig.~8 for real traces): one spot's intensity {as} a function of time, corresponding to the arrival of RNA polymerases in (D) and the promoter state in (B).
 \label{Fig1} }
\end{center}
\end{figure}

In order to quantitatively describe the events involved in transcription initiation,  we need to have a signature of this process in the form of time dependent traces of RNA production. Recently, live imaging techniques have been developed to simultaneously track the RNA production in all nuclei throughout the developmental period from nuclear cycle 11 to cycle 14  \cite{Lucas2013, Garcia2013}.
In these experiments, an MS2-binding cassette is placed directly under the control of an additional copy of a proximal {\it hunchback} promoter. As this reporter gene is transcribed, mRNA loops are expressed that bind fluorescent MCP proteins. Their accumulation at the transcribed locus gives an intense localized signal above the background level of unbound MCP proteins (Fig.~\ref{Fig1}C) \cite{Ferraro2016}. By monitoring the developing embryo, we  obtain for each nucleus a time dependent fluorescence trace that is indicative of the dynamics of transcription regulation at the {\it hunchback} promoter (Fig.~\ref{Fig1}B, D and F). 

However the fluorescent time traces inevitably provide an indirect observation of the transcription dynamics. The signal is noisy, convoluting both experimental and intrinsic noise with the properties of the probe: the jitter in the signal is not necessary indicative of actual gene switching but could simply result from a momentarily decrease in the recording of the intensity.  To obtain a sufficiently strong intensity of the signal to overcome background fluorescence, a long probe with a large number of loops is needed, which introduces a  buffering time. In the current experiments the minimal buffering time \AW{is the time needed to transcribe a fluorescent probe made of $24$ loops. It} is $\tau^{\rm buff}_{min}=72$s and it prevents direct observation of the activation events \cite{Ferraro2016}. 

To understand the details of the regulatory process that controls mRNA expression we need to quantify the statistics of the activation and inactivation times, as has been performed in cell cultures \cite{Suter2011,Zoller2015,Kandhavelu2012, Muthukrishnan2012}. However the very short duration of the cell cycles (6-15 minutes for cell cycles 11-13)  in early fly development prevents accumulation of statistics about the inactivation events and interpretation of these distributions. Direct observation of the traces suggests that transcription regulation is not static but displays bursts of activity and inactivity. However the eye can often be misleading when interpreting stochastic traces. In this paper we develop a statistical analysis of time dependent gene expression traces based on specially designed autocorrelation functions to investigate the dynamics of transcription regulation. This method overcomes the analysis difficulties resulting from naturally short traces caused by the limited duration of the cell cycles that make it impossible to infer the properties of regulation directly from sampling the activation and inactivation time statistics. Combining our analysis technique with models of transcription initiation\AW{, estimates of the precision of the transcriptional readout} and high resolution microscopy imaging of the MS2 cassette under the control of the {\it hunchback} promoter \AW {in heterozygous flies}, we find evidence suggesting bursty transcription initiation in cell cycles 12-13. \AW{For the switching timescales we observe experimentally, the autocorrelation function analysis alone is not able to reliably distinguish between different models for promoter activation and we use information about the precision of the transcriptional readout to conclude that transcription is most likely bursty. Based on the analysis of the time traces, we show that the precision of the transcriptional readout in each cell cycle is relatively imprecise compared to the expected precision of the mRNA measurement obtained from fixed samples, both in terms of cell-to-cell variability \cite{Porcher2010}  and embryo-to-embryo variability \cite{Little2013}. We discuss the limitations of the inference for models of different complexity in different parameter regimes.} 

\section*{Results}

\subsection*{Characterizing the time traces}

\AW{Before we present our results, we first analyze the traces and present a new analysis technique.} We study the transcriptional dynamics of \AW{ the {\it hunchback} promoter (depicted in Fig.~\ref{Fig1}A and B)} by generating embryos that express an MS2 reporter cassette under the control of the proximal {\it hunchback} promoter (Fig.~\ref{Fig1}C), using previously developed tools \cite{Lucas2013, Garcia2013}, with an improved MS2 cassette \cite{Lucasinprep} (see Materials and Methods for details). The MS2 cassette was placed towards the 3' end of the transcribed sequence and contained  $24$ MS2 loop motifs. While the gene is being transcribed, each newly synthesized MS2 loop binds MCP-GFP (expressed at low levels and freely diffusing in the embryo). In each nucleus, where transcription at this reporter is ongoing, we observe a unique bright fluorescent spot, which corresponds to the accumulation of several MCP-GFP molecules at the locus (Fig.~\ref{Fig1}C). We assume that the fluorescent signal from a labeled mRNA disappears from the recording spot when the RNAP reaches the end of the transgene. With this setup we image the total signal in four fly embryos using confocal microscopy, simultaneously in all nuclei (Fig.~\ref{Fig1}D)  from the beginning of cell cycle (cc) 11 to the end of cell cycle 13. In each nucleus we obtain a signal that corresponds to the temporal dependence of the fluorescence intensity of the transcriptional process, which we refer to as the time trace of each spot. A cartoon representation of such a trace resulting from the polymerase activity (Fig.~\ref{Fig1}E) dictated by the promoter dynamics (Fig.~\ref{Fig1}B) \AW{is shown in Fig.~\ref{Fig1}F}. We present examples of the traces analyzed in this paper in Fig.~8 and the signal preprocessing steps in the Materials and Methods and  Section~A. 

\begin{figure}[h!]
\begin{center}
\noindent\includegraphics[width=\linewidth]{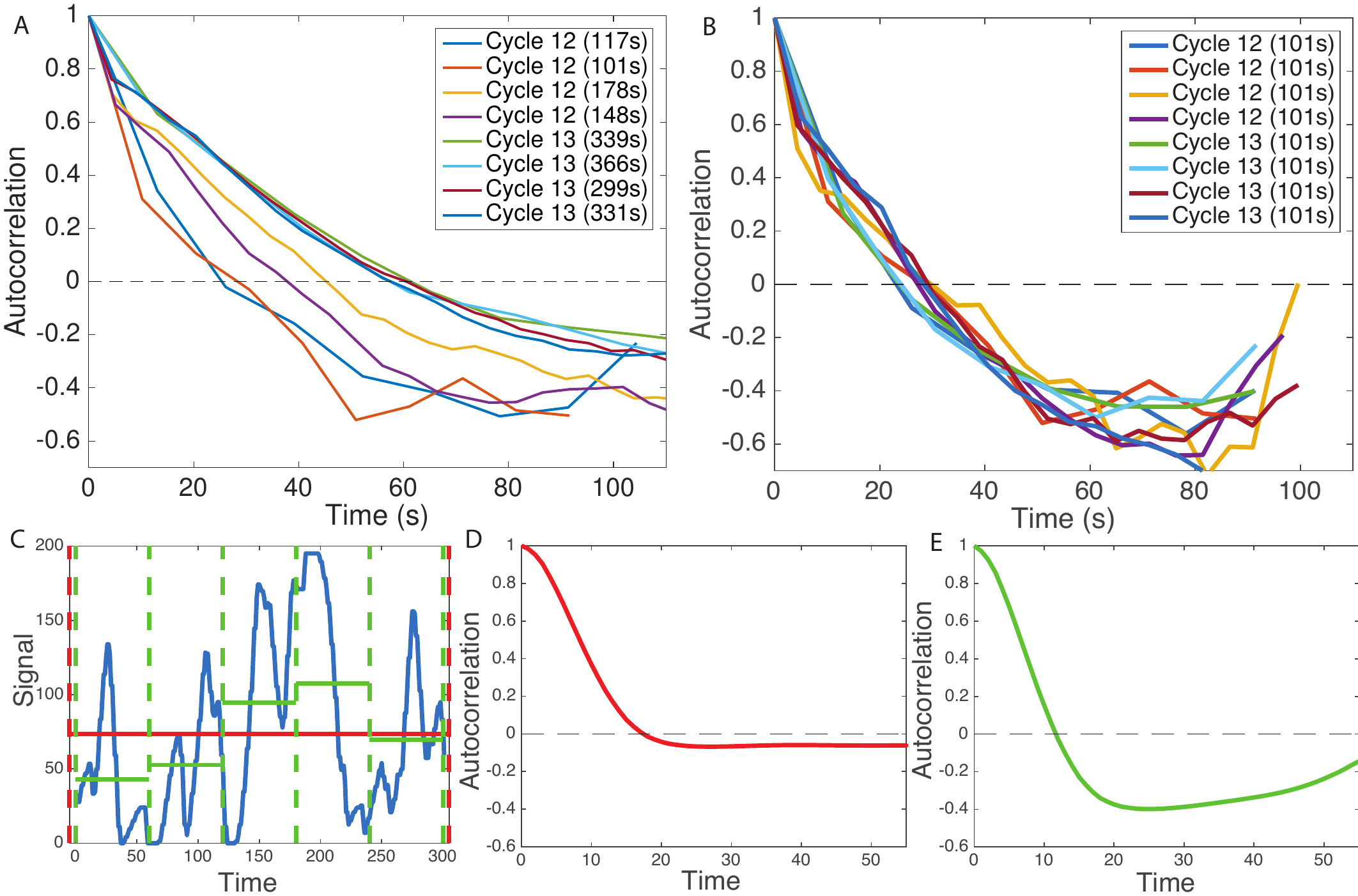}
\caption{{\bf Autocorrelation analysis of fluorescent traces from cell cycles 12-13.} 
(A) Autocorrelation functions for traces of different length caused by the variable duration of the cell cycle. Each autocorrelation function is calculated from one embryo and one cell cycle from traces in the anterior region of the embryos.   Reading off the autocorrelation time as the time at which the autocorrelation function decays by a value of $e$ would give different values for each trace. \AW{ The analysis is restricted to the steady state part of the traces (as defined in the text and Fig.~9). The durations of the steady state windows are given in Table~I.}
(B) Autocorrelation functions calculated for the same traces reduced to having equal trace lengths, all equal to the trace length  of the shortest trace \AW{($101s$)}, shows that the differences observed in panel A are due to finite size effects. \AW{ In the curtailed traces all sequential time points until the $101s$ time point were used.}
(C) An example of a signal simulated using the process described in Fig.\ref{Fig1} for $300$ seconds (blue curve) for a two state model. Taking the whole $300$ second interval (red dashed lines) gives a good approximation of the average signal (red line) and the effect of finite size on the autocorrelation function is small (D) Reducing the time window to $60$ seconds (green dashed lines) correlates the average with the signal much more and the effect of the finite size on the autocorrelation is strong ({E}). \AW{The sampling rates of the four embryos are: $13.1s$, $10.2s$, $5.1s$ and $4.3s$, respectively}. Parameters for the simulation in (C-E) are: $k_{\rm on}=k_{\rm off} =0.06 s^{-1}$, sampling time $dt=4s$, for the red curve $T= {300}s$ and $M=2000$ nuclei, for the green curve $T={60}s$ and  $M=10000$ nuclei (same total amount of data). \AW{These parameters were chosen for illustrative purposes.}
\label{teresa} }
\end{center}
\end{figure}

To characterize the dynamics of the {\it hunchback} promoter we need to describe its switching rates between ON states, when the gene is transcribed by the polymerase at an enhanced rate and the OFF states when the gene is effectively silent with only a small basal transcriptional activity (Fig.~\ref{Fig1}A and B). Estimating the ON and OFF rates directly from the traces is problematic due to the buffering time and to the high background fluorescence levels coming from the unbound MCP-GFP proteins that make it difficult to distinguish real OFF events from noise. To overcome this problem, we consider the autocorrelation function of the signal. To avoid biases from different signal strengths from each nucleus, we first subtract the mean of the fluorescence in each nucleus, $F(t_i)- \avg{F(t_i)}$ and then calculate the steady state connected autocorrelation function  of the fluorescence signal (equivalent to a normalized auto-covariance), $C(\tau)$, at two time points separated by a delay time $\tau$, $F(t_i)$ and $F(t_i+\tau)$, normalized by the variance of the signal over the traces, according to Eqs.~\ref{insidecr} and \ref{meanCemp} in Materials and Methods. \AW{We limit our analysis to the constant expression part of the interphase (which we call ''steady state" -- we discuss this assumption at the end of the Simulated data Results section) by taking a window in the middle of the trace to avoid the initial activation and final  deactivation of the gene between the cell cycles (see Materials and Methods and Fig.~9).} We will always work with the {\it connected} autocorrelation function, which indicates that the mean of the signal is subtracted from the trace.
 The autocorrelation function is a powerful approach since it averages out all temporally uncorrelated noise, such as camera shot noise or the instantaneous fluctuations of the fluorescent probe concentrations. 

Fig.~\ref{teresa}A compares the normalized connected autocorrelation functions calculated for the steady state expression in the anterior of the embryo (excluding the initial activation and final deactivation times after and before mitosis) in cell cycles 12 and 13 of varying durations: $\sim 3$ and $\sim 6$ minutes. \AW{Fig.~\ref{teresa}B shows the same functions for traces that have been curtailed to all have equal length.} The steady state signal from cell cycle 11 did not have enough time points to gather sufficient statistics \AW{to calculate the autocorrelation function}. As expected, the functions decay showing a characteristic correlation time, then {reach} a \AW{valley} at negative values before increasing again. Since the number of data points separated by large intervals is small the uncertainty increases with $\tau$. Autocorrelation functions calculated for very long time traces have neither the negative \AW{valley} nor the increase at large $\tau$. For example, the long-time connected autocorrelation functions \AW{ calculated from the simulated traces (Fig.~\ref{teresa}C) of  the process described in Fig.~\ref{Fig1} that are shown} in Fig.~\ref{teresa}D, differ from the short time connected autocorrelation function in Fig.~\ref{teresa}E calculated from the same trace (see  Section~G for a description of the simulations). As the traces get longer the connected autocorrelation function approaches the longtime results (Fig.~11). The connected autocorrelation function of a finite duration trace of a simple correlated brownian motion (an Ornstein-Uhlenbeck process) displays the same properties (see Fig.~12). The dip is thus an artifact of the finite size of the trace. We also see that the autocorrelation functions shift to the left for short cell cycles (Fig.~\ref{teresa}A), resulting, for earlier cell cycles, in shorter {directly read-off} correlation times, defined as the value of $\tau$ at which the autocorrelation function decays by $e$. However, calculating the autocorrelation functions for time traces of equal lengths for all cell cycles (Fig.~\ref{teresa}B) shows that the shift was also a bias of the finite trace lengths, and after taking it into account, the transcription process in all the cell cycles has the same dynamics (although we note that the dynamics directly read out from this truncated trace is not the true long time dynamics).

\begin{figure}
\begin{center}
\noindent\includegraphics[width=\linewidth]{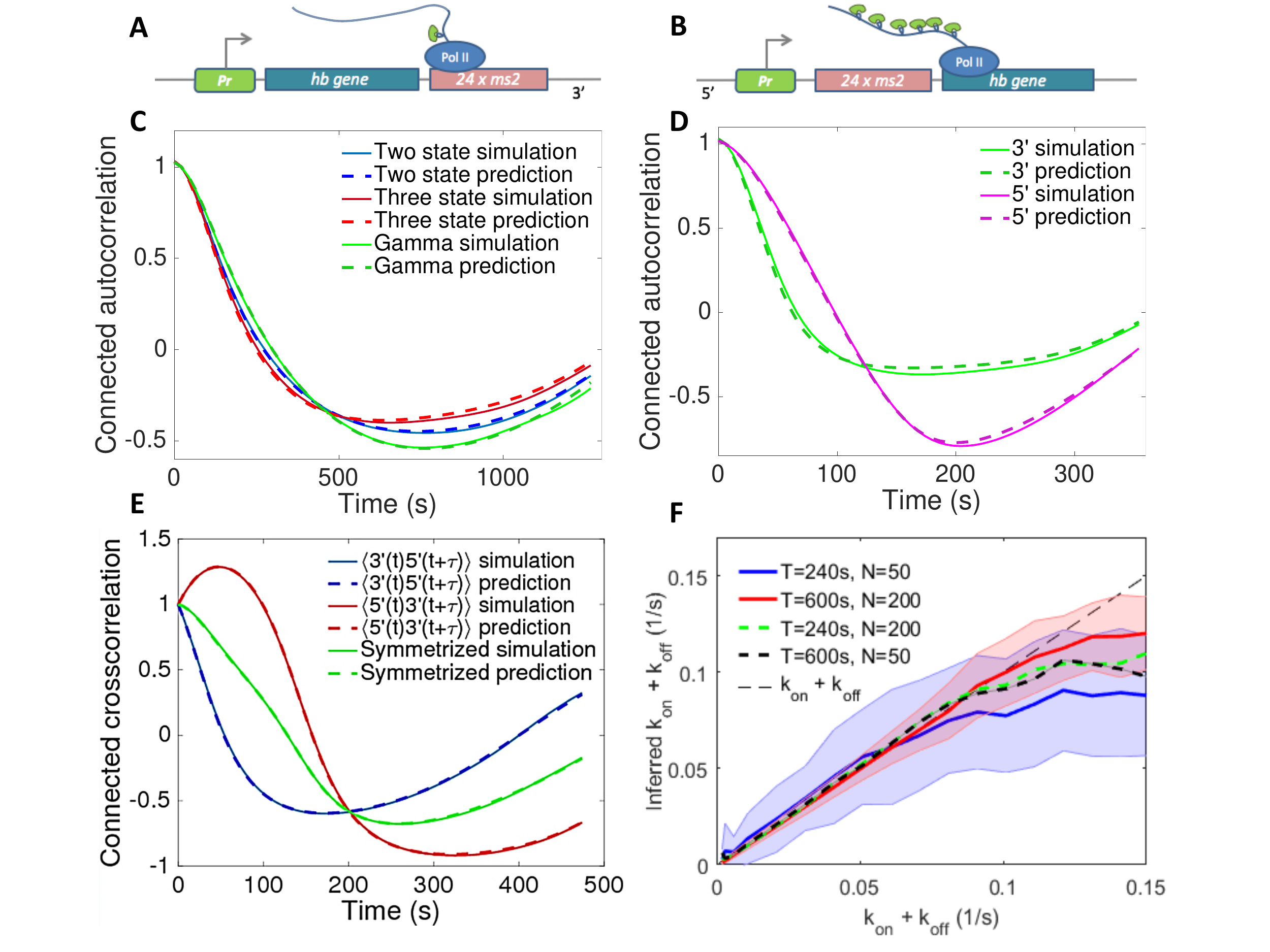}
\caption{{\bf The autocorrelation prediction and autocorrelation based inference analysis performed on short trace simulated data for models of various complexity and positioning of the MS2 probe.}   \AW{A cartoon of the construct with the MS2 cassette placed (A) after the gene (3') and (B) before the gene (5'). Examples of the autocorrelation function's analytical predictions compared to ones calculated from simulated traces (according to the Gillespie simulations described in  Section G) show perfect agreement for 3' MS2 insertions assuming a two state (telegraph) model, three state model and gamma function bursty model (C), as well as for the 3' and 5' constructs in the two state model (D).}
(E)  \AW{ Comparison between prediction and simulation for} the cross-correlation between the signal coming from two different colored fluorescent probes positioned at the 3' and 5' ends. 
(F) The inference procedure for the two state model correctly finds the parameters of transcription initiation in a wide parameter range. The inference range grows with trace length and the number of nuclei. Error bars shown only for $T=240s$, $N=50$ nuclei (blue line) and $T=600s$, $N=200$ nuclei (red line) for clarity of presentation.
Parameters for the simulations and predictions are, (C) for the two state model $k_{\rm on}= 0.005$ s$^{-1}$, $k_{\rm off}= 0.01$ s$^{-1}$, sampling time $dt=6$ s, $T=360$ s and number of cells $M=20000$,  the same parameters for the three state cycle model with $k_{\rm off}=0.01$ s$^{-1}$, $k_{1}=0.01$ s$^{-1}$ and $k_{2}= 0.02$ s$^{-1}$, the same  parameters for the $\Gamma$ model with $k_{\rm off}=0.005$ s$^{-1}$ and  $\alpha=2$ and $\beta = 0.01$ s$^{-1}$; (D) $k_{\rm on}= 0.02$ s$^{-1}$, $k_{\rm off}= 0.01$ s$^{-1}$, sampling time $dt=6$ s, $T=600$ s and number of cells $M=20000$; (E) $k_{\rm on}= 0.01$ s$^{-1}$, $k_{\rm off}= 0.01$ s$^{-1}$, $dt=6$ s, $T=480$ s and $M=20000$. The $5$' construct is modeled as by adding a $3000$bp non-MS2 binding sequence to the $3'$ end of the MS2-binding cassette. 
(F) $P_{\rm on} = 0.1$
\label{fitex} }
\end{center}
\end{figure}

This preliminary analysis shows that to extract information about the dynamics of transcription initiation we will need to account for the finite time traces.  \JD{Additionally, a direct readout of even effective rates from the correlation time is difficult, because the autocorrelation coming from the underlying gene regulatory signal (Fig.~\ref{Fig1}B) is obscured by the autocorrelation due to the timescale needed for the 
transcription of the sequence containing the MS2 cassette (Fig.~\ref{Fig1}D) -- the gene buffering time, $\tau_{\rm buff}$. The observed time traces are a convolution of these  inputs (Fig.~\ref{Fig1}F). The analysis is thus limited by the buffering time of the signal ($tau_{\rm buff} =72 s$ in our system), given as the length of the transcribed genomic sequence that carries the fluorescing MS2 loops divided by the polymerase velocity. A direct readout of the switching rates is only possible if the autocorrelation time of the promoter is larger than the buffering time. 

The form of the autocorrelation function and our ability to distinguish signal from noise also depends on the precise positioning and length of the fluorescent gene \cite{Ferraro2016}.  A construct with the MS2 transgene placed at the 3' end of the gene (Fig.~\ref{fitex}A) gives a differentiable readout of the promoter activity even for two sets of fast switching rates between the active and inactive states. However, in this case the weak signal is hard to distinguish from background fluorescence levels. Conversely, a 5' positioning of the transgene (Fig.~\ref{fitex}B) is insensitive to background fluorescence. However it  only differentiates autocorrelation functions calculated from very slow switching processes\cite{Ferraro2016}. } \AW{In summary, a construct with the MS2 placed at the 3' end of the gene allows for a direct readout of the transcriptional kinetics in a much wider range of switching rates than a 5' construct, although the autocorrelation function of a 3' construct is more sensitive to background fluorescence.} 

\subsection*{Method development -- Promoter switching models}

The promoter activity we are interested in inferring can in principle be described by models of varying complexity (see Fig.~\ref{Fig1}A). \AW{We consider and compare three types of models in this paper. We note this is a small subset of possible models. In particular, we do not consider models with multiple levels of transcription as was considered in \cite{Bothma2014} or reversible promoter cycles.} In the simplest case, the gene is consecutively yet noisily expressed. The RNAP starts transcribing following a Poisson distribution of \AW{ discrete} ON-activation (or firing) events -- this has previously been called a static promoter (not represented in Fig.~\ref{Fig1}A). \AW{After the polymerase binds, the next polymerase cannot bind before the promoter is cleared (a timescale estimated to be $\tau_{\rm block} \sim 6s$ in our experiments). The effective firing rate of this model is the Poisson rate, $r$, shifted by a deterministic $\tau_{\rm block} \sim 6$s, $r_{\rm eff}= (\tau_{\rm block} +r^{-1})^{-1}$, and we call this discrete time model a Poisson-like promoter.} Although the promoter dynamics would be uncorrelated in this case, the gene buffering would still produce a finite correlation time (see  Section~F). Alternatively, the promoter could have two well defined expression states: an ON state during which the polymerase is transcribing at an enhanced level and {an} OFF state when it transcribes at a basal level. This situation can be modeled by stochastic switching between the two states with rates $k_{\rm on }$ and $k_{\rm off}$ (left panel in Fig.~\ref{Fig1}A and Materials and Methods). However, as was previously observed in both eukaryotic and prokaryotic cell cultures \cite{Suter2011,Zoller2015,Kandhavelu2012, Muthukrishnan2012}, once the gene is switched off the system may have to progress through a series of OFF states before the gene can be reactivated. Recently these kinds of cycle models have been discussed for the  {\it hunchback} promoter \cite{Estrada2016}. The intermediate states can correspond to, for example, the assembly of the transcription initiation complex, opening of the chromatin or transcription factor cooperativity. These kinds of situations can either  be modeled by a promoter cycle (middle panel in Fig.~\ref{Fig1}A and Materials and Methods), with a number of consecutive OFF states, or by an effective two state model that accounts for the resulting non-exponential, but gamma function distribution of waiting times in the OFF state (right panel in Fig.~\ref{Fig1}A and Materials and Methods). The time the polymerase spends transcribing the DNA does not dependent on the promoter model. 

\AW{In both the two-state and promoter cycle model the gene switches from the ON to the OFF state with exponentially distributed waiting times described by a rate $k_{\rm off}$ (Fig.~\ref{Fig1}A). In the two-state model the jumps from the OFF to the ON state are also exponentially distributed with a switching rate $k_{\rm on}$ (Fig.~\ref{Fig1}A). In the three state cycle model considered in this paper, an inactive gene can be in two different OFF states. The gene leaves these states with different switching rates,  $k_1$ and $k_2$, respectively. The ordering of $k_1$ and $k_2$ is impossible to detect in the current experiment. In the three state cycle model we can define an effective on-switching rate $k_{\rm on}^{\rm eff} = (1/k_1+1/k_2)^{-1}$.  $k_{\rm on}^{\rm eff}$ corresponds to the inverse of the average waiting time in the overall OFF state, and the waiting times for exiting this effective OFF state are not exponentially distributed. The gamma function distributed switching time is an approximation of this effective rate.} 
We present our method for all of these models and consider all but the gamma function distributed switching time model to learn about the dynamics of {\it hunchback} promoter dynamics. 

\subsection*{Method development -- Autocorrelation approach}

\begin{figure}
\begin{center}
\noindent\includegraphics[width=\linewidth]{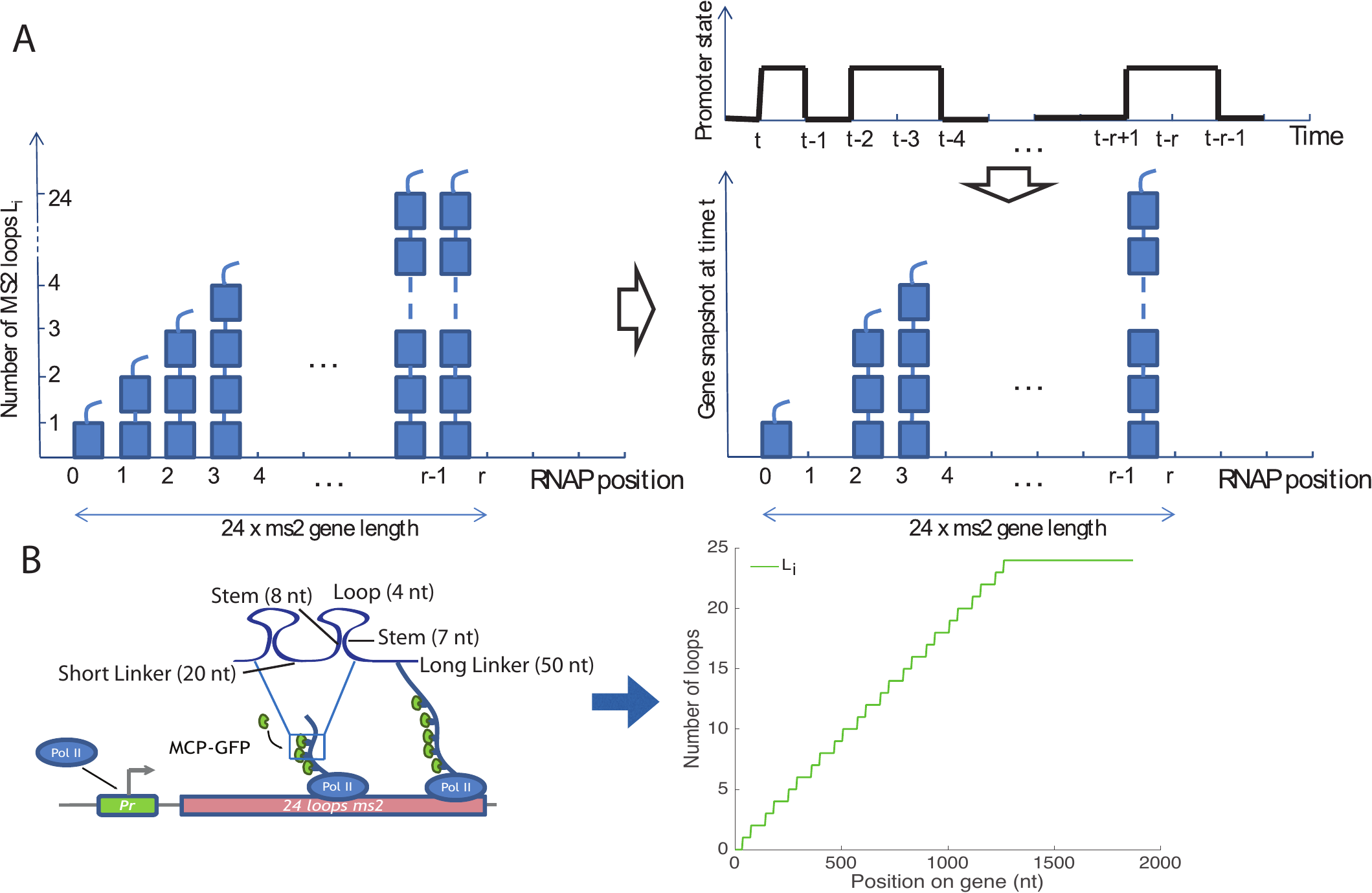}
\caption{{\bf The gene expression model used in the autocorrelation function calculation.} The autocorrelation inference approach is based on the idea that the stochastic transcriptional dynamics can be deconvoluted from the signal coming from the deterministic fluorescent construct, if we know the gene construct design.  (A) A concatenation of snapshots of the gene from $r$ consecutive time steps. A polymerase covers a length on the gene corresponding to its own length in one time step, producing {about two MS2 loops}. The gene has total length $r$ and at any position $i$ along the gene $L_i<24$ loops have been produced. (Top left) The promoter state as a function of time and  (center left) an instantaneous snapshot of the gene corresponding to  transcription from this promoter. (B) The construct design is encoded in the loop function $L_i$. As the polymerase moves along the gene it produces MS2 loops. $L_i$ is an average representation in terms of polymerase time steps of how many loops have been produced by a single polymerase. It is based on the experimental design shown on the left of the panel. 
\label{model setup}}
\end{center}
\end{figure}

To infer the transcription dynamics from the data we built a mathematical model that calculates the  autocorrelation functions accounting for the experimental details of the probes, incorporating the MS2 loops at various positions along the gene and correcting for the finite length of the \AW{ time traces}. 
The basic idea behind our approach is that while the initiation of transcription is stochastic and involves switching between the ON and possibly a number of OFF states ($X(t)$ in Fig.~\ref{Fig1}B denotes the binary  gene expression state), \AW{if we assume a constant elongation velocity} the obscuring of the signal by the probe design is completely deterministic \cite{Coulon2014,Garcia2013}, which results in the random variable $a(i,t) \in\{0, 1\}$ that describes the presence or absence of the polymerase at position $i$ at time $t$ (Fig.\ref{Fig1}D).  We count the progression of the polymerase in discrete time steps, where one time step corresponds to the time it takes the polymerase to cover a distance of $150$ base pairs equal to its own length (Fig.~\ref{model setup}A). The promoter dynamics can thus be learned from the noisy autocorrelation function of the fluorescence intensity normalized by the intensity coming from one MS2 loop, $F(t) = \sum_{i=1}^r L_i a(i,t)$ (Fig.\ref{Fig1}F), \AW{even for switching timescales smaller than the fluorescent probe buffering time $\tau_{\rm buff}$}, provided the parameters of the probe design encoded in the loop function $L_i$ (positioning of the probe etc.) are known (Fig.~\ref{Fig1}C) and the intensity signal is calibrated knowing the fluorescence intensity coming from one MS2 loop \cite{Garcia2013}. 

Broadly, our model assumes that once the promoter is in an ON state the polymerase binds and deterministically travels along the gene producing MS2 loops containing mRNA that immediately bind MCP and result in a strong localized fluorescence (Fig.~\ref{model setup}).  
The presence or absence of a polymerase at position $i$ at time $t$, $a(i,t)$ is simply a delayed readout of the promoter state at time $t-i$, $a(i, t)=X(t-i)$ where $t$ is measured in polymerase time steps (Fig.~\ref{Fig1}B). We assume that polymerase is abundant and that at every time step a new polymerase starts transcribing, provided the gene is in the ON state (Fig.~\ref{Fig1}B and D). The amount of fluorescence produced by the gene at one time point is determined by the number of polymerases on the gene (Fig.~\ref{model setup}A). The amount of fluorescence from one polymerase that is at position $i$ on the gene depends on the cumulated number of loops that the polymerase has produced $L_i$, where $1 \leq i \leq r$. $r$ corresponds to the \AW{\it maximum} number of polymerases that can transcribe the gene at a given time
 and $L_i=1$  corresponds to one loop fluorescing, as depicted in the cartoon in Fig.~\ref{model setup}B. The known loop function $L_i$  depends on the build and the position of the MS$2$ cassette on the gene, it is input to the model and does not necessarily take an integer value since the polymerase length and the loop length do not coincide (Fig.~\ref{model setup}B). Given the average fraction of time the transcription initiation site is occupied by the polymerase, $P_{\rm on}$, the average fluorescence in the steady state is: 
\beq
\avg{F} = P_{\rm on} \sum_{i=1}^{r} L_i.
\label{meanF}
\eeq
Since we assume the polymerase moves deterministically along the gene, seeing a fluorescence signal both at time $t$ and position $i$, and at time $s$ and position $j$ means the gene was ON at time $t-i$ and $s-j$, which is determined by how many loops ($i$ and $j$) the polymerase has produced. 
Taking the earlier of these times, we need to calculate the probability that the gene is also ON at the later time. The autocorrelation function of the fluorescence can thus be written as:
\begin{eqnarray}
\avg{F(t) F(s)}= & \sum_{i=1}^{r} \sum_{j=1}^{r} L_i L_j P \left( \text{gene was ON at time}  \right. \nonumber  \\ &  \left. \min(t-i,s-j) \right) \cdot A(| t-i - s+j |),
\label{autoF}
\end{eqnarray}
where $A(n)$ is the probability that the gene is ON at time $n$ given that it was ON at time $0$, and time is expressed in polymerase steps. The precise form of $P_{\rm on}$, $P (\text{gene was ON at time} \min(t-i,s-j))$ and $A(| t-i - s+j |)$ depends on the type of the promoter switching model.  We assume that the polymerase moves at constant speed along the gene  and that there is no splicing throughout the transcription process. We give explicit expressions for all the models used in the Materials and Methods section and the Supplementary Sections. \AW{Knowing the design of the construct (length of the probe and number of loops that have been transcribed at each position) and calibrating the signal, we use Eq.~\ref{meanF} to directly learn $P_{\rm on}$ from the data. 
In the two and multi-state models $P_{\rm on}$ provides us with the ratio of switching rates and we then use Eq.~\ref{autoF}  to obtain their particular values (see Materials and Methods). }

To avoid biases coming from nucleus to nucleus variability, we calculated the normalized connected correlation function defined in Eqs.~\ref{insidecr} and \ref{meanCemp} in Materials and Methods.  The theoretically calculated connected autocorrelation function, $C_r$ (Eq.~\ref{Ex_conncorr_largeK}, which corresponds to the longtime correlation function in Fig.~\ref{teresa}C and D), differs from the empirically calculated connected autocorrelation function from the traces, $c(r)$  (Eqs.~\ref{insidecr} and \ref{meanCemp} in Materials and Methods, which correspond to the short time correlation function in Fig.~\ref{teresa}C and E), due to finite size effects coming from spurious correlations between the empirical mean and the data points. Since by definition the mean of a connected autocorrelation function is zero (see  Eqs.~\ref{insidecr} and \ref{meanCemp} in Materials and Methods), the area under the autocorrelation function must be zero. For short traces this produces the artificial dip discussed in Fig.~\ref{teresa}, which for long traces is not visible, as it is equally distributed over long times. To compare our theoretical and empirical correlation functions we explicitly calculate the finite size correction and include this correction in our analysis (Materials and Methods and  Section~H and I). 

In this paper, we have analyzed data from fly embryos with 3' promoter constructs only, limiting ourselves to the steady state part of the trace (see Fig.~9). However the method can also be applied to  non-steady state systems (see  Section~C) and to other constructs, including cross-correlation functions calculated from signals of different colors inserted at different positions along the gene (see  Section~J). 
\AW{We use simulated data to show that prediction and inference are possible for cross-correlation functions of a two-colored signal (see Fig.~13), but that the accuracy of inference is limited by the use of the 5' probe.}

\subsection*{Simulated data}

\AW{To check that the inference method correctly infers the parameters of the model,} we first tested the autocorrelation based inference on simulated short-trace data with underlying molecular models with different levels of complexity ( Fig.~\ref{fitex}C) for a construct with the MS2 probe placed at the 3' end of the gene (Fig.~\ref{fitex}B). In Fig.~\ref{fitex}D we compare autocorrelation functions for the three state model for constructs with the MS2 loops positioned at the beginning of the transcribed region (5', Fig.~\ref{fitex}B) and at the end of the transcribed region (3', Fig.~\ref{fitex}A), and the cross-correlation function calculated from a two-colored probe construct (Fig.~\ref{fitex}E). The analytical model correctly calculates the short trace autocorrelation function  and is able to infer the dynamics of promoter switching for all models. It can also be adapted to infer the promoter switching parameters for any intermediate MS2 construct position, given the limitations of each of the constructs discussed above \cite{Ferraro2016}. 

The autocorrelation function based inference reproduces the underlying parameters of the dynamics with great accuracy not just for switching timescales \AW{longer} than  the gene buffering time, $\tau_{\rm buff}$, that obscures the signal (Fig.~\ref{fitex}F){, but also for smaller timescales that are within an order of  magnitude of the gene buffering time}. In Fig.~\ref{fitex}F we show the results of the inference for the 3' two state model for different values of the ON and OFF rates, $k_{\rm on}$ and $k_{\rm off}$.  For switching \AW{timescales {much shorter}}  than the gene buffering time, the autocorrelation function coming from the length of the construct dominates the signal and the precision of the inference goes down. For very fast switching rates ($\AW{k_{\rm on}+k_{\rm off}}>0.12 s^{-1}$), increasing the length of the traces or the number of nuclei (red vs blue curve for values of $k_{\rm on}+k_{\rm off}$ larger than $0.1s^{-1}$ in Fig.~\ref{fitex}F) does not help estimate the properties of transcription. \AW{In this regime, the inferred value of $k_{\rm on}+k_{\rm off}$ disagrees with the  true parameters even when the inference uses  long time traces and  a large number of nuclei.} For intermediate switching rates ($0.07 - 0.12 s^{-1}$), increasing the trace length or increasing the number of nuclei extends the inference range (black and green dashed lines vs blue solid line in Fig.~\ref{fitex}F), and in all cases increasing the number of  nuclei decreases the uncertainty as can be seen from the smaller error bars (shown only for the red and blue lines for figure clarity).

Using two colored probes attached at different positions along the gene gives two measurements of transcription 
 \AW{and} allows for an independent measurement of the speed of the polymerase -- one of the parameters of the model that currently must be taken from other experiments. While the estimates of polymerase speed in the fly embryo are reliable \cite{Garcia2013}, this parameter has been pointed out as a confounding factor in other correlation analyses \cite{Coulon2016}.  

The autocorrelation approach also correctly infers the parameters of transcriptional processes when applied to traces that are out of steady state (see  Section~C). However, since the process is no longer translationally invariant more traces are needed to accumulate sufficient statistics. For this reason, in the current analysis of fly embryos we do not analyze the transient dynamics at the beginning and end of each cycle and we restrict ourselves to the middle of the interphase assuming steady state is reached \AW{(see Fig.~9 for details). We do not know whether the underlying dynamics is completely in steady state. We limit our analysis to a time frame window where the intensity of the fluorescence signal plateaus (see Fig.~9 for an example). We can motivate the steady state assumption {\it a posteriori}:  the inferred switching timescales (smaller than $50$s) are small enough for the system to relax to steady state within one cell-cycle. However we cannot fully rule out  other mechanisms that could keep the system out of steady state (such as changes in the Bicoid concentration).} 

\subsection*{Fly trace data analysis}

\begin{figure}
\begin{center}
\noindent\includegraphics[width=\linewidth]{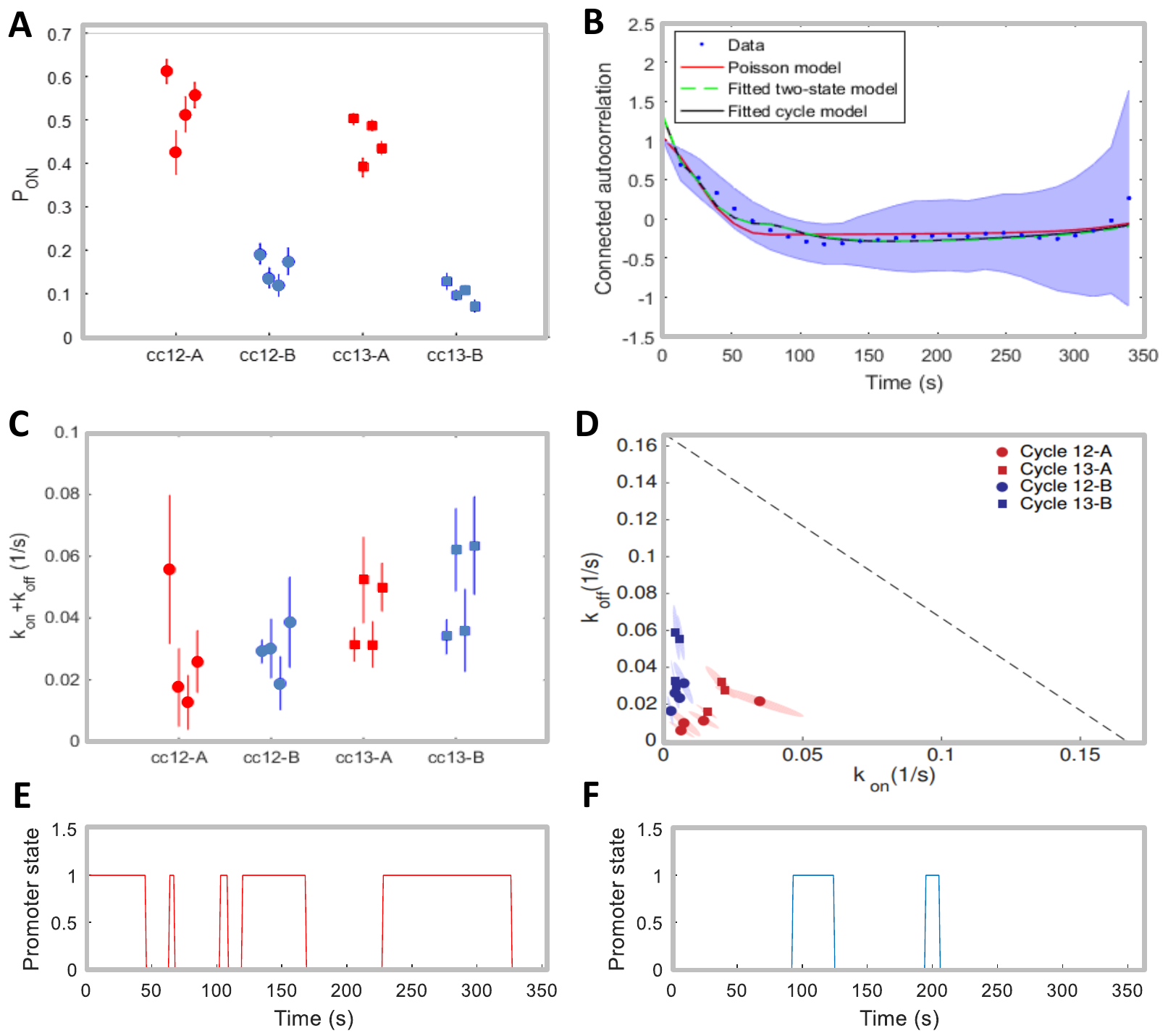}
\caption{\textbf{Inference results for fly data}. {(A)} Inferred values of $P_{\rm on}$ for different nucleus positions (A-Anterior, B-Boundary) and cell cycles. {(B)} Example of the mean connected autocorrelation function of the traces in cell cycle 13 \AW{from the boundary region of embryo 1} (blue dots, with shaded error region) and of the fitted Poisson-like (red), two-state (green) and cycle (black) promoter models. The fitted curves generated from the two-state and three state cycle model are almost superimposed. \AW{See Fig.~10 for fits of all autocorrelation functions in both cell cycles and regions.} {(C)} Inferred values of $k_{\rm on}+k_{\rm off}$ using the two-state model. In (A) and (C), the standard error bars are calculated by performing the inference on 20 random subsets that take 60\% of the original data. {(D)} Inferred values of $k_{\rm on}$ and $k_{\rm off}$ in the Anterior (red) and Boundary (blue) \AW{ for the two-state model}, in cell cycle 12 (circle) and cell cycle 13 ({square}). For each condition, 4 inferred values for 4 movies are shown. \AW{ The dotted black line depicts the limit to inference coming from the time of $\tau_{\rm block} \sim 6s$ it takes the polymerase to leave the transcription initiation site ($k_{\rm on}+k_{\rm off}=1/(6s)$). The shaded areas represent the standard deviational ellipse of $k_{\rm on},k_{\rm off}$ for each cycle and each embryo. The axes of the ellipses are the eigenvalues of the covariance matrix, represented in the directions of the eigenvectors.
}(E-F) Two \AW{simulated} trajectories of the promoter state with the inferred parameters in the Anterior (red) and Boundary (blue).
\label{fig4} }
\end{center}
\end{figure}

We divided the embryo into the anterior region, defined as the region between $0 \%$ and $35 \%$ of the egg length
(the position at $50 \%$ of the egg length marks the embryo midpoint), where {\it hunchback} expression is high, and the boundary region, defined as the region between  $45 \%$ and $ 55 \%$ egg length, where {\it hunchback} expression decreases. The mean probability for the gene to be ON during a given cell cycle $P_{\rm on}$ (restricted to the times excluding the initial activation and deactivation of the gene, which we will call the steady state regime), given by Eq.~\ref{meanF}, is \AW{consistent} between the four embryos in cell cycle 12 and 13, both in the anterior region and at the boundary (Fig.~\ref{fig4}A). The probability for the gene to be ON is over three fold higher in the anterior region than in the boundary and does not change with the cell cycle. $P_{\rm on}\sim0.5$ in the anterior indicates that in each nucleus the polymerase spends about half the steady state expression time transcribing the observed gene. At the boundary the gene is transcribed on average during about $10 \%$ of the steady state part of the cell cycle. The estimates for $P_{\rm on}$ in the earlier cell cycles were not reproducible between the four embryos, likely because the time traces were too short to gather sufficient statistics \AW{to accurately calculate the {maximum and average of the signal.} }
We concentrated on cell cycle 12 and 13 for the remainder of the analysis. 

\subsubsection*{Initial comparison of the promoter models}

Based on the different behavior at the boundary and in the anterior, we separately inferred the transcriptional dynamics parameters in the two regimes, using the autocorrelation approach that corrects for finite time traces for different models. 
The Poisson-like promoter model, the two and three state cycle models all provide reasonably good fits to all the traces in both regions (see Fig.~\ref{fig4}B for an example and Fig.~10 for the fits in both regions in all embryos). Although the 
fit of the Poisson-like promoter model (red line) only captures the short time behavior of the measured autocorrelation function, \AW{there is not enough statistical evidence from the autocorrelation analysis to exclude this model.} 
The two and three state model fits are indistinguishable (Fig.~\ref{fig4}B) and the two state fit is reproducible between cell cycles and embryos (Fig.~\ref{fig4}C and D). 
\AW{ The form of the autocorrelation function in the Poisson-like promoter model is completely determined by the autocorrelation signal of the fluorescent construct (the loop function $L_i$), since the random firing is itself uncorrelated. In the two and three state models, the autocorrelation signal from the fluorescent  construct is convoluted with the autocorrelation signal of the promoter. The fact that the Poisson-like promoter model fits the data so well, indicates that the autocorrelation time of the promoter is comparable to the autocorrelation time of the fluorescent construct. In Fig.~10 we plot the autocorrelation functions for simulated two state models with different correlation times ($k_{\rm on}+ k_{\rm off}$) and a Poisson-like promoter with the same $P_{\rm on}$. For short correlation times, the Poisson-like promoter model and two state model have indistinguishable autocorrelation functions, just like in the analyzed data. For long autocorrelation times, the difference between the two models is clear.  }

The three state fit is reproducible at the level of the sum of the effective ON and OFF rates (same fit as shown for the two state model in Fig.~\ref{fig4}C), 
but gives fluctuating values for $k_1/k_2$, the parameter ratio determining how well it is approximated by a two state model (see 
Fig.~15, $k_1/k_2<1$ describes one fast reaction between the OFF states, effectively giving a two state model, while $k_1/k_2=1$ gives equal weights to the two reactions, clearly distinguishing two OFF states). Since the two state model is reproducible, and has lesser complexity we will further consider \AW{only the two state and Poisson-like promoter models.} 

\subsubsection*{Discussion of the two state model}

For the two state model, the inference procedure independently fits the characteristic timescale of the process \AW{from the autocorrelation function}, defined as the inverse of the sum of two rates, $k_{\rm on}+k_{\rm off}$ (Fig.~\ref{fig4}C)
, and then uses an independent fit of the probability of the gene to be ON, $P_{\rm on} = \AW{ k_{\rm on}/ (k_{\rm on}+k_{\rm off})}$, 
to disentangle the two rates (Fig.~\ref{fig4}D). 
Examples of the \AW{simulated} promoter state over time with the rates' inferred values are shown in Fig.~\ref{fig4}E (for the anterior region) and Fig.~\ref{fig4}F (for the \AW{boundary} region). Assuming the two state model we find that the characteristic timescale, \AW{$(k_{\rm on} + k_{\rm off})^{-1}$}, in most embryos is {slightly} shorter at the boundary ($\sim25s$) than in the anterior region ($\sim 33s$) and the variability between the two cell cycles is comparable to the embryo-to-embryo variability (Fig.~\ref{fig4}C). Both timescales are much larger than the \AW{polymerase blocking time}, $\tau_{\rm block}\sim 6s$, during which a second polymerase cannot bind because the first one has not cleared the binding site (shown as the gray dashed line in Fig.~\ref{fig4}D), which sets a natural scale for the timescales we can infer. We find that in the anterior region of the embryo the two switching rates, $k_{\rm on}$ and $k_{\rm off}$, show variability from embryo to embryo (between $0.009 s^{-1}$ to $0.078 s^{-1}$ -- see Table~I and II), but always scale together, which gives the observed one-half probability of the gene to be ON in a given nuclei during the steady state part of the interphase. Since the polymerase in the anterior on average spends half the steady state interphase window transcribing the gene, the inferred rates suggests a clear bursting behavior of the transcription process, with switching between an identifiable active and inactive state of the promoter \AW{if the two state promoter is correct, and rare firing if the Poisson-like promoter model is correct}. 

At the boundary $k_{\rm on}$ is much smaller than in the anterior with very little embryo to embryo variability, while $k_{\rm off}$ has a similar range in the anterior and at the boundary. This behavior is expected since high Bicoid concentrations in the anterior upregulate the transgene whereas lower concentrations at the boundary result in smaller activation rates. The ratio of the average $k_{\rm on}$ rates at the boundary and anterior is $\sim 5$, which can be compared to the $4$ fold decrease expected from pure Bicoid activation, assuming the Bicoid gradient decays with a length scale of {$100\mu m$} \cite{Gregor2007} and comparing the activation probabilities in the middle of the anterior and boundary regions. Given the crudeness of the argument stemming from the variability of the Bicoid gradient in the boundary region and the uncertainty of the inferred rates, these ratios are in good agreement and suggest that a big part of the difference in the transcriptional process between the anterior and boundary is due to the change in Bicoid concentration. Of course other factors, such as maternal Hunchback, could also affect the promoter, leading to discrepancies between the two estimates.

\begin{figure}
\begin{center}
\noindent\includegraphics[width=\linewidth]{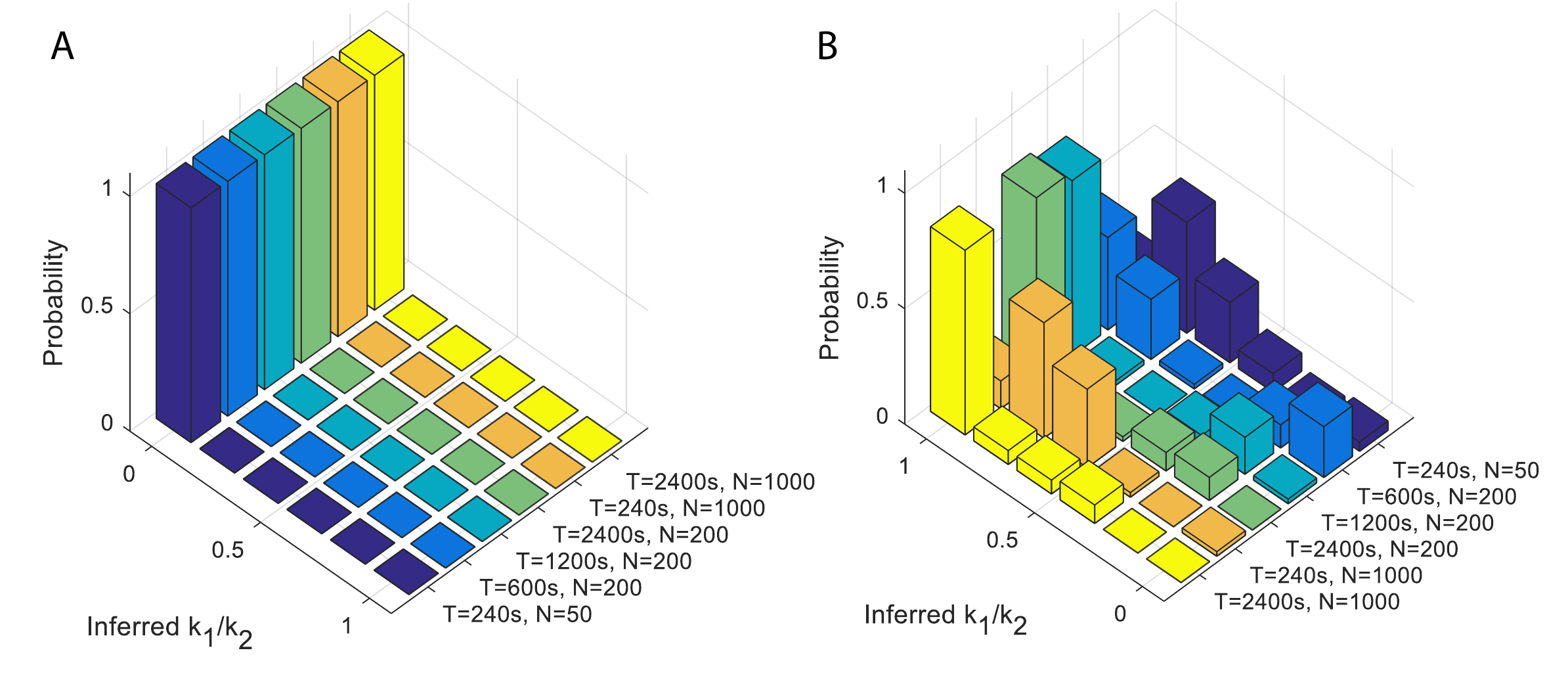}
\caption{{\bf Longer time traces help distinguish between two state and three state cycle models.} A. Inference from data generated by a two state model, which corresponds to $k_{\rm 1}/k_{\rm 2}=0$, from traces of different lengths $T$ and using different numbers of nuclei $N$ shows that longer traces help increase the probability to correctly learn the model type. Increasing the number of nuclei for short traces shows little improvement. The inference is repeated $50$ times per condition. The experimental conditions studied in this paper are closest to the $T=240s$ and $N=50$ nuclei panel. B. The same numerical experiment but assuming a three state cycle model, which corresponds to $k_{\rm 1}/k_{\rm 2}=1$. Parameters of the simulations: $P_{\rm on} = 0.1$, $k_{\rm off}+ 1/( 1/k_1 +1/k_2  ) =0.02 s^{-1}$ and $k_1/k_2= 0$ in A, and $k_1/k_2=1$ in B.
\label{fig5}}
\end{center}
\end{figure}

\subsubsection*{Discriminating between two and three state models}

The current data coming from four embryos and $\sim 50$ nuclei in each region with trace lengths of $\sim 300 s$ does not make it possible to distinguish between the two and three state models. We asked whether having longer traces or more nuclei could help us better characterize the bursty properties. We performed simulations with characteristic times similar to those inferred from the data ($k_{\rm on}+k_{\rm off}^{\rm eff}=0.01$) assuming a two \AW{state ($k_{\rm off}^{\rm eff}= k_{\rm off}$}, Fig.~\ref{fig5}A) and three state cycle model (Fig.~\ref{fig5}B). We then inferred the sum of the ON and OFF rates ($k_{\rm on}+k_{\rm off}^{\rm eff}$) and the ratio of the two OFF rates ($k_1/k_2$). If the two OFF rates are similar ($k_1/k_2\sim1$), we infer a three state model. If one of the rates is much faster ($k_1/k_2\sim0$, \AW{which implies $k_{\rm off}^{\rm eff}= k_{\rm off}$}), we infer a two state model. We find that having more nuclei, which corresponds to collecting more embryos, would not significantly help our inference. However looking at longer traces would allow us to disambiguate the two scenarios, if the traces were $4$ times longer, or $\sim 20$ minutes long. Since  cell cycle 14 lasts $\sim 45$ minutes, analyzing these traces could inform us about the effective structure of the OFF states. However in cell cycle 14, several other direct Bicoid targets (mostly transcription factors) are likely to be expressed,
so additional regulatory elements could be responsible for the observed transcriptional dynamics compared to cell cycle 12 and 13. Our results suggest  that with our current trace length we should be able to identify a two state model with large certainty, but we could not clearly identify a three state model. Our data may point towards a more complex model than two state, but  different kinds of multistate models or a two state model obscured by other biases cannot be ruled out.

The error bars for the autocorrelation functions \AW{(see Fig.~\ref{fig4}B and  Fig.~3)} describe the variability between nuclei coming from both natural variability and  measurement imprecision. While the autocorrelation function is insensitive to white noise, it does depend on correlated noise. The noise increases for large time differences $\tau$, as the number of pairs of time points that can be used to calculate the autocorrelation function decreases and in our inference we reweigh the points according to their sampling, so that the noise does not impair the precision of our inference. The error bars on the inferred parameters \AW{(e.g. Fig.~\ref{fig4}A, C and D)} are due to variability between nuclei and are obtained from sampling different subsets of the data in each region and cell cycle. Additionally to the inter-nuclei and experimental noise, there is natural variability between embryos. Since each nucleus transcribes independently and we assume similar Bicoid concentrations in each of the regions, the inter-embryo variability is of a similar scale as the inter-nuclei variability (Fig.~\ref{fig4}C), as one expects given that the Bicoid gradient is reproducible between embryos \cite{Gregor2007}. \AW{Additonally, variability between Bicoid gradients in different embryos, for example due to their different lengths \cite{Cheung2014},  could also contribute to the observed variability.}

\subsection*{Accuracy of the transcriptional process}

At the boundary, neighboring nuclei have dramatically different expression levels of the Hunchback protein. From measurements of the Bicoid gradient, Gregor and collaborators estimated that for two neighboring nuclei to make different readouts, they must be able to distinguish Bicoid concentrations that differ by $10 \%$ \cite{Gregor2007b}. Following the Berg and Purcell \cite{Berg1977} argument for receptor accuracy, and using measurements of diffusion constants for Bicoid proteins from cell cycle 14, the authors showed that, based on protein concentrations, the {\it hunchback} gene is not able to read-out the differences in the concentrations of Bicoid proteins to the required $10 \%$ accuracy in the time that cell cycle 14 lasts. Even considering the revised higher values of Bicoid's diffusion coefficient measured in a subsequent study \cite{Porcher2010}, the precision of the Bicoid gradient read-out remains difficult to explain. The authors invoked spatial averaging of Hunchback proteins as a possible mechanism that achieves this precision. Spatial averaging can increase precision, but it can also smear the boundary. Erdmann et al calculated the optimal diffusion constant Hunchback proteins must have for the averaging argument to work \cite{Erdmann2009} and showed it is consistent with experimental observations \cite{Porcher2010, Gregor2007}. However precision can already be established at the mRNA level and, using  measurements \AW{on fixed embryos}, Little and co-workers found that the relative intrinsic nuclei-to-nuclei variability of the mRNA transcribed from a {\it hunchback} locus is $\sim 50 \%$ \cite{Little2013}. Measurements of cytoplasmic mRNA reduced this variability to $\sim 10 \%$ \cite{Little2013}. 

Here we go one step further and use our direct measurements of transcription from the {\it hunchback} gene to directly estimate the precision with which the {\it hunchback} promoter makes a readout of its regulatory environment in a given cell cycle in a given region of the embryo, $\delta P_{\rm on}/ P_{\rm on}$. $\delta P_{\rm on}/ P_{\rm on}$ is the relative error of the probability of the gene to be ON averaged  over the steady state part of a cell cycle. Since the total number of mRNA molecules produced in a given cycle is proportional to $P_{\rm on}$ (shown in Fig.~16A as a function of embryo length), the precision at the level of {\it produced} mRNA in a given cycle is equal to the precision in the expression of the gene,  $\delta {\rm mRNA}/{\rm mRNA}= \delta P_{\rm on}/ P_{\rm on}$. The accuracy of transcription activation is encoded in the stochasticity of gene activation. 

\begin{figure}
\begin{center}
\noindent\includegraphics[width=\linewidth]{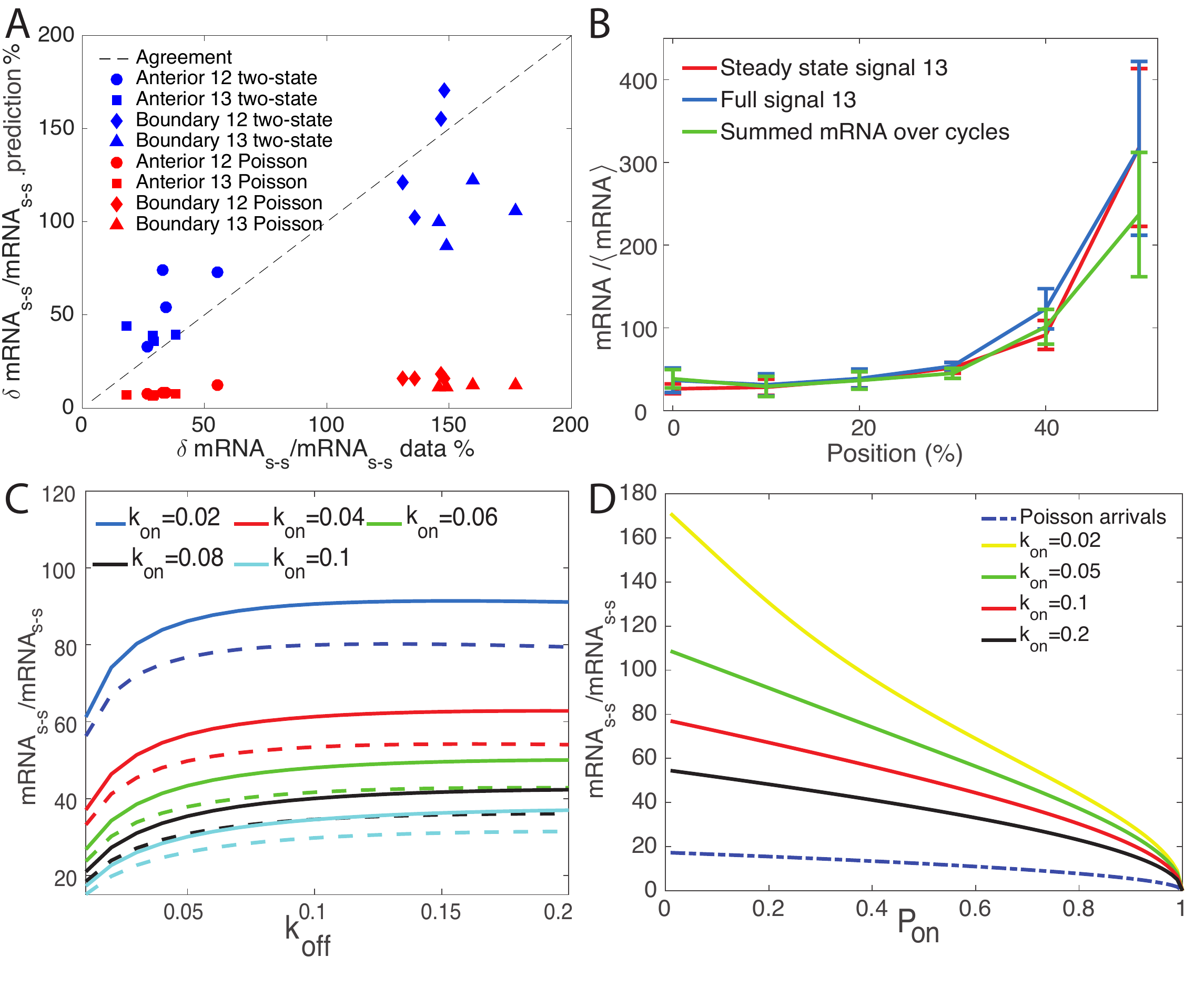}
\caption{{\bf Precision of the {\it hunchback} gene transcription readout.} A. Comparison of the relative error in the mRNA produced during the steady state of the interphase estimated empirically from data (abscissa) and from theoretical arguments in Eq.~\ref{binomial} \AW{for a two state switching promoter (blue symbols) using the inferred parameters in Fig.~\ref{fig4}C (ordinate), and theoretical arguments in Eq.~\ref{Poisson_error} for a Poisson-like static promoter (red symbols) using the inferred parameters in Fig.~\ref{fig4}A (ordinate),  in the anterior (circles and squares) and the boundary (diamond and triangles) regions. The theoretical prediction for the two state promoter shows very good agreement with the data, whereas the Poisson-like promoter shows poor agreement, especially in the boundary region.} 
B. The relative error in the total mRNA produced in cell cycle 13 directly estimated from the data as the variance over the mean of the steady state mRNA production (red line, same data as in A), sum of the intensity over the whole duration of the interphase (blue line) and the  total mRNA produced during cell cycles $11$ to $13$ (green line) for equal width bins equal to $10 \%$ embryo length at different positions along the AP axis. Each line describes an average over four embryos (see Fig.~16C for the same data plotted separately for each embryo) and the error bars describe the variance. 
 To calculate the total mRNA produced over the cell cycles, we take all the nuclei within a strip at cell cycle $13$ and trace back their lineage through cycle $12$ to cycle $11$. We then sum the total intensity of each nuclei in cell cycle $13$ and half the total intensity of its mother and $1/4$ of its grandmother. C. Comparison of the relative error in the mRNA produced during the steady state for a two state, $k_1/k_2=0$, (solid lines) and three state cycle  model, $k_1/k_2=1$, (dashed lines) with the same $k_{\rm on}+k^{\rm eff}_{\rm off}$, for different values of $k_{\rm on}$ and $k^{\rm eff}_{\rm off}$, shows that the three state cycles system allows for greater readout precision. \AW{D. A comparison of the theoretical prediction of the steady state relative error rate for the Poisson-like and two state promoter as a function of $P_{\rm on}$ shows that the Poisson-like promoter is always more accurate. Different values of $k_{\rm on}$  are considered for the two state model.} 
  \label{precision}}
\end{center}
\end{figure}

\AW{In the two state model,} the gene randomly switches between two states:  active and inactive, making a measurement about the regulatory factors in its environment and indirectly inferring the position of its nucleus. Since no additional information is provided by a measurement that is strongly correlated to the previous one, the cell can only base its positional readout on a series of independent measurements. Two measurements are statistically independent, if they are separated by at least the expectation value of the time $\tau_i$ it takes the system to reset itself: 
\beq
\tau_{i} \sim \frac{1}{k^{\rm eff}_{\rm on} + k^{\rm eff}_{\rm off}},
\label{eqtaui}
\eeq
where in a two state model $k^{\rm eff}_{\rm on}=k_{\rm on}$ and $k^{\rm eff}_{\rm off}=k_{\rm off}$. A more detailed estimate obtained by computing the variance of the time spent ON by the gene during the interphase (see  Section~K) shows that Eq.~\ref{eqtaui} underestimates the time needed to perform independent measurements. We find that for a two state model the accuracy of the readout of the total mRNA produced is  limited by the variability of a two state variable divided by the estimated number of independent measurements within one cell cycle: 
\beq
\frac{\delta {\rm mRNA}}{{\rm mRNA}} =\sqrt{ 2\frac{\tau_i (1-P_{\rm on})} { T P_{\rm on}}},
\label{binomial}
\eeq
where $T$ is the duration of the cell cycle and the factor $\sqrt{2}$ is a prefactor correction to the naive estimate. Eq.~\ref{binomial} is valid in the limit of $T>>\tau_i$ (the exact result if given in  Section~K). Using the rates inferred from the autocorrelation analysis (Fig.~\ref{fig4}D) we see that the precision of the gene readout is much lower at the boundary than in the anterior, does not change with the cell cycle and is reproducible between embryos \AW{(blue and red points on the ordinate in Fig.~\ref{precision}A)}. In the anterior part of the embryo it reaches $\sim 50 \%$, while at the boundary, it is  very large, $\sim 150 \%$, even at cell cycle 13.

\AW{In the Poisson-like promoter model, to calculate the relative error in  the total mRNA produced, the polymerase arrival times are described by an effective firing rate of $r^{\rm eff}=(\tau_{\rm block}+1/r)^{-1}$. 
Within this model, the fraction of the total time the polymerase cannot bind, because the binding site is occupied is $P^{\rm P}_{\rm on}= \tau_{\rm block} \cdot r^{\rm eff}$ (see  Section~F). The total produced mRNA is then proportional to the time the gene is transcribed, and the relative error in the total produced mRNA depends on the relative error of the firing times of this modified Poisson process and the number of independent measurements, $n_P=T/(\tau_{\rm block} +1/r)$ (see  Section~K):
\beq
\frac{\delta {\rm mRNA}}{{\rm mRNA}} =\sqrt{  \frac{\tau_{\rm block} (1-P^{\rm P}_{\rm on})}{T} }.
\label{Poisson_error}
\eeq
Using the rates for $P^{\rm P}_{\rm on}$ inferred from the data (Fig.~\ref{fig4}A), the relative error in the total mRNA produced (green and purple points on the ordinate in  Fig.~\ref{precision}A) is slightly higher in the boundary region ($\sim 15 \%$) than in the anterior of the embryo ($\sim 10 \%$) and does not change with the cell cycle. }

We can compare \AW{both of} these theoretical estimates with direct estimates of the relative error of the total mRNA produced during a cell cycle, ${\delta {\rm mRNA}}/{{\rm mRNA}}$, from the data. We divide the embryo into anterior and boundary strips, as we did for the inference procedure 
and calculate the mean and variance of $P_{\rm on}$. These empirical estimates of the gene measurement precision agree with the theoretical estimates (Fig.~\ref{precision}A) \AW{ for the two state model, but disagree with the predictions of the Poisson-like promoter model. For completeness, we also calculated the relative error for the three state model (see Fig.~16B), which shows better agreement than the Poisson-like promoter model but slightly worse than a two state model. We verified that our conclusions about the scale of our empirical estimates are the same for all embryos (Fig.~16C) and do not depend on the definition of the boundary and anterior regions (Fig.~16D). Since the predicted relative error for the Poisson-like promoter model is lower than the relative error calculated directly from the data, we could imagine that the data estimate is more susceptible to additional sources of experimental noise. However the very large disagreement between the Poisson-like promoter prediction and the data at the boundary suggests the Poisson-like promoter model is not an accurate description. This higher experimental variability also cannot be explained by variable expression levels within the regions we are considering: Fig.~16D shows that the experimental relative error does not significantly decrease if we take smaller windows and taking the $P_{\rm on}$  in the boundary region to range from $0.35$ to $0.5$ (Fig.~\ref{fitex}A) would translate into a, at best, two fold increase in the relative error predicted from the Poisson-like promoter model, which is not enough to reach the experimentally observed relative error. While we are unable to rule out the Poisson-like promoter model based on the fit to the autocorrelation function, a different statistic - the relative error in the produced mRNA - suggests that the promoter is most likely well described by a two state model, and possibly a three state cycle.}

To see whether temporal integration of the mRNA produced can increase precision, we compared the empirical estimate of the steady state mRNA production (red line in Fig.~\ref{precision}B) to the relative error of the total mRNA produced in cell cycle 13 (blue line in Fig.~\ref{precision}B) and the total mRNA produced from cell cycle 10 to 13 (green line in Fig.~\ref{precision}B) averaged over embryos. \AW{ Assuming that the mRNA molecules  are equally divided between daughter cells during division, and they are all kept in the cell throughout cell-cycles $10-13$ (which is incorrect but provides a best case estimate), then} each nuclei has the total mRNA produced in cell cycle 13, $1/2$ of the total mRNA produced by its mother in cell cycle 12, $1/4$ of the mRNA produced by its grandmother in cell cycle \AW{ etc}. While we see about a $1/3$ increase in the precision at the boundary from integrating the mRNA produced in different cell cycles, the estimate in the anterior region is not helped by integration over the cell cycles. 

\AW{For completeness of the discussion of the relative errors in the different models,} we calculated the relative error assuming the same $k_{\rm off}+k^{\rm eff}_{\rm on}$ for a three state cycle ($k^{\rm eff}_{\rm on}= k_1+k_2$) as for a two state model ($k^{\rm eff}_{\rm off}=k_{\rm off}$) for different values of $k_{\rm off}$ and $k^{\rm eff}_{\rm on}$ (Fig.~\ref{precision}C). We found that the relative error is always lower for the three state cycle model and the error decreases regardless of the duration of the cell cycle. As expected from Eq.~\ref{binomial}, the relative error is decreased by increasing $k_{\rm on}$  and decreasing $k_{\rm off}$. However the increase in precision from a three state cycle model in the parameter regime we inferred for the two state model in the the fly embryo is relatively modest \AW{(from $\sim 74 \%$ for the two state model to $\sim 67\%$ for the three state model)}. \AW{Similarly, in Fig.~\ref{precision}D we compared the prediction for the relative errors for the Poisson-like promoter model to two state models with the same probability of the gene to be transcribed, $P_{\rm on}$, but different switching rates between the two states ($k_{\rm on}$ and $k_{\rm off}$). Faster switching increases the precision of the two state promoter, since the number of independent measurements increases. The Poisson-like promoter is always more accurate than the two state promoter.}

Many previous analysis of precision from static images calculated the relative error of the distribution of a  binary variable, which in each nucleus was $1$ if the nucleus expressed mRNA in the snapshot, and $0$ if it did not express\cite{Porcher2010a,Perry2012}. We analyzed our data using this definition of activity (see Fig.~16E for mean activity as a function of position)  and found that for most embryos the relative error in the anterior drops to zero (Fig.~16F), indicating that all nuclei in a given region show the same expression state, but at the boundary the precision is still $\sim 50\%$, in agreement with previous reports about the total mRNA in the nucleus \cite{Little2013}. This provides additional evidence for the bursty nature of transcription in the anterior of the embryo, \AW{ in agreement with previous results that showed a relationship between Bicoid concentration and transcriptional burst of downstream genes \cite{He2011}.}

\section*{Discussion}

In contrast to previous studies \cite{Lucas2013,Garcia2013}, including ours, which failed to show evidence for bursty switching of the {\it hunchback} promoter, by developing more advanced analysis techniques 
we show that the promoter has distinct periods of enhanced polymerase transcription followed by identifiable periods of basal polymerase activity. Our conclusions are based on \AW{combining a new autocorrelation based analysis approach, applied to live imaging MS2 data to infer switching parameters, with an analysis of the precision of readout of promoter}. The data we used in this paper was generated with a modified MS2 cassette \cite{Lucasinprep} \AW{(see the Experimental procedures section in Materials and Methods)} compared to the previously published data \cite{Lucas2013}. However the difference in our conclusions mainly comes from a detailed analysis of the traces. 

Quantification of transcription from time dependent fluorescent traces in prokaryotes and mammalian cell cultures has shown that the promoter states cycle through at least three states \cite{Suter2011,Zoller2015}. In one of these states the polymerase transcribes at enhanced levels, while in most of the remaining states the transcription machinery gets reassembled or the chromatin remodels. 
We find that in the living developing fly embryo, the {\it hunchback} promoter also cycles through at least two states, although \AW{based on the parameter inference alone} we cannot conclusively rule out the possibility of \AW{a static promoter with a Poisson-like firing rate or of a more complex promoter with}  more \AW{effective} states when the gene is inactive. \AW{Only a combination of the inferred parameters using the autocorrelation function with another statistic (the relative error in the produced mRNA) allows us to favor the two state model (or more complex models) over the other considered mode of transcription activation. } 

The main impediment to distinguishing different types of  \AW{transcriptional models} comes from the very short durations of the interphase in the early cell cycles when the {\it hunchback} gene is expressed. We showed \AW{ using simulations} that increasing the number \AW{ of} embryonic samples would not help us distinguish between two and three state models, however looking at longer time traces would be informative (Fig.~\ref{fig5}). Since cell cycle 14 lasts about 45 minutes, our analysis shows that the steady state part of the interphase provides enough time to gather statistics that can inform us about the detailed nature of the bursts. Unfortunately, other transcription factors, such as the other gap genes regulating {\it hunchback} expression in cell cycle 14, could possibly change the nature of the transcriptional dynamics in a time dependent manner. We showed that the transcriptional dynamics is constant and reproducible in the earlier cell cycles (12-13) (Fig.~\ref{teresa}), so independently of the question of the nature of the bursts, it would be very interesting to see whether and how it changes when the nature of regulation changes.  

\AW{In the parameter regime of relatively fast switching that we inferred from our data, the autocorrelation function for the two state model and the Poisson-like promoter model are very similar. In this parameter regime, the form of the autocorrelation function is governed by the autocorrelation of the fluorescent probe. So while the autocorrelation function approach is able to disentangle the real promoter switching from the buffering of the construct to determine the parameters assuming an underlying model, we cannot conclusively discriminate between these two models, without looking at other statistics. Using simulations (Fig.~14), we showed that for promoters with slower switching characteristics, this discrimination task is possible and the autocorrelation function approach alone can reliably discriminate between different models. In the parameter regime inferred for the {\it hunchback} promoter, having longer traces would not be helpful for this discrimination task and we have to look for other statistics (Fig.~17). However using new constructs with MS2 binding sites that have higher binding affinity to MCP and decrease the noise from the binding/unbinding of MCP to the RNA would make it possible to use shorter MS2 cassettes without increasing background fluorescence. These cassettes would decrease the buffering time and extend the parameter regime in which we can distinguish between the Poisson-like and two state promoter models. }

Alternatively to focussing on longer traces, a construct with two sets of MS2 loops placed at the two ends of the gene that bind different colored probes could be used to learn more about transcription dynamics \cite{Fukaya2016}. We do not have access to data coming from such a promoter
, but our analysis approach can be extended to calculate the cross-correlation function between the intensities of the two colored probes. Such cross-correlation analysis {has} previously been used to study transcription in cell cultures \cite{Wang2011a}, transcriptional noise \cite{Lin2015} and regulation in bacteria \cite{Dunlop2008a, Munsky2013} . 
 Our theoretical prediction for such a cross-correlation function agrees with simulation results (Fig.~\ref{fitex}C). Unfortunately, the cross-correlation function with one set of probes inserted at the 5' end and the other at the 3' shares the same problems of a 5' construct. 
For fast switching rates, such a cross-correlation function suffers from the large buffering time ($\tau_{\rm buff}\sim 300s$ in\cite{Garcia2013}) drawback of the 5' design and can only be used for inferring large switching rates \cite{Ferraro2016a} (see Fig.~13). 
\AW{Similarly, the cross-correlation function cannot discriminate between a two state and Poisson-like promoter for relatively fast switching.}
However,  it does gives us access into dynamical parameters of transcription such as the speed of polymerase and it is able to characterize whether mRNA transcription is in fact deterministic and identify potential introns. Possibly, cross-correlations from two colored probes both inserted closer to the 3' end could be optimal designs.

Our method requires knowing the design of the experimental system (number and position of the loops),  the speed of polymerase as input and calibrating the maximal fluorescence from one gene. 
While the polymerase speed is an important parameter and erroneous assumption could influence the inference, we have shown that our inference is relatively insensitive to polymerase speeds (see Fig.~18). 
In the current experiments we do not have an independent calibration of the maximal fluorescence coming from one gene, which could introduce potential errors in our analysis. However the reproducibility of our results suggests that these potential errors are small. 

We assumed an effective model that describes the transcription state of the whole gene and does not explicitly take into account the individual binding sites. As a result all the parameters we learn are effective and describe the overall change in the expression state of the gene and not the binding and unbinding of Bicoid to the individual binding sites. For concreteness we presented our model assuming a change in the promoter state and constitutive polymerase binding, but our current model does not discriminate between situations where the transcriptional kinetics are driven by polymerase binding and unbinding and promoter kinetics. The presented formalism can be extended to more complex scenarios that describe the kinetics of the individual binding sites and random polymerase arrival times. Since we already have little resolution power to discriminate between these effective models, we chose to interpret the results of only these effective models. The exact contribution of the individual transcription binding sites could be inferred from the activity of promoters with mutated binding sites. \AW{Similarly, other more complex models, such as a reversible three state model, or a model with many ON states, have not been ruled out by our current analysis but are possible within the current framework. }

The time traces we had to analyze are very short and finite size effects are pronounced. Unlike in cell culture studies, where long time traces are available, we could not collect enough ON and OFF time statistics to characterize the promoter dynamics from the waiting time distributions. In this paper we show that simple statistics, the auto- and cross-correlation functions are powerful general tools that can be used in these kinds of challenging circumstances. \AW{To reach our final conclusion we had to combine different kinds of statistics, which is also a useful strategy when limited by data. }

The approach we propose is a general method that can be used for any type of time trace analysis. However it becomes very useful when studying {\it{in vivo}} biological processes, where the biology naturally limits the available statistics. In our case the number of ON and OFF events is naturally limited by the short duration of the cell cycles. Our method explicitly calculates correlation functions for short traces, correcting for the finite size effects, and can be also used without making steady state assumptions about the dynamics (although this requires collecting sufficient statistics about two time points, which may be hard for short traces). With these corrections we see that while an effective two state model of the underlying dynamics of transcription regulation holds in the anterior and boundary regions of the embryo in all of the early cell cycles, the rates are different in the boundary and anterior regions, showing a strong dependence on position dependent factors such as Bicoid or maternal and zygotic Hunchback concentrations. More statistics will make it possible to build more explicit models of Bicoid dependent activation. 

\AW{While our method is able to deconvolute the effects of the fluorescent probe and infer rates below the buffering limit of the probe (in our case $\tau_{\rm buff} \sim 72$s, see Fig.~\ref{fitex}F),} in all cases, the rates that we can infer from time dependent traces are naturally limited by the timescales at which the polymerase leaves the promoter, which in our case is estimated to be $\tau_{\rm block} \sim 6s$. If the switching rates are faster than this scale, even a perfect, noiseless and infinitely accurate sampling of the dynamics will not be able to overcome this natural limit.

The inferred rates are reproducible between nuclei and embryos and the inter-embryo variability is similar to the inner-embryo variability (Fig.~\ref{fig4}A, C and D). \AW{ The embryo-to-embryo variability can come from Bicoid variability, which is $\sim10\%$ \cite{Gregor2007b}, so we do not expect the observed expression variability to be less, variability in growth rate and RNAP availability and external environmental factors. Additional sources of noise are experimental noise and most importantly problems with data calibration of what is the maximal level of fluorescence intensity}.

We used the obtained results to estimate the precision of the transcriptional process from the {\it hunchback} promoter. We found that even in the anterior region, the variability in the mRNA produced in steady state by the different nuclei is large, with a relative error of about $50\%$ (Fig.~\ref{precision}A). This variability further increases to $150\%$ of the mean mRNA produced at the boundary. These empirical estimates are completely explained  \AW{for a two state promoter model} by theoretical arguments, which treat the gene as an independent measuring device that samples the environment, correcting for the number of independent measurements during a cell cycle. In both cases, the precision at the level of the gene readout is not sufficient to form the precise Hunchback boundary up to half a nuclear width \cite{GregoreGAsperinprep}. Even extending our argument to the total mRNA produced in the early cell cycles (Fig.~\ref{precision}B) does not help. 
 Having an irreversible promoter cycle could increase the theoretical precision, but only slightly in the parameter regime we have inferred and it would not change the quantitative conclusions about low precision backed by  the empirical results. \AW{A Poisson-like promoter, while not compatible with the observed error rates, does have {a} significantly smaller error.}

The construct we used here was limited to the $500$ bp of the proximal {\it hunchback} promoter, which  is known to recapitulate the {\it hunchback} endogenous expression observed
 in Fluorescent In Situ Hybridization (FISH) \cite{Lucasinprep}. 
It is possible that the boundary phenotype is recovered 
  by averaging of mRNAs and proteins produced by the real gene or the transgenes in other nuclei. In the latter case, this would point towards a robust "safety" averaging mechanism that relies on the population. Alternatively, we have to be aware that the sharp boundaries were only detected on fixed samples and that having access to the dynamics of the transcription process likely provides a more accurate view on the process. We calculated and estimated from the data the precision of the gene readout based on the variability of the transcription process between nuclei. We find that the transcriptional process at a given position is quite noisy. Previous estimates of precision were based on  data from fixed samples and did not consider the probability of the gene to be ON, but assumed a binary representation where each nuclei is either active or inactive. By analyzing the full dynamic process we show that the gene is bursty and the transcriptional process itself is much more variable. Reducing the information contained in our traces to binary states, we find precise expression in the anterior, but still large variability at the boundary, similarly to previous results from Fluorescent In Situ Hybridization (FISH) aiming to detect all mRNAs \cite{Little2013}.

Assuming that the precision in determining the position of the nuclei is encoded in the precision of the gene readout, a gene with the dynamics characterized in this paper needs to measure the signal $\sim 200$ times longer at the boundary to achieve the observed $\sim10\%$ precision. A gene in the anterior would need to integrate only $\sim 25$ times longer. These results again suggest that the precision in determining the position of the nuclei is not only encoded in the time averaged gene readout, but probably relies either on spatial averaging mechanisms \cite{Gregor2007b, Okabe-Oho2009, Erdmann2009} or \AW{more detailed features of the temporal information encoded in the full trace \cite{He2011}}.

In summary, the early developing fly embryo provides a natural system where we can investigate a functional setting the dynamics of transcription in a living organism. In our data analysis we are confronted {with} the same limitations that natural genes face: an estimate of the environmental conditions must be made in a very short time. Analysis of dynamical traces suggests that transcription is a bursty process with relatively large inter-nuclei variability, suggesting that simply the templated one to one time-averaged readout of the Bicoid gradient is unlikely. {Comparing} mutant experiments can shed light on exactly how the decision to form the sharp {\it hunchback} mRNA and protein boundary {is} made.

\section*{Materials and Methods}
\label{methods}

\section*{Experimental procedures}
\subsection*{Constructs}

For live monitoring of \textit{hb} transcription activity in Drosophila embryos, we used the MS2-MCP system, which allows fluorescent labeling of RNAs as they are being transcribed \cite{Bertrand1998,Lucas2013,Ferraro2016a}. To implement the reporter system in embryos, we generated flies transgenic for single insertions of a P-element carrying the \textit{hb} proximal promoter upstream of an iRFP-MS2 cassette carrying 24 MS2 repeats \cite{Janicki2004,Lucas2013},  from which Zelda binding motifs have been removed \cite{Lucasinprep}. The flies also carry the P$\{$mRFP-Nup107.K$\}$ \cite{Katsani2008} transgenic insertion on the 2$^{nd}$ chromosome and the P{w[+mC]=Hsp83-MCP-GFP} transgenic insertion on the 3rd chromosome. These allow the expression of the Nucleoporin-mRFP (mRFP-Nup) for the labeling of the nuclear envelopes and the MCP-GFP required for labeling of nascent RNAs \cite{Bertrand1998}. All stocks were maintained at 25$^{\circ}$C.

\subsection*{Live Imaging}

Embryo collection, dechorionation and imaging have been done as described in \cite{Lucas2013}. Image stacks ($\sim$19Z $\times$ 0.5$\mu$m, 2$\mu$m pinhole) were collected continuously at 0.197$\mu$m XY resolution, 8bits per pixels, 1200x1200 pixels per frame. A total of 4 movies capturing 4 embryos from nuclear cycle 10 to nuclear cycle 13 were taken. Each movie had different scanned fields along the embryo's width, which results in different time resolutions: 13.1 \textit{s}, 10.2 \textit{s}, 5.1 \textit{s} and 4.3 \textit{s}.

\subsection*{Image analysis}

Nuclei segmentation, tracking and MS2-MCP loci analysis were performed as in \cite{Lucas2013} and recapitulated here. All steps were inspected visually and manually corrected when necessary. Nuclei segmentation and tracking were done by analyzing, frame by frame, the maximal Z--projection of the movies' mRFP-Nup channel. Each image was fit with a set of nuclei templates, disks of adjustable radius and brightness comparable with those of raw nuclei, from which the nuclei's positions were extracted. During the cycle's interphase, each nucleus was tracked over time with a simple minimal distance criterion. For 
MS2-MCP loci detection and fluorescent intensity quantification, the 3D GFP channel (MS2-MCP) were masked with the segmented nuclear images obtained in the previous step. This procedure also helps associating spots to nuclei. We then applied a threshold equal to $\sim2$ times the background signal to the masked images and selected only the connected regions with an area larger than 10 pixels. The spot positions were set as the position of the centroids of the connected regions. The intensity of each spot was calculated by summing up the total pixel intensity in the vicinity of the centroids (a region of 1.5$\mu$m x 1.5$\mu$m x 1$\mu$m) subtracted from the background intensity, which was extracted from the nearby region but excluding the spots. \AW{The flies were heterozygous for the MS2 reporter, so one spot was visible at a time.} In the (rare) case of multiple spots detected per nucleus, the biggest spot was selected.

For each nucleus, we collected the nucleus' position and the spot intensity over time (here referred to as "traces"). The traces were then classified according to their respective embryos (out of 4 embryos), cell cycle (10 to 13) and position along the AP axis (either anterior or boundary). See  Section~A and Fig.~8 for examples of traces.

\subsection*{Trace preprocessing}

Before the autocorrelation function can be calculated the traces need to be preprocessed.  To ensure that the data captures the dynamics of gene expression in its steady state, for each embryos and each cell cycle, we observed the spot intensity only in a specific time window. The beginning and the end of this window is determined as the moment the mean spot intensity over time of all traces (both at the anterior and the boundary) reaches and leaves an expression plateau (see example in Fig.~9). 

\subsection*{The two state model}

The detailed form of the autocorrelation function in Eq.~\ref{autoF} depends on the underlying gene promoter switching model. For the two state -- telegraph switching model (left panel in Fig.~\ref{Fig1}A) the 
jumping times between the two states are 
both exponential and the dynamics is Markovian. The mean steady state probability for the promoter to be ON is $P_{\rm on}= {k_{\rm on}}/({k_{\rm on}+k_{\rm off}})$, which combined with Eq.~\ref{meanF} gives the form of the mean   fluorescence $\avg{F}$. 
The probability that the gene is ON at time $n$ given that it was on at time $0$ is $A_{n} = P_{\rm on}+ e^{(\delta-1)n} (1-P_{\rm on}) $, where $\delta = 1- k_{\rm on} - k_{\rm off}$. The steady state connected correlation function depends only on the time difference (see  Section~B):
 \begin{equation}
 \tilde C_{\tau}=\avg{F(t) F(t+\tau)}- \avg{F(t)}^2 = \sum_{i,j}  L_i L_j  P_{\rm on}(1-P_{\rm on}) e^{(\delta-1)|\tau-j +i|}.
\label{connected2state}
 \end{equation}
 
\subsection*{The \AW{Cycle} model}
In the cycle model \AW{(also called the three state model in the text)} (center panel in Fig.~\ref{Fig1}A) the OFF period is divided into different sub-steps that correspond to $K$ intermediate states with exponentially distributed jumping times from one to the next. The transition matrix $T$ encodes the rates of this irreversible chain. The probability of the promoter to be in the ON state is: 
\begin{equation}
{P_{\rm on} = \frac{k_{\rm off}^{-1}}{ k_{\rm off}^{-1} + \sum_{m=1}^{K}k_{m}^{-1}},}
\label{multistatepon}
\end{equation}
and that the  steady state connected autocorrelation function is (see  Section~D):
\begin{eqnarray}
\label{multistateauto}
 \tilde C_{\tau}&=&\avg{F(t)F(t+\tau)} - \avg{F(t)}^2  \\ \nonumber
 &=&  { \sum_{i=1}^{r} \sum_{j=1}^{r} L_i L_j P_{\rm on}  \left[\begin{pmatrix}
 1 & 0 & ... & 0 
\end{pmatrix} e^{(T- \text{Id})*|i-j- \tau| } \begin{pmatrix}
 1 \\ 0 \\ ...\\ 0
\end{pmatrix} -
P_{\rm on} \right], }
\end{eqnarray}
where $\tau$ is counted in polymerase steps {, Id is the identity matrix of dimension $K+1$, and the two unit vectors are of dimension $K+1$ }. In the simple case of a two state model Eq.~\ref{multistateauto} reduces to Eq.~\ref{connected2state}.

\subsection*{The \AW{$\Gamma$} waiting time model}
An alternative description of a promoter cycle relies on a reduced description to an effective two state model where we use the fact that the transitions between the states are irreversible. The distribution of times spent in the effective OFF  state $\tau$, is no longer exponential, as it was in the two state model, but it has a peak at nonzero waiting times, which can be approximated by a Gamma distribution
\begin{equation}
\Gamma(\tau)  = \frac{\beta^{\alpha}}{\Gamma(\alpha)} x^{\alpha-1} e^{- \beta x},
\end{equation}
with mean $\alpha/\beta$, where $\beta$ is the scale parameter, $\alpha$ is the shape parameter and $\Gamma (\alpha)$ is the gamma function. The true distribution of waiting times in a cycle model approaches the $\Gamma$ distribution if {the rates $k_i$ } are all the same and $k_{i} <<1$. In this limit $\beta \approx k_{{i}}$, and $\alpha$ describes the number of intermediate OFF states. In the more general case it correctly captures the effective properties of the process. The mean probability of the promoter to be in {the} ON state in the $\Gamma$ waiting time model is given by 
\begin{equation}
P_{\rm on}= (1+\frac{\alpha k_{\rm off}}{\beta})^{-1}.
\label{gammaPon}
\end{equation}
The autocorrelation function cannot be computed directly analytically. The steady state  Fourier transform of the steady state autocorrelation is (see  Section~E):
\begin{eqnarray}
&& \mathcal{F}  ( \avg{F(t) F(t+ \tau)}  - \avg{F(t)}^2) (\xi) 
 =\int_{- \infty}^{+ \infty} d\tau  e^{ - 2 i \pi \tau } ( \avg{F(t)F(t+\tau)} -  \avg{F(t)^2} )\\ \nonumber
  &&= \sum_{k,j} L_k L_j P_{\rm on} 2 \Re \Big[ e^{-2 i \pi (i-j)}  \big[  \Big( k_{\rm off} +2 i \pi \xi  - 
   k_{\rm off} ( 1+\frac{2 i \pi \xi \beta}{\alpha k_{\rm off}} )^{-\alpha} \Big)^{-1} - \frac{P_{\rm on}}{ 2 i \pi \xi }    \big]  \Big].
 \label{fourierauto}
\end{eqnarray}

\subsection*{Finite cell cycle length correction to the connected autocorrelation function}
Due to the short duration of the cell cycle, the theoretical connected correlation functions need to be corrected for finite size effects when comparing them to the empirically calculated correlation functions. 
When analyzing the data we calculate the autocorrelation function from  $M$ traces $\{\boldsymbol{v_{\alpha}} \}_{1 \leq \alpha \leq M}$ of the same length $K$, $ \boldsymbol{v_{\alpha}} = \{ v_{\alpha j} \}_{1 \leq j \leq K} $.
We calculate the connected autocorrelation function for each trace and normalize it to $1$ at the second time point to avoid spurious nucleus to nucleus variability:
\begin{equation}
c_{\alpha} (r) =  \frac{ \sum\limits_{(i,j), |i-j| = r} \left\{  \left( v_{\alpha i} - \frac{1}{K} \sum\limits_{l=1}^{K} v_{\alpha l} \right) \left( v_{\alpha j} - \frac{1}{K} \sum\limits_{l=1}^{K} v_{\alpha l}  \right)  \right\} }{   \dfrac{K-r}{K} \sum\limits_{j=1}^{K} \left( v_{\alpha j} - \frac{1}{K} \sum\limits_{l=1}^{K} v_{\alpha l}  \right)^2  },
\label{insidecr}
\end{equation}
and then average over all $M$ traces to obtain the final connected autocorrelation function: 
\begin{equation}
\label{meanCemp}
c (r) = \frac{1}{M} \sum_{\alpha = 1}^{M} c_{\alpha} (r).
\end{equation}

We define $\bar v=\avg{ v_{i}}$ --  the steady state true theoretical average of the random fluorescence intensity over random realizations of the process, and $\bar{v^2}=\avg{  v^2_{i}}$ -- the true theoretical second moment of the fluorescence signal. When $K \rightarrow \infty$ the average over time points is equal to the theoretical average, $1/K\sum\limits_{i=1}^{K}v_{\alpha i} =\bar{v}$. Using time invariance in steady state the autocorrelation function becomes:
\begin{equation}
C_r = \frac{  \avg{v_{i} v_{ i+r} } - \bar{v}^{2}}{  \bar{v^2} - \bar{v}^{2} },
\label{Ex_conncorr_largeK}
\end{equation}
where $\avg{\cdot}$ is an average over random realizations of the process. Eq.~\ref{Ex_conncorr_largeK} corresponds to the limit we calculated in the theoretical models. To account for the finite size effects that arise due to short time traces, we need to correct for the fact that for short traces $\frac{1}{K} \sum\limits_{i=1}^{K}v_{\alpha i} \neq \bar{v}$ and $\frac{1}{K} \sum\limits_{i=1}^{K}v^2_{\alpha i} \neq \bar{v^2}$, instead both the mean and the variance 
 are functions of $K$. We note that for short traces the definitions of autocorrelation and autocovariance differ:
\begin{multline}
\sum\limits_{(i,j), |i-j| = r} \left\{  \left( v_{\alpha i} - \frac{1}{K} \sum\limits_{l=1}^{K} v_{\alpha l} \right) \left( v_{\alpha j} - \frac{1}{K} \sum\limits_{l=1}^{K} v_{\alpha l}  \right) \right\} \neq 
 \sum\limits_{(i,j), |i-j| = r}  \left(  v_{\alpha i}  v_{\alpha j} - \frac{1}{K^2} \sum\limits_{l=1}^{K} v_{\alpha l} \sum\limits_{m=1}^{K} v_{\alpha m} \right) {.}
\end{multline}

In practice for the analyzed dataset we found that the finite size effects for the variance can be neglected,  
however the mean over time points is a bad approximation to the ensemble mean. We present the finite size correction to the mean below. For completeness we include the finite size correction for the variance in  Section~I, although we do not use it in the analysis due to its numerical complexity and small effect.

If the variance of the normalized fluorescence intensity over random realizations of the process is well approximated by the average over the $K$ time points, we can replace the denominator in Eq.~\ref{insidecr} by $\bar{v^2} - \bar{v}^{2}$ and in steady state evaluate the mean connected autocorrelation function (see  Section~H for details):
\begin{eqnarray}
\label{finalcorrected}
c(r) &=& \frac{1}{ \bar{v^2} - \bar{v}^{2}} \Big[ \tilde C_r + \frac{1}{K} \left( \frac{1}{K} - \frac{2}{(K-r)} \right) \left(K \tilde C_0 + \sum_{k=1}^{K-1} 2 (K-k) \tilde C_k \right)  \\ \nonumber
&&+ \frac{2}{K(K-r)} \big(  r \tilde C_0 + \sum_{k=1}^{r-1} 2 (r-k) \tilde C_k +
 \sum_{m=1}^{K-1} \tilde C_m [ \min(m+r,K) - \max(r,m)]  \big)   \Big]
\end{eqnarray}
where $\tilde C_k = \avg{ v_iv_{i+k}}$ is the {infinite-size} steady state non-connected correlation function of the process \AW{(given in Eq.~\ref{connected2state} for the two state model, Eq.~\ref{multistateauto} for the cycle model and as the Fourier transform of Eq.~\ref{fourierauto} for the $\Gamma$ model)} and the average is over random realizations of the process. \AW{The mean and variance of the signal, $\bar{v}$ and $\bar{v^2}$, provide a normalization factor that is constant for all time differences $r$. We normalize the autocorrelation function setting the second term to $1$ and these terms are not needed for the inference.} If $v_i = X(i)$ then $C_k$ is proportional to $A(k)$. 

\subsection*{Inference}

The inference proceeds in three steps:

{Step 1. Signal calibration.} The intensity of the measured signal depends on a constant trace{-}dependent offset value $I_0$, $I(t) = \sum_{i=1}^{r} I_{0} {a(i,t)} L_i$. To calibrate this offset we take the maximum expression to be the mean of the maximun expression over all traces in a given region 
$I_{\rm max} = \avg{\max_{t} I(t)} = I_{0} \sum_{i=1}^{r} L_i$. \AW{ We take the mean of the maxima of the intensities rather than the absolute maximum of all signals to avoid errors from overestimating the maximum.} The calibrated fluorescence signal used in the analysis is then $F(t) = {I(t)}/{I_{0}} = \sum_{i=1}^{r} \Jon{a(i,t)} L_i$. $P_{\rm on}$ is directly calculated using Eq.~\ref{meanF}. \AW{$L_i$ is a known function.}

{Step 2. Estimating parameter ratios.} \AW{For the two state,  three state cycle and $\Gamma$ models, the ratios of the rates can be estimated directly from the steady state mean fluorescence values using Eqs.~\ref{multistatepon} and \ref{gammaPon}. The Poisson model is uniquely defined by $P_{\rm on}$ and does not require further parameter inference beyond Step 1.}

{Step 3. Estimating parameters.} Using the estimate for the ratio of the rates, the ON and OFF rates are found by minimizing the mean squared error between \AW{the autocorrelation function calculated from the data (Eq.~\ref{meanCemp}),} and the model \AW{(Eq.~\ref{finalcorrected} with the theoretical prediction for the appropriate model: Eq.~\ref{connected2state} for the two state, Eq.~\ref{multistateauto} for the cycle model and the Fourier transform of Eq.~\ref{fourierauto} for the $\Gamma$ model). } 

\section*{Acknowledgments}
We thank T. Mora for helpful discussions. This work\ was supported by a Marie Curie MCCIG grant No. 303561(AMW), PSL IDEX REFLEX (ND, AMW, MC), ARC PJA20151203341 (ND), ANR-11-LABX-0044 DEEP Labex (ND), ANR-11-BSV2-0024 Axomorph (ND and AMW) and PSL ANR-10-IDEX-0001-02.

 % sections/SItextnodata.tex

\section{Autocorrelation analysis}
\subsection{Basic setup and data preprocessing}
\label{basicsetup}

The raw data produced experimentally is a fluorescent signal $I(t)$ measured at discrete times corresponding to the sampling time frame of the movie (see Fig.~\ref{trace_example} for examples of traces). 
At each locus and at each time point it is the sum of the background signal and a number of fluorescent molecules attached to loops formed by the mRNA. Each loop contributes to the signal by a constant $I_0$. This constant is unknown and can vary from trace to trace due to noise in the experimental setup and the variability in the locations of the nuclei in the embryo. All models are written for the renormalized signal $F(t) = I(t)/I_0$. 

Because the fluorescent signal is produced by discrete polymerases that travel down the gene, we divide the gene into chunks of $150$ base pairs, a length that corresponds to the irreducible space occupied by a polymerase on the gene (Fig.~{4} in the main text). The positions the polymerase can occupy on the gene are labeled by an index $1 \leq i \leq r$. 
The number of MS2 loops that have been formed by a polymerase that has reached a given position depends only on the MS2 gene construct and we define a deterministic function $L_i$ for the whole length of the gene that describes the number of MS2 loops that have been produced by a polymerase at position $i$. In practice the exact number of loops is not an integer and varies from base pair to base pair so we take $L_i$ as the average number of loops at this polymerase position (see Fig.~{4} in the main text). 

When the gene is fully loaded with polymerases (the number of polymerases is equal to the length of the gene divided by $150$ bp), the fluorescence intensity is $I(t) = I_0 \sum_i^r L_i$. Assuming that the maximum of the signal over the whole trace is a good approximation for the fully loaded value we can determine $I_0$ and renormalize the data. In practice, since we see variability in the expressed signal in  different nuclei at the same position, we are not sure the fully loaded polymerase scenario occurs in each nuclei, so we take the mean of the maximum intensity values in the anterior. We use this renormalized fluorescence signal to infer the parameters of the dynamics.

The experimental data is analyzed assuming the system is in steady state and does not take into account the initial activation period after mitosis, and the end of the trace when the gene is deactivated before mitosis. \AW{We take only a window of the traces where the mean spot intensity in all traces is stable (see Fig.~\ref{data_calibration} for an example). As the duration of the interphase differs slightly between embryos, we use a different steady state window for each embryo as summarized in Table \ref{TableWindow}}.

In all models based on a stochastic gene switching (so all models except the Poisson-like model) we assume that the gene can be in several states with only two effective transcription rates: a non zero transcription rate in the ON state and an basal production rate equal to zero in the  OFF state. When the gene is ON the polymerase loads at a maximal rate set by clearing of the binding site by the previous polymerase, which is one polymerase every $6$ seconds (calculated as the irreducible polymerase length along the gene $150$ bp divided by the polymerase speed, $v=25 bp/s$). The state of the gene is described by a stochastic process $X(t)$ that is equal to $1$ when the gene loads polymerase (i.e is ON) and $0$ when the gene is OFF (see Fig.~1B in the main text). Once the polymerase is loaded its path is assumed to be deterministic with constant speed.

The gene can be described by the locations where there is a polymerase: we define $a(i,t)$ as a function of time $t$ and position $1\leq i \leq r$ that is equal to $1$ if there is polymerase at position $i$ at time $t$ and $0$ otherwise (see Fig.~1D in the main text). The fluorescence signal is then a convolution of the polymerase position, $a(i,t)$, and the details of the loop design of the MS2 construct, $L_i$:
\beq
F(t) = \sum_{i=1}^r L_i a(i,t),
\eeq
and the polymerase position can easily be translated back to the gene state through the deterministic relation, $a(i,t)=X(t-i)$ (see Fig.~3D in the main text for the form of $L_i$).
This disruption is exact for a system with a discrete regulatory process and a discrete time step equal to the polymerase time step. Unfortunately, the moments in time when the gene switches are not necessarily multiples of the natural coarse graining steps of the system (the polymerase time step and its equivalent length) 
so it is necessary to introduce a continuous time in the system. We will present results for both the discrete and continuous time models. The continuous description is valid in the limit where the typical time spen{t} by the gene in each state is long compared to the polymerase step or equivalently the gene switching constants are small compared to $1/6$ s$^{-1}$. See  Section~\ref{apptwo} for a more detailed argument. 

\begin{figure}[h]
\includegraphics[width=0.45\textwidth]{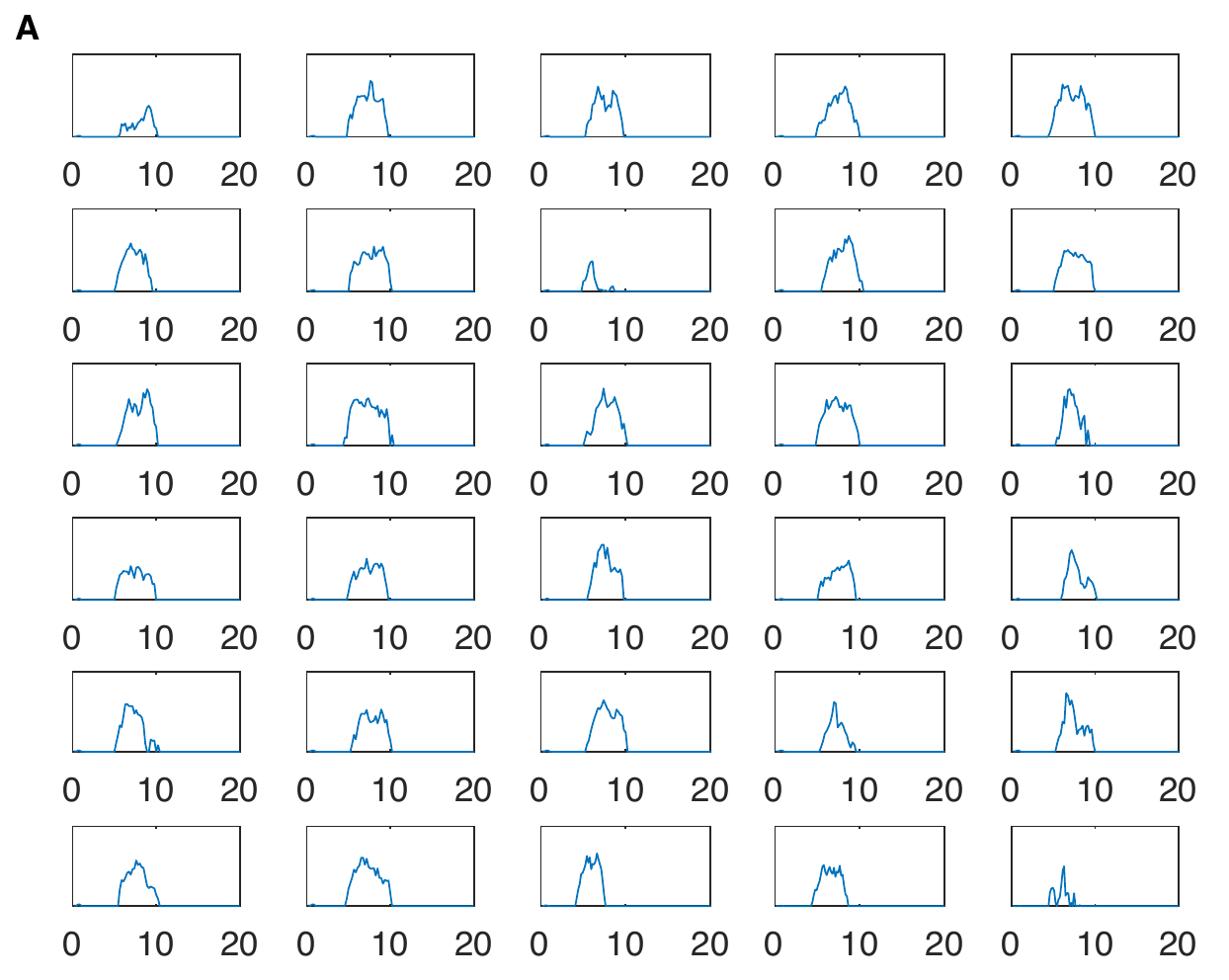} 	
\includegraphics[width=0.45\textwidth]{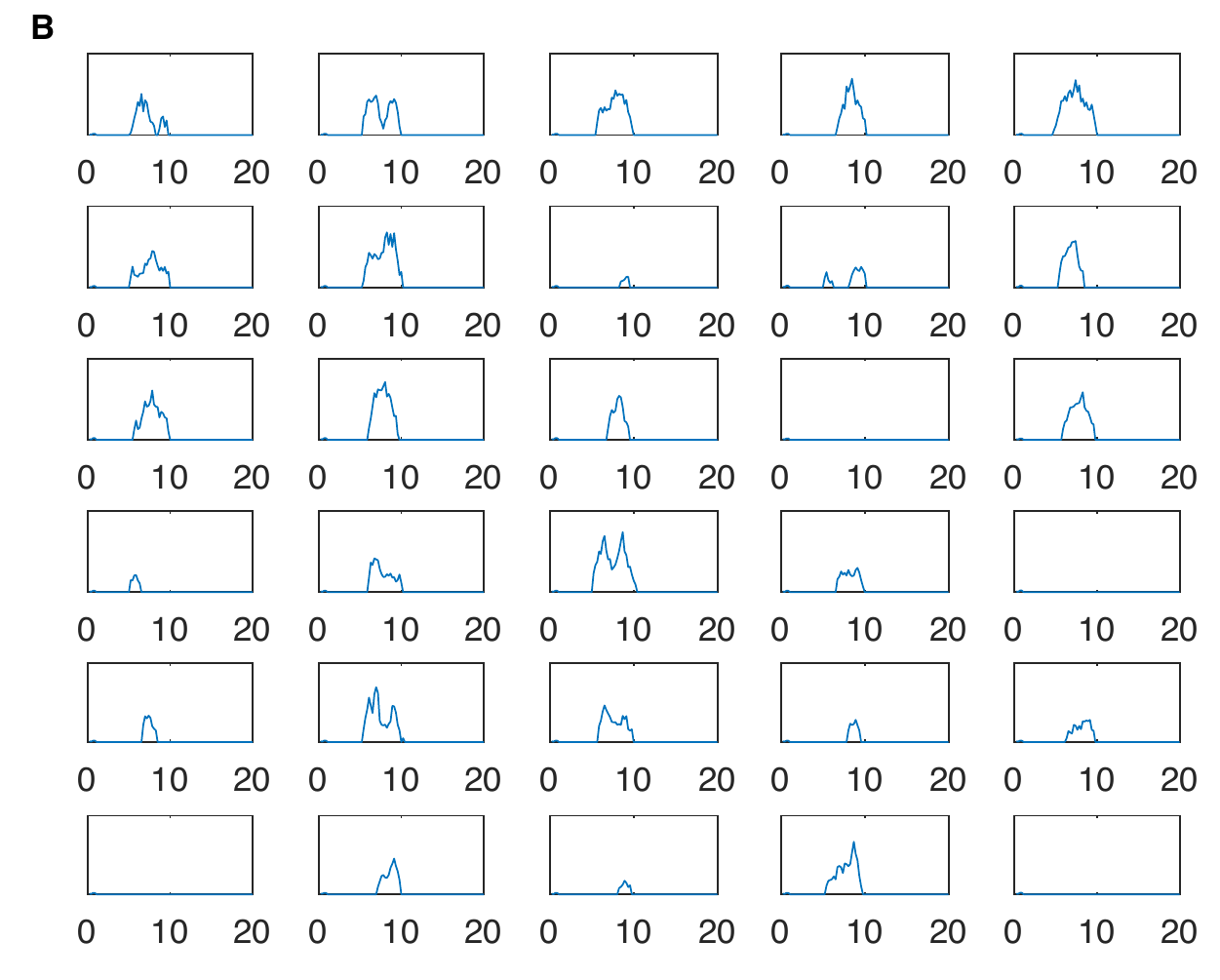}	
\includegraphics[width=0.45\textwidth]{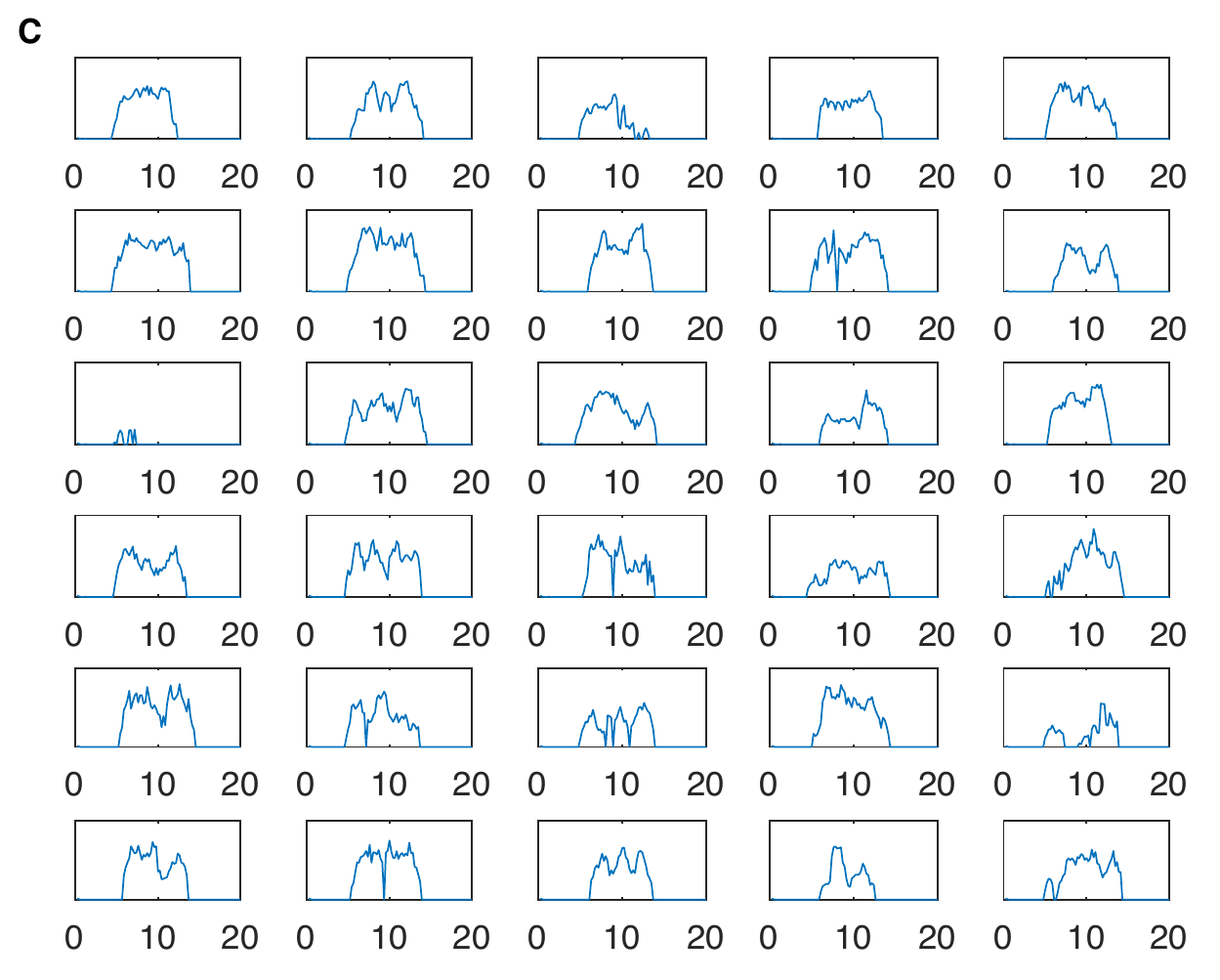} 	
\includegraphics[width=0.45\textwidth]{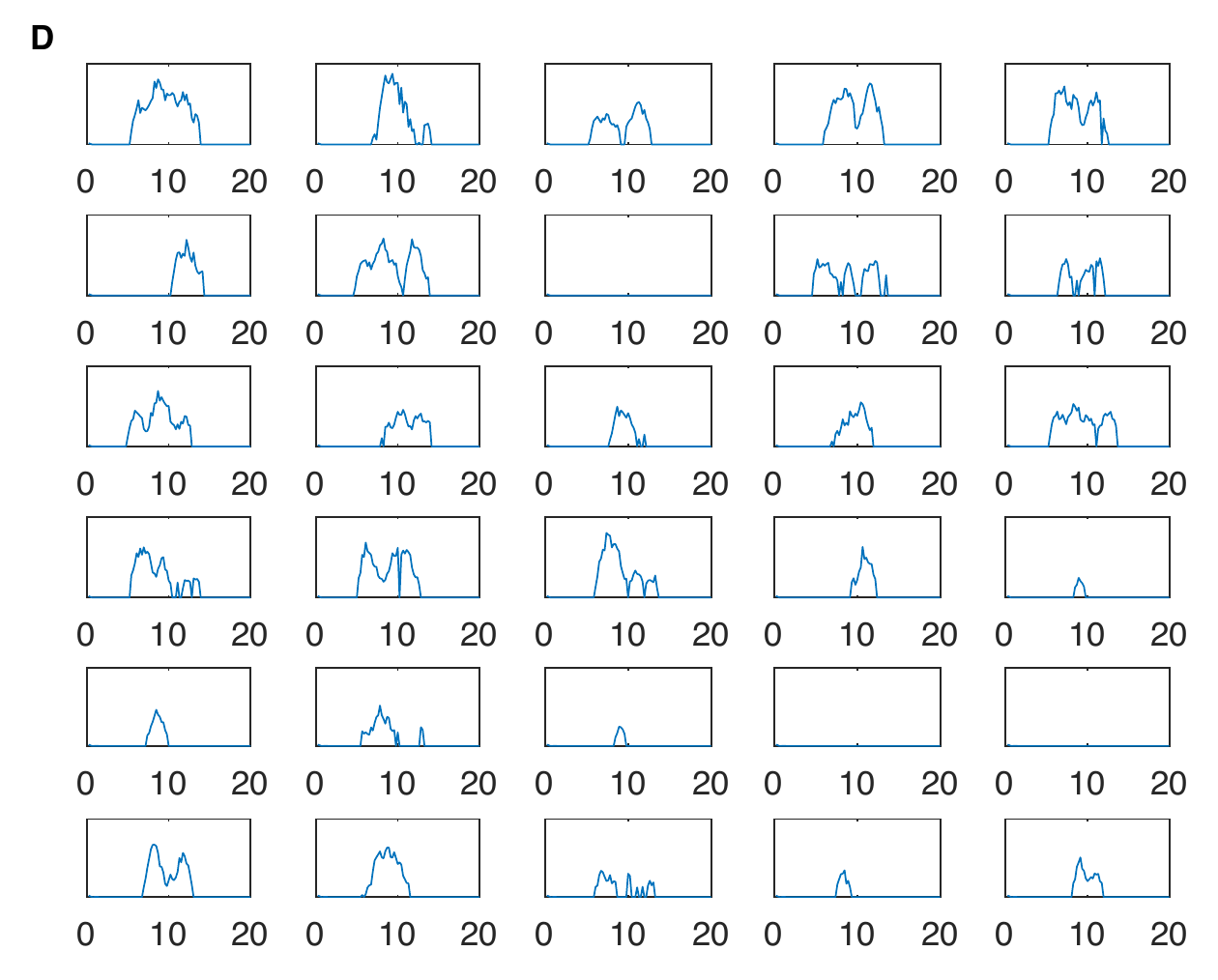}	
\caption{{\bf Examples of individual spot intensity over time}. Consecutively shown are the traces in {(A)} Cycle 12, Anterior, {(B)} Cycle 12, Boundary {(C)} Cycle 13, Anterior, {(D)} Cycle 13, Boundary. The x axis is time in minute and y axis is the spot intensity in AU. \label{trace_example}}
\end{figure}

\begin{figure}[h]
\includegraphics[width=0.8\textwidth]{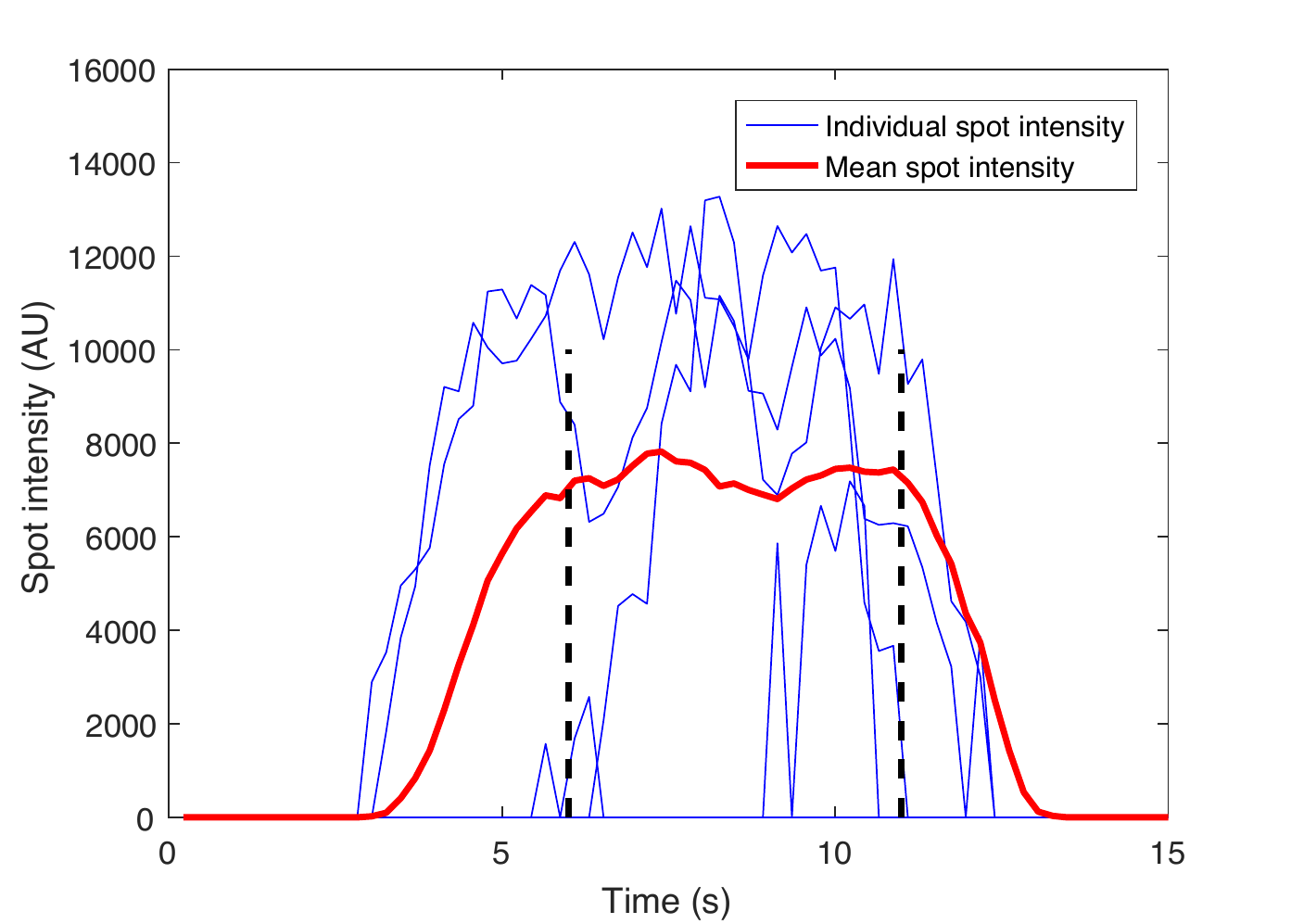}	
\caption{\textbf{Data calibration}. Shown are examples of 5 (out of 154) individual traces (blue) taken from embryo 1, cycle 13. Also shown is the mean spot intensity over time of all traces (red). The steady state window is chosen to be from the 6$^{th}$ minute to the 11$^{th}$ minute (dashed lines).\label{data_calibration}}
\end{figure}

\HT{
\begin{table}
\begin{center}
\begin{tabular}{ c | c | c | c | c}
  		 			& beginning (s)	& end (s)	& interphase duration (s)		\\ \hline
  Embryo 1 - cc 12 	& 278		& 391 		& 652		\\
  Embryo 1 - cc 13 	& 318 		& 546 		& 796		\\
  Embryo 2 - cc 12 	& 281		& 393 		& 570		\\
  Embryo 2 - cc 13 	& 278 		& 484 		& 695		\\
  Embryo 3 - cc 12 	& 297		& 408 		& 676		\\
  Embryo 3 - cc 13 	& 330 		& 587 		& 824		\\
  Embryo 4 - cc 12 	& 261		& 515 		& 616		\\
  Embryo 4 - cc 13 	& 321 		& 529 		& 751		\\
\end{tabular}
\end{center}
\caption{The steady-state window for the autocorrelation analysis: Shown for each embryo and cell cycle are the beginning and ending times of the steady state window, and the duration of interphase.\label{TableWindow}}
\end{table}
}

\begin{figure}[h]
\includegraphics[width=0.45\textwidth]{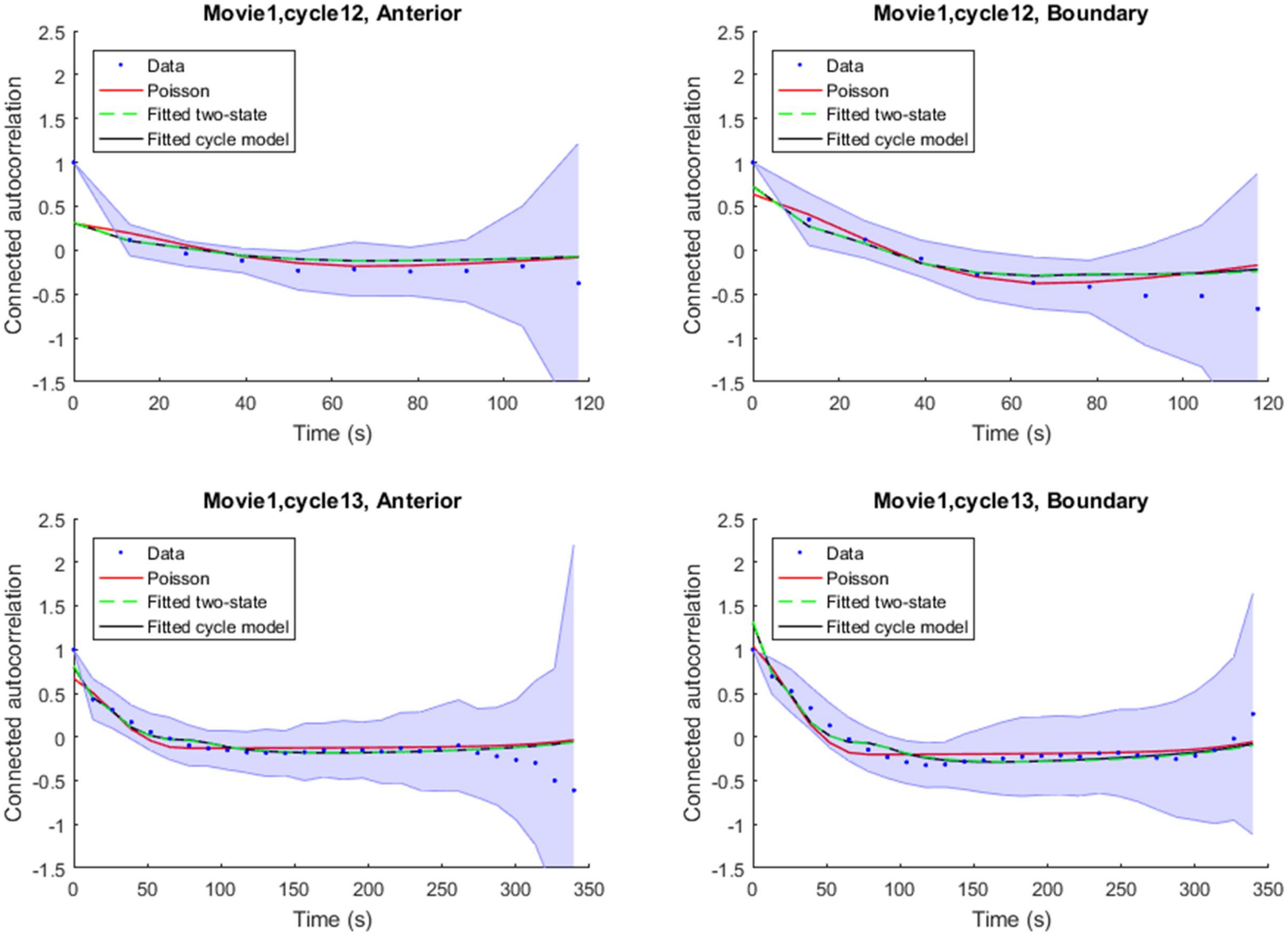} 	
\includegraphics[width=0.45\textwidth]{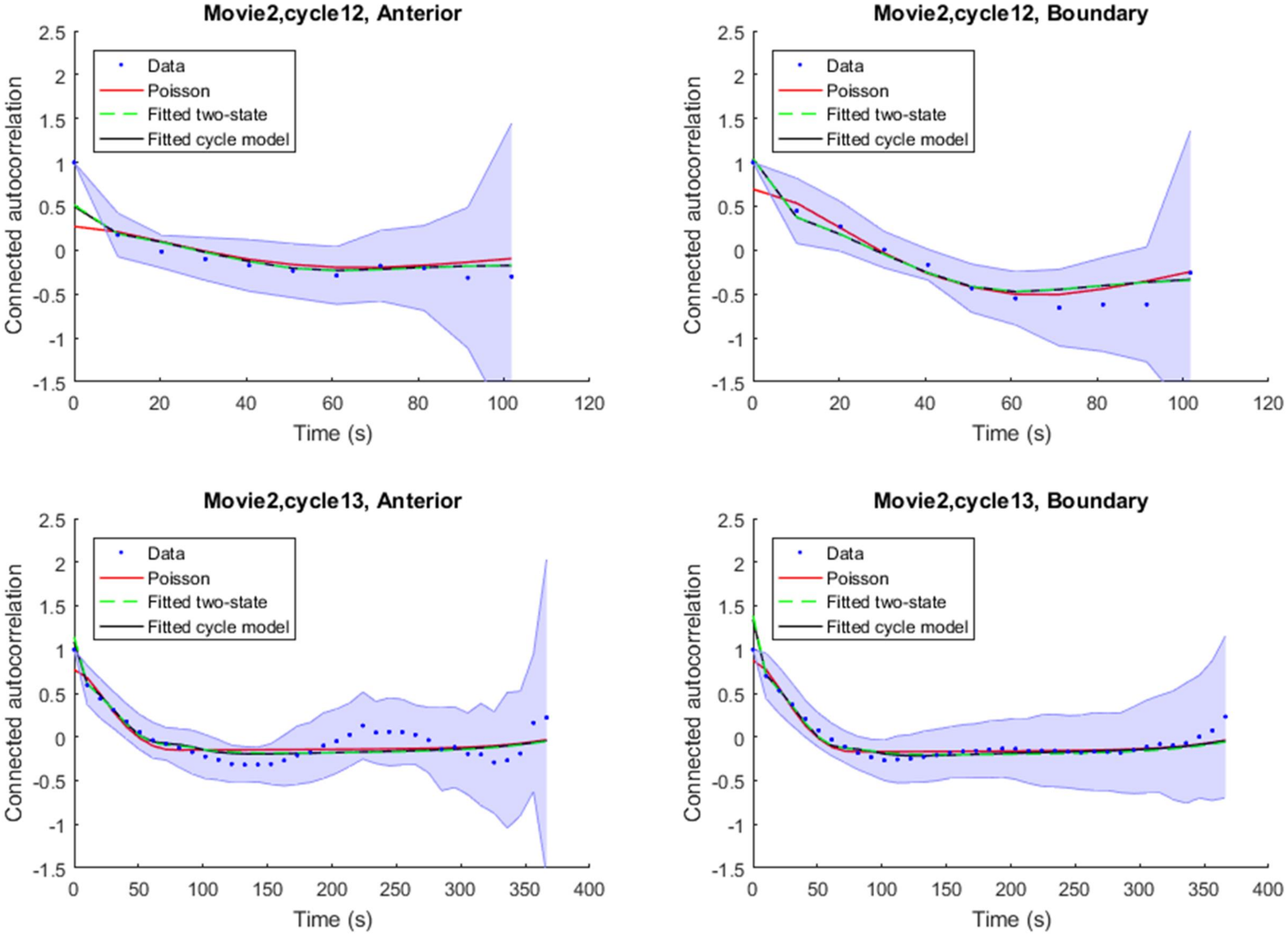}	
\includegraphics[width=0.45\textwidth]{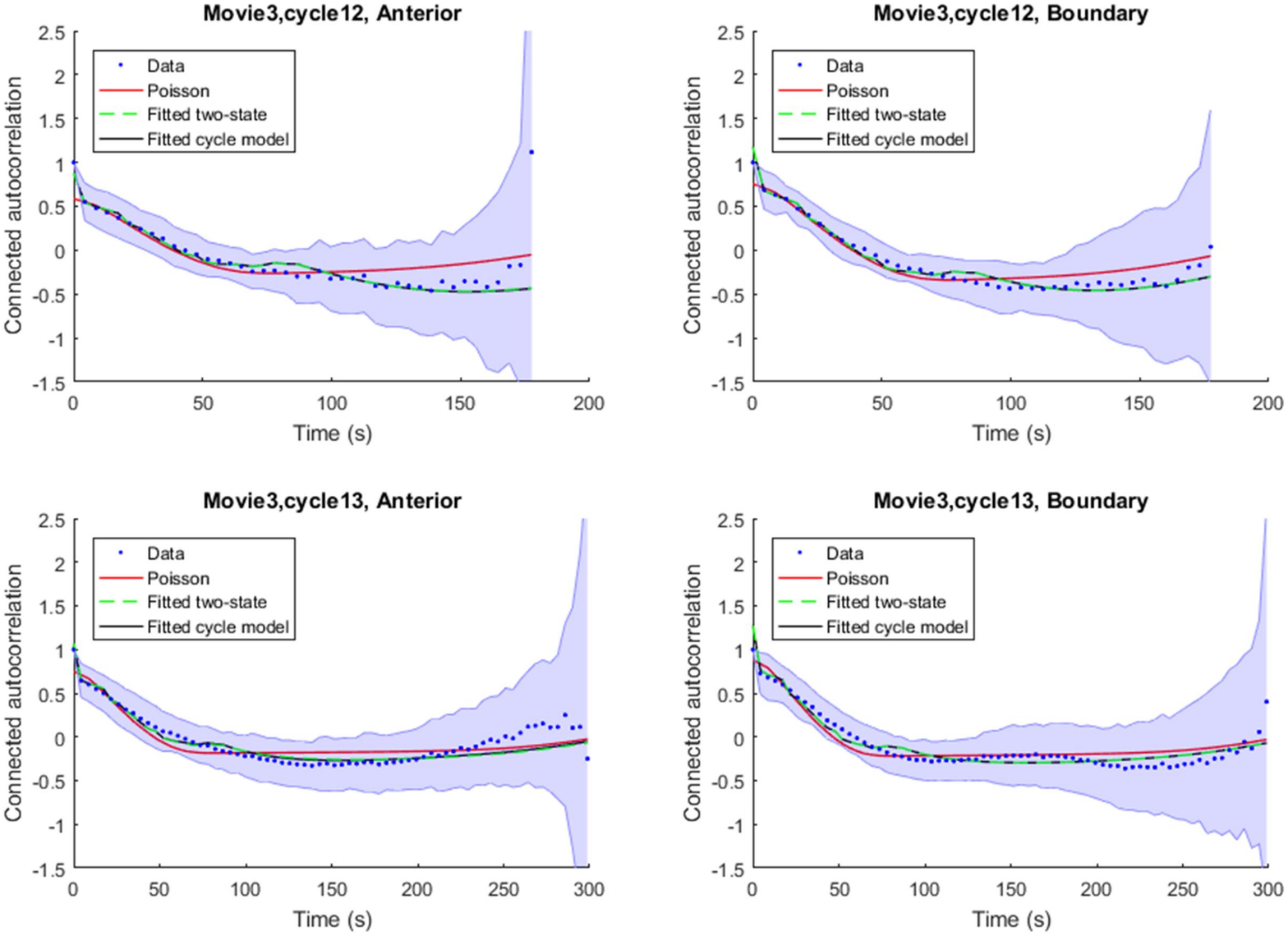} 	
\includegraphics[width=0.45\textwidth]{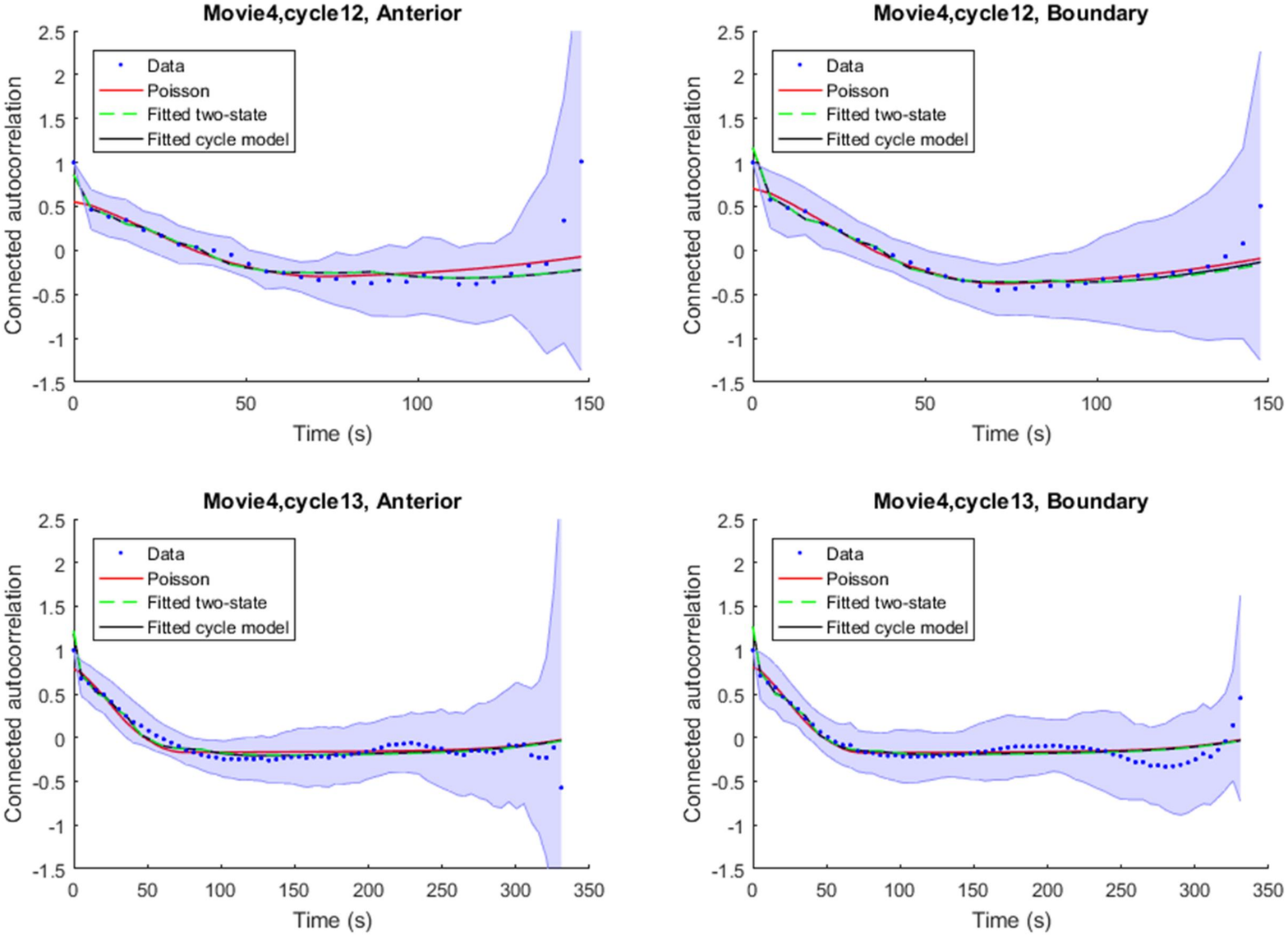}	
\caption{\textbf{Fits of the autocorrelation function}. The empirical autocorrelation function (blue dots) for both the anterior and boundary regions in all four embryos is fit using the autocorrelation function with the  finite size corrections for the Poisson-like model (red lines),  two{-}state model (green lines) and three{-}state cycle model (black lines). \label{allfits}}
\end{figure}

\begin{figure}
\begin{center}
\noindent\includegraphics[width=0.75\linewidth]{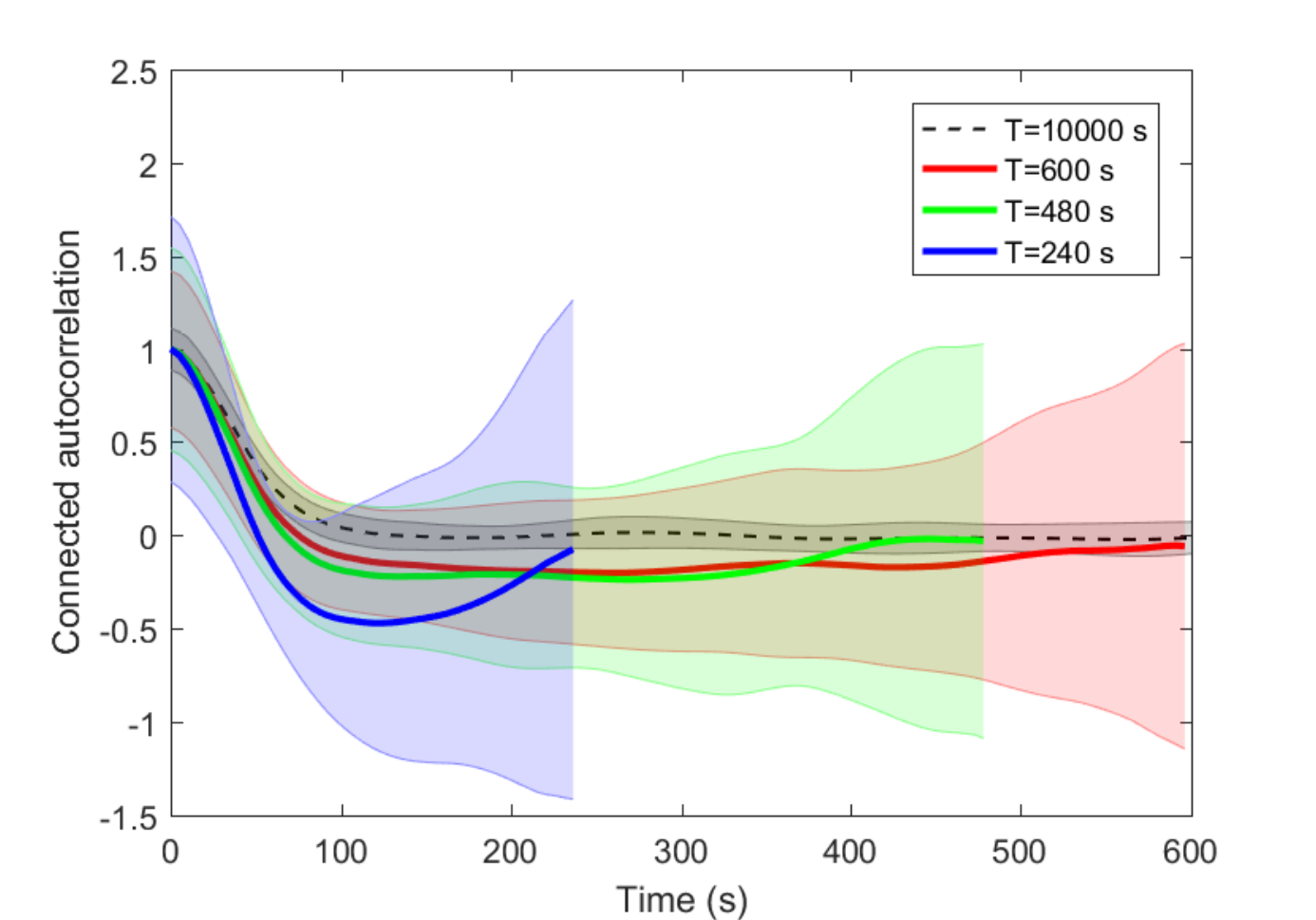}
\caption{\textbf{ Example of the connected autocorrelation function for the two state model calculated for different trace lengths T.} The shaded areas denote the standard variation over 500 simulated traces. The switching rates $k_{\rm on}= k_{\rm off}= 0.01 s^{-1}$.   \label{figSI3} }
\end{center}
\end{figure}

\begin{table}
\begin{center}
\begin{tabular}{ c | c | c | c | c}
  kon($1/s$)	& mov1 		& mov2		& mov3 		& mov4  \\ \hline
  12A 			& 0.078		& 0.056 	& 0.009		& 0.023 \\
  12B 			& 0.004 	& 0.005 	& 0.003		& 0.011	\\
  13A 			& 0.017		& 0.020 	& 0.014		& 0.021 \\
  13B 			& 0.004		& 0.006		& 0.004		& 0.005 \\
\end{tabular}
\end{center}
\caption{The inferred $k_{\rm on}$ rates from the autocorrelation approach assuming a two{-}state model for the four embryos and cell cycle 12 and 13, in the anterior and boundary.\label{Tablekon}}
\end{table}

\begin{table}
\begin{center}
\begin{tabular}{ c | c | c | c | c}
  koff($1/s$)	& mov1 		& mov2		& mov3 		& mov4  \\ \hline
  12A 			& 0.060		& 0.088 	& 0.008		& 0.019 \\
  12B 			& 0.020 	& 0.034 	& 0.021		& 0.051	\\
  13A 			& 0.018		& 0.031 	& 0.016		& 0.027 \\
  13B 			& 0.031		& 0.054		& 0.031		& 0.064 \\
\end{tabular}
\end{center}
\caption{The inferred $k_{\rm off}$ rates from the autocorrelation approach assuming a two{-}state model for the four embryos and cell cycle 12 and 13, in the anterior and boundary.\label{Tablekoff}}
\end{table}

\subsection{The two{-}state model}
\label{apptwo}

In this section we derive the equations required for the inference of the dynamics under the assumption that the gene can be in two states: ON or OFF{,} represented by a two dimensional vector $x(t)=[x_{\rm on}(t), x_{\rm off}(t)]$. $x_{\rm on}(t)$ is the probability of the gene to be ON and $x_{\rm off}(t)$ is the probability of the gene to be OFF. $x_{\rm on}(t)$ is the average over traces of the random variable $X(t)$ depicted in Fig.~1B of the main text. 
We assume that the switching times between the two are exponentially distributed:
\beq
\partial_t \begin{pmatrix}
x_{\rm on}   \\
x_{\rm off}
 \end{pmatrix}
=
\begin{pmatrix}
  -k_{\rm off}  & k_{\rm on}  \\
  k_{\rm off} &  -k_{\rm on}
 \end{pmatrix}
  \begin{pmatrix}
 x_{\rm on}   \\
x_{\rm off}
 \end{pmatrix}.
 \label{xeqexp}
\eeq
The steady state probability to be ON is $P_{\rm on}=x_{\rm on}(t=\infty)=1/T\sum_{t} x_{\rm on}(t)$, where $T$ is the duration of the steady state window in Fig.~\ref{data_calibration}, and is:
\begin{equation}
\label{ratiotwostate}
\frac{k_{\rm on}}{k_{\rm off}}= \frac{P_{\rm on}}{1-P_{\rm on}}.\end{equation}
 We learn $P_{\rm on}$ from Eq.~1 in the main text:
\beq
\avg{F} = P_{\rm on} \sum_{i=1}^{r} L_i.
\label{SImeanF}
\eeq
and use it to obtain the ratio of the switching rate{s} from Eq.~\ref{ratiotwostate}.

The autocorrelation function is:
\begin{equation}
\avg{F(t) F(s)} = \sum_{i=1}^{r} \sum_{j=1}^{r} L_i L_j \avg{a(i,t) a(j,s)},
\label{firstauto}
\end{equation}
where the brackets are an average over traces (different realizations of the random process). We define $A(t-i,s-j) = 1/x_{\rm on}(s-j) \avg{a(i,t) a(j,s)}$ --  the probability that the polymerase is at position $i$ and time $t$ given that there was a polymerase at position $j$ at time $s$ (here we assume that $ t-i \geq s-j $).  Using the deterministic relation between the polymerase position at a given time $a(i,t)$ and the probability to be on at an earlier time $X(t-i)$, $A(t-i,s-j)$ is equivalent to the probability that the gene in ON at time $t-i$ given that is was ON at time $s-j$:
\begin{equation}
\label{Adef}
A(t-i,s-j)= x_{\rm on}(t-i  | \text{ ON at time} \ s -j).
\end{equation}
Plugging the expression into Eq.~\ref{firstauto} we obtain Eq.~2 in the main text: {
\begin{eqnarray}
\avg{F(t) F(s)}= & \sum_{i=1}^{r} \sum_{j=1}^{r} L_i L_j x_{\rm on}(\min(s-j,t-i)) A(t-i,s-j).
\label{secondauto}
\end{eqnarray}
}

In steady state the system is translationally invariant  $A(t-i,s-j) = A(\abs{t-i-s-j})$ and for brevity we will denote it as $A(n)$ - the probability that the gene is ON at time $n$, given that it was ON at time $0$. To find $A_n$ we need to solve for $x(t)$:
\begin{equation}
\partial_t x(t) =(T-\mathbb{1} ) x(t),
\end{equation}
where $T-\mathbb{1}$ is given by Eq.~\ref{xeqexp} and calculate the expectation value that the gene is ON at time $t$ given in was ON initially:  
\begin{equation}
\label{atwostate}
A_{n} = 
\begin{pmatrix}
  1 & 0
 \end{pmatrix}
e^{n(T-\mathbb{1})}
 \begin{pmatrix}
  1 \\
  0
 \end{pmatrix}.
\end{equation}
Eq.~\ref{atwostate} is correct in a continuous time model. Its discrete time equivalent is 
\begin{equation}
\label{atwostatediscrete}
A_{n} = 
\begin{pmatrix}
  1 & 0
 \end{pmatrix}
T^{n} 
 \begin{pmatrix}
  1 \\
  0
 \end{pmatrix}.
\end{equation}
In the limit of $k_{\rm on}$ and $k_{\rm off}$ much smaller than the polymerase step they are also much smaller than $1$ and 
$e^{n(T-\mathbb{1})} \simeq \mathbb{1} + n(T-\mathbb{1}) \simeq (\mathbb{1} + (T-\mathbb{1}))^n $. In this limit the continuous and discrete time descriptions of Eq.~\ref{atwostate} and Eq.~\ref{atwostatediscrete} are equal.

The eigenvalues of $T-\mathbb{1}$ are $[1, \delta]$, where $\delta = 1-k_{\rm on}-k_{\rm off}$ with corresponding eigenfunctions:
\beq
\begin{pmatrix} P_{\rm on} \\ P_{\rm off} \end{pmatrix}, \begin{pmatrix} 1 \\ -1 \end{pmatrix}.
\eeq
The transition matrix $T$ is
\begin{equation}
T= \frac{1}{P_{\rm on} + P_{\rm off}}
\begin{pmatrix}
P_{\rm on} & 1 \\ P_{off } & -1
\end{pmatrix}
\begin{pmatrix}
1 & 0 \\ 0 & \delta
\end{pmatrix}\begin{pmatrix}
1 & 1 \\ P_{off } & -P_{\rm on}
\end{pmatrix}
\end{equation}
and  
\begin{equation}
e^{n(T-\mathbb{1})}= 
\begin{pmatrix}
P_{\rm on} + e^{n(\delta-1)} P_{\rm off} & P_{\rm on} - e^{n(\delta-1)} P_{\rm on} \\ P_{off } - e^{n(\delta-1)} P_{\rm off} & P_{\rm off} + e^{n(\delta-1)} P_{\rm on}
\end{pmatrix}
\end{equation}
resulting in
\begin{equation}
\label{onontrans}
A_{n} = P_{\rm on}+ e^{n(\delta-1)} P_{\rm off}. 
\end{equation}

In steady state $x_{\rm on}(s-j)=P_{\rm on}$ and the connected autocorrelation is:
\begin{eqnarray}
\label{aut2s}
\tilde C_{\tau}=<F(t) F(t+\tau)> -<F(t)>^2= \sum_{i=1}^r \sum_{j=1}^r  & L_i   L_j P_{\rm on}     P_{\rm off} e^{|\tau-j +i|(\delta-1)}.
 \end{eqnarray}
 Since we already know the ratio of the rates from $P_{\rm on}$, inferring $\delta$ using Eq.~\ref{aut2s} determines $k_{\rm on}$ and $k_{\rm off}$.

 \subsection{Computing out of steady state}
The autocorrelation approach can be generalized to a case when the system is out of steady state, when the  autocorrelation function explicitly depends on the two time points and not only on their difference. During mitosis the gene is OFF and then gets turned ON in early interphase. Motivated by the {{\it hunchback}} expression we will present the calculation assuming the gene is initially {OFF}, but it is generalizable to any other initial condition. Assuming $t-i<s-j$, we want to calculate the probability that the polymerase is at position $i$ at time $t$, given that it was at position $j$ at time $s$. Since the gene is initially OFF, we need to calculate the probability that the gene is ON at time $t-i$. The autocorrelation function of the polymerase position is:
\begin{equation}
\avg{a_{i}(t) a_{j}(s)} = 
\begin{pmatrix}
  1 & 0
 \end{pmatrix}
e^{(s-t+i-j)(T-\mathbb{1})} 
 \begin{pmatrix}
  1 \\
  0
 \end{pmatrix}
\begin{pmatrix}
  1 & 0
 \end{pmatrix}
e^{(t-i)(T-\mathbb{1})} 
 \begin{pmatrix}
  0 \\
  1
 \end{pmatrix}.
\end{equation}
Using Eq.~\ref{onontrans} and
\begin{equation}
\begin{pmatrix}
  1 & 0
 \end{pmatrix}
e^{n(T-\mathbb{1})} 
 \begin{pmatrix}
  0 \\
  1
 \end{pmatrix}
=P_{\rm on}(1 - e^{n(\delta-1)}),
\end{equation}
we obtain:
\beq
\avg{F(t) F(s)}   = \sum_{i=1}^r \sum_{j=1}^r  L_i L_j P_{\rm on} (1- e^{(\delta-1)\min(t-i,s-j)}) (P_{\rm on} + P_{\rm off} e^{|s-j -t+i|(\delta-1)}) {.}
\eeq

 \subsection{Multiple off states}
 
 The calculations presented in Appendix \ref{apptwo} can be extended to models that include more OFF or ON states as long there are only two production states for the mRNA: one enhanced and one basal production state. The transition matrix $T$ will then be of higher dimension and in practice should be (and has to be for dimensions larger than 3) diagonalized numerically. The exact analytical solution for the autocorrelation function is still valid written in terms of the powers of $T$.

\subsection{Generalized multi step model}
A gene with many OFF states can also be described using a reduced model with two effective gene expression states ON and OFF, where the times of transitions between these two state are not exponential but follow a {peaked} distribution approximated by a Gamma distribution. The Gamma distribution describes an effective transition over many irreversible transitions between a series of OFF states:
\beq
\label{GammaSI}
 \Gamma_{\alpha,\beta} (x) = \frac{\beta^\alpha}{\Gamma (\alpha)} x^{\alpha -1} e^{-\beta x},
\eeq
where $\beta$ is the scale parameter, $\alpha$ is the shape parameter, and $\Gamma (\alpha)$ is the gamma function. The mean time spent in the OFF state is $ 1/k^{\rm eff}_{\rm on}=\alpha/\beta$, so the probability for the gene to be in the ON state is:
\begin{equation}
 P_{\rm on}  = \frac{k^{\rm eff}_{\rm on}}{k^{\rm eff}_{\rm on}+k_{\rm off}}=  \frac{1}{1+{\alpha k_{\rm off}}/\beta}.
 \end{equation}
 This model has three parameters, regardless of the number of OFF states, and using Eq.~1 of the main text reduces the number of parameters to two, which greatly simplifies the inference. The remaining two parameters are learned from the  autocorrelation function in Eq.~\ref{firstauto}, which formally has the same form as Eq.~\ref{secondauto}: {
 \begin{eqnarray}
\avg{F(t) F(s)}= & \sum_{m=1}^{r} \sum_{n=1}^{r} L_m L_n x_{\rm on}(\min (t-m , s-n)) A_{\Gamma}(|s-n-t+m|),
\label{gammaauto}
\end{eqnarray}
}but $ A_{\Gamma}(|s-t+m-n|)$ {is now the two-point correlator of a non-Markovian process.}
  We limit our presentation to the steady state, but the calculation generalizes to out of steady {state} systems.

We cannot solve the problem in real space, but we compute the Fourier transform of the autocorrelation function of the fluorescence signal:
\beq
 \hat{C}(\xi)  = \int_{-\infty}^{+\infty} d \tau (\avg{F(t) F(t+\tau)} - \avg{F(t)}^2)e^{- 2 i \pi \tau \xi},
\eeq
which using Eq.~\ref{firstauto}
\beq
              \hat{C}(\xi)        = {P}_{\rm on} \sum_{m,n} L_m L_n 2 \Re \left[ e^{-2 i \pi (m-n)} \hat{A}_{\Gamma}^*(\xi) \right] 
              \eeq
we reduce to calculating 
\beq
 \hat{A}_{\Gamma}^*(\xi) = \int_{0}^{+ \infty} dt e^{- 2 i \pi t \xi}  ( A_{\Gamma}(t) -P_{\rm on}).
 \eeq

We decompose $A_{\Gamma}(t)$ into a sum over full cycles of the gene turning from ON to OFF, with the constraint that at time $t$ the gene is ON: 
\beq
  A_{\Gamma}(t)  = \sum_{k=0}^{\infty} A_{\Gamma k} (t),
  \eeq
where 
\beq
A_{\Gamma k}(t) = x_{\rm on}(t | \text{ ON at time} \ {0} \  \text{\& process has gone though  k cycles}  ) . 
\eeq
Since the first jump is from the ON to OFF, which is exponential it contributes $A_{\Gamma 0} (t) = e^{-k_{\rm off} t}$.

First we compute an auxiliary probability distribution function of the time it takes the process to go through a full ON-OFF cycle $\eta(t)$ of taking an exponential jump out of the ON state followed by a Gamma distributed jump out of the OFF state:
\beq \eta(t)  = \int_{0}^t dx k_{\rm off} e^{-k_{\rm off}x}  \frac{\beta^\alpha}{\Gamma (\alpha)} (t-x)^{\alpha -1} e^{-\beta(t- x)}.
\eeq
The Fourier transform of this distribution is:
\beq
 \hat{\eta} (\xi) = \int_{0}^{+ \infty} dt e^{-2 i \pi \xi t} \eta(t) = \frac{k_{\rm off}}{2 i \pi \xi +k_{\rm off}} \frac{\beta^{\alpha}}{(2 i \pi \xi +\beta)^\alpha}.
 \eeq

To compute $\hat{A}_{\Gamma}^*(\xi)$  we need to sum over all the possible times at which the cycles could have occurred, with the constraint that at time $t$ the gene is ON: 
\beq
\label{agmmainte}
\hat{A}_{\Gamma}^*(\xi) = \int_{0}^{+\infty} dt e^{- 2 i \pi  \xi t} \left[\sum_{k=0}^{\infty} \left(  \int_{ t_{i} >0,\sum_{i=1}^k t_{i} < t } e^{-k_{\rm off} (t- \sum_{i} t_{i})}  
                  \prod_{i=1}^k \eta(t_{i}) dt_{i}  \right) - P_{\rm on} \right].
\eeq

We can rewrite the  last term in Eq.~\ref{agmmainte}:
\begin{align}
\hat{A}_{\Gamma}^*(\xi) &  = \int_{0}^{+\infty} dt e^{- 2 i \pi  \xi t} \left[ \sum_{k=0}^{\infty} \left(  \int_{ t_{i} >0,\sum_{i=1}^k t_{i} < t } 
e^{-k_{\rm off} (t- \sum_{i} t_{i})} \prod_{i=1}^k \eta(t_{i}) dt_{i}  \right)- P_{\rm on} \sum_{k=0}^{\infty} \right. \nonumber \\
&  \left. \int_{\sum_{i} t_{i} < t} (k_{\rm off})^k e^{- k_{\rm off} \sum_{i} t_{i}} e^{- k_{\rm off}(t - \sum_{i} t_{i})} \right] ,
\end{align}
using the  expansion of unity:
 \begin{align}
1 & = \sum_{k=0}^{\infty} e^{-k_{\rm off} t} \frac{(k_{\rm off} t)^k}{k!} \\
   & = \sum_{k=0}^{\infty} \int_{\sum_{i} t_{i} < t} (k_{\rm off})^k e^{- k_{\rm off} \sum_{i} t_{i}} e^{- k_{\rm off}(t - \sum_{i} t_{i})},
\end{align}
with the convention for the $k=0$ term:
\beq
  \int_{\sum_{i} t_{i} < t} (k_{\rm off})^k e^{- k_{\rm off} \sum_{i} t_{i}} e^{- k_{\rm off}(t - \sum_{i} t_{i})} = e^{- k_{\rm off} t}.
\eeq
Collecting terms:
\beq
 \sum_{k=0}^{\infty} \left[ \int_{ t_{i} >0 } \prod_{i=1}^{k} dt_{i}  \left[ \left(  \prod_{i=1}^{k} \eta ( t_{i}) - P_{\rm on} (k_{\rm off})^{k} 
 e^{- k_{\rm off} \sum_{i} t_{i} }  \right) \int_{t>\sum_{i} t_{i}}  dt e^{- 2 i \pi \xi t} e^{- k_{\rm off} (t- \sum_{i} t_{i})} \right] \right]
\eeq
and setting $u = t -\sum_{i} t_{i}$ in the last integral:
\beq
\hat{A}_{\Gamma}^*(\xi)  = \sum_{k=0}^{\infty} \left[ \int_{ t_{i} >0 } \prod_{i=1}^{k} dt_{i}  \left[ \left(  \prod_{i=1}^{k} \eta ( t_{i}) - P_{\rm on} (k_{\rm off})^{k} e^{- k_{\rm off} \sum_{i} t_{i} })  \right) 
\int_{0}^{+\infty}  du e^{- 2 i \pi \xi (u+\sum_{i} t_{i})} e^{- k_{\rm off}u} \right] \right]
\eeq
we obtain:
\beq
 \hat{A}_{\Gamma}^*(\xi)  = ( k_{\rm off} + 2 i \pi \xi - k_{\rm off} ( 1+ \frac{2 i \pi \xi }{\beta} )^{-\alpha  })^{-1} - \frac{P_{\rm on}}{2 i \pi \xi}.
\eeq
Using Eq.~\ref{gammaauto} we recover Eq.~11 in Materials and Methods of the main text.  For $\alpha =1$ we recover results of the two state model.

 \subsection{The \AW{average occupancy} {and} autocorrelation of a Poisson-like polymerase firing model}

\AW{We compared the autocorrelation function calculated for the two and three state cycle promoter models to the autocorrelation of the Poisson-like polymerase firing model. In this model we assume that the gene expression rate is memoryless and the transcription interval follows an exponential distribution with rate $r$:
\begin{equation}
\label{Poisson}
P(t) = \frac{1}{\tau_P} e^{- r t}.
\end{equation}
However, once a polymerase is loaded and starts transcribing, the gene must wait $6$ seconds for the polymerase to leave the transcription initiation site before another polymerase can start transcribing. For this reason, the process is not a simple Poisson counting process, but includes a constant delay for every firing event. On average, there is a polymerase binding event every $T_{ \rm eff} = 1/r +6$ s. $T_{\rm eff}$ is the effective time and $r_{\rm eff}=1/T_{\rm eff}$ is the effective rate of this delayed Poisson process.

The quantity $P_{\rm on}$ corresponds to the average occupancy of polymerase binding sites over the duration of the cell cycle. 
The size of the polymerase is $150$ bp and its speed is {$\sim 25$ bp/second}, so the maximum loading rate of the polymerase is one every $6$ second. Since the polymerase cannot load faster than once every $6$ seconds, we calculate the average occupancy of the gene as the ratio of the average number of polymerase events within a given time window to the maximum number of polymerase events that could happen:   
\begin{equation}
P_{\rm on} =\frac{6}{ T_{\rm eff}} = \frac{6}{6+1/r}.
\label{poisspon}
\end{equation}
Note that, for the Poisson-like model, $P_{\rm on}$ does not correspond to an average time the gene spends in the ON state (since an OFF and ON state is not part of this model).

Using Eq.~\ref{poisspon}, we find that within the Poisson-like firing model,  polymerase arrival rates are very heterogeneous across the embryo. In the anterior polymerases arrive at a rate $r=1/6 s^{-1}$, and in the boundary region $r=1/54 s^{-1}$.}

Since the process is memoryless and the Poisson-like firing process is uncorrelated, its connected autocorrelation is close to a delta function $\delta(\tau=0)$. However, due to the gene lengthy elongation time, there is a non-flat autocorrelation function of the fluorescence signal. 
At steady state, the connected autocorrelation function is:
\begin{equation}
\label{poisscor}
\langle F(t) F(t+\tau)   \rangle - \langle F(t)^2 \rangle = P_{\rm on} \sum_{i,j} L_i L_j A_P(j-\tau {+} i) -  \left(P_{\rm on}  \sum_{i} L_i\right)^2,
\end{equation}
where $A_P({j-\tau+i})$ is the probability of the polymerase to be at position $i$ at time $\tau$, given it was at position $j$ at time $0$ in the Poisson-like firing model. 

If  $\tau<6s$ then the two positions on the gene, $i$ and $j$, share the same polymerase with a probability proportional to $|6-\tau|$, taking equally distributed polymerase positions. If $\tau>6s$, $A_P(\tau)$ is given by the probability that there is a polymerase at the second site, which is independent of what happened at the first site. The two cases give:
\begin{equation}
\label{poisA}
A_P(\tau) = \frac{\theta(6-|\tau|)}{6}  \left[  (6-|\tau|) + P_{\rm on} |\tau| \right] + \theta(|\tau| - 6) P_{\rm on}, 
\end{equation}
{where $\theta$ is the Heavyside function.} This function is flat for $\tau>6s$ and the first part of the right hand side of Eq.~\ref{poisA} has little effect on the autocorrelation function over a cell cycle (as cell cycle duration is much bigger than $6s$). For this reason we use a flat function as a very good approximation for $A_P$ in our analysis.

From the form of Eqs.~\ref{poisscor} and \ref{poisA} and the flat approximation of $A_P$ we see that $P_{\rm on}$ is only a normalizing constant and the shape of the function is completely determined by the loop function $L_i$, which is known. We can compare the expected autocorrelation function of a Poisson-like model to data and find that  \AW{it agrees quite well.}

\subsection{Numerical simulations}

To simulate the time evolution of MCP-GFP loci's intensity, we used the Gillespie algorithm \cite{Gillespie1977, Bratsun2005} to predict the time it takes for the gene to switch between the states, the active ON state and the inactive OFF states. In all models we assume that the time of the transition from the active to the inactive states, $\tau_{\rm on}$ is exponentially distributed with rate $k_{\rm off}$. The time of the transition from the inactive OFF states to ON state, $\tau_{\rm off}$ depends on the model considered:

\begin{itemize}
\item for the two-state model  $\tau_{\rm off}$ is exponentially distributed with rate $k_{\rm on}$.
\item for the three-state model $\tau_{\rm off}$ is a sum of two exponential processes with rates $k_{\rm 1}$ and $k_{\rm 2}$ that describe the transitions between the two OFF states.
\item for the Gamma model $\tau_{\rm off}$ is {sampled} from the $\Gamma(\alpha,\beta)$ distribution defined in Eq.~\ref{GammaSI}.
\end{itemize}

To generate the traces of length $T$ from $N$ nuclei, we first simulate a long trajectory of length $N\times T$, denoted as $X(t)$. 
To account for the incompressibility of the polymerase, we divide the traces into $6$s intervals, which is the time the polymerase needs to cover a region of the gene equal to its own lengths. We assume that at each $6$s time point, if the gene is in the ON state, there is a transcription initiation event by a single RNA polymerase with a full transcription rate, defined as the length of the gene divided by the polymerase velocity, defined in  section~\ref{basicsetup}. Following this event, the RNA polymerase will slide along the target gene segment and synthesize a nascent RNA. At time $i$ into this elongation process, the nascent RNA has $L_i$ MS2 binding sites as depicted in Fig.~3 
of the main text. To impose $P_{\rm on}=k^{\rm eff}_{\rm on}/(k^{\rm eff}_{\rm on}+k_{\rm off})${, if} the gene switches into the OFF state before a full $6$s interval, the polymerase transcribes the gene at a reduced rate proportional to the fraction of the $6$s interval for which the gene was ON. The number of MS2 binding sites at the transcription locus site is therefore given by the convolution of the gene state and the promoter construct design function $L$ (see Fig.~1 in the main text):
\begin{equation} \label{convolution}
F(t) =X(t)*L.
\end{equation}

We assume that the number of MCP-GFP molecules in the nuclei is sufficient to bind to all newly transcribed MS2 binding sites and that the binding process is infinitely fast. The spot intensity is calculated as the number of binding sites produced at the loci (given the intensity of each MPC-GFP dimer equal to $1$). Lastly, the long spot intensity traces are divided equally into $N$ smaller traces of length $T$.

\subsection{Correction to the autocorrelation function for finite trace lengths}
\label{meancorrection}

\begin{figure}
\begin{center}
\noindent\includegraphics[width=0.5\linewidth]{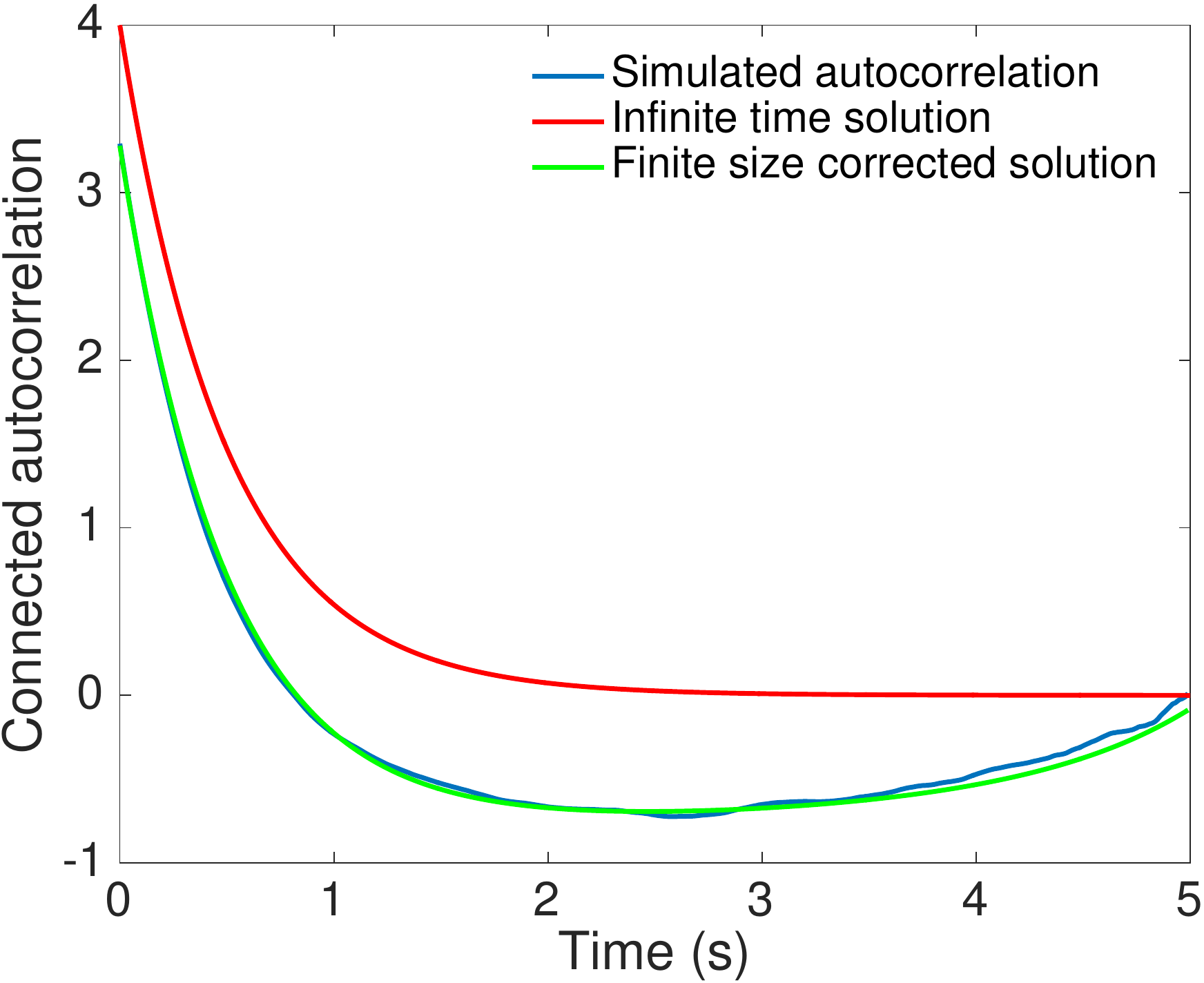}
\caption{{\bf The finite trace effect for the Ornstein-Uhlenbeck process.} The connected autocorrelation function $C_r=\exp(-t/\tau)$ (red line) compared to the connected autocorrelation function calculated from short time traces as described in Section~\ref{meancorrection} (blue line) and the corrected connected autocorrelation function (Eq.~\ref{corrfinal} green line). $\lambda = 2 $s$^{-1}$, $\gamma=4$s$^{-1/2}$ and the short trace length is $5$s where the Ornstein-Ulhenbeck process is $\partial_t x= -\lambda x + \gamma \xi $ and $\xi$ is Gaussian white noise. }
\label{figornstein} 
\end{center}
\end{figure}

The short duration of the experimental traces,  $v_{\alpha, i}$, where $1 \leq \alpha \leq M$ describes the identity of the trace and $0<i<{N}$ denotes the sampling times, coupled with the need to correct for experimental biases by calculating the connected correlation function introduces finite size effects.  The true  connected correlation function between time points at a distance $r$, $C_r$ (red line in Fig.~\ref{figornstein}), is not equal to the empirical connected correlation function calculated as an average over the $M$ traces, $c(r)$ (blue line in Fig.~\ref{figornstein}), of the  autocorrelation functions of the  finite traces.  
The theoretical connected autocorrelation function calculated in our model is:
\begin{equation}
C_r = \frac{\avg{ v_{  i} v_{ i+r}}- \bar{v}^{2}}{  \bar{v^2} - \bar{v}^{2} },
\label{SIEx_conncorr_largeK}
\end{equation}
where $\avg{\cdot}$ denotes an average over random realizations of the process and we assume steady state $\bar {v^k}=   \avg{ v^k_{ i}}= \avg{  v^k_{i+j}}$. The empirical connected correlation function of each finite trace of length ${N}<<\infty$ has the form:
\begin{equation}
\label{SIinsidecr0}
c_{\alpha}(r) =    \left[  \frac{ \sum\limits_{(i,j), |i-j| = r} \left\{  \left( v_{\alpha i} - \frac{1}{N} \sum\limits_{l=1}^{N} v_{\alpha l} \right) \left( v_{\alpha j} - \frac{1}{N} \sum\limits_{l=1}^{N} v_{\alpha l}  \right)  \right\} }{   \dfrac{N-r}{N} \sum\limits_{j=1}^{N} \left( v_{\alpha j} - \frac{1}{N} \sum\limits_{l=1}^{N} v_{\alpha l}  \right)^2  } \right]
\end{equation}
and the empirical connected correlation function calculated averaged over  $M$ traces is
\begin{equation}
c (r) = \frac{1}{M} \sum_{\alpha = 1}^{M} c_{\alpha} (r).
\label{SIcremp}
\end{equation}
$C_r$ requires  knowing the true second moment of the fluorescence signal $\bar{v^2}$. In our data we find that the true variance of the normalized fluorescence signal, $\bar{v^2} - \bar{v}^2$ is well approximated by the average over traces, so we approximate  Eq.~\ref{SIinsidecr0} by:
\begin{equation}
c_{\alpha}(r) =  \frac{ \sum\limits_{(i,j), |i-j| = r} \left\{  \left( v_{\alpha i} - \frac{1}{N} \sum\limits_{l=1}^{N} v_{\alpha l} \right) \left( v_{\alpha j} - \frac{1}{N} \sum\limits_{l=1}^{N} v_{\alpha l}  \right)  \right\} }{  {(N-r)}  (\bar{v^2} - \bar{v}^{2})  }.
\label{SIinsidecr}
\end{equation}
The difference between the theoretical and empirical connected correlation function is independent of our model and arises for the connected correlation function of any random process, as shown in Fig.~\ref{figornstein} for the simplest random process -- the Ornstein-Uhlenbeck process. The difference is due to the fact that the short time average induces spurious correlations when calculating averages of the signal taken at different times. When analyzing the data, to avoid describing nucleus-to-nucleus variability that is not connected to the signal, we first subtract the mean steady state fluorescence signal of each trace, normalize this connected autocorrelation function to $1$ at time $t=0$, and then average over traces (Eq.~\ref{SIinsidecr}) before averaging over the trace ensemble (Eq.~\ref{SIcremp}). In steady state, the infinite trace mean equals the ensemble average,  $\lim_{{N}\rightarrow \infty} \frac{1}{{N}} \sum\limits_{i=1}^{{N}}v_{\alpha i} = \bar{v}$. However, as shown in Fig.~2 of the main text, the short trace mean is not a good approximation to the long term (or ensemble) average, $\frac{1}{{N}} \sum\limits_{i=1}^{{N}}v_{\alpha i} \neq \bar{v}$. The points located in the center of the trace are much more correlated with the mean than the points at the beginning and end of the time interval.
The correction for each value of $r$ is different and must be separately computed. 

In analyzing our data we use the finite size correction for the mean derived below that expresses the empirical connected correlation function $c(r)$ in terms of the theoretical connected correlation function $C_r$. For ${N}\rightarrow \infty$ the empirical connected correlation function becomes the infinite time connected correlation function, however our traces are very short. These corrections are valid for all time dependent data sets so for completeness the finite size correction for the variance is derived in  Section~\ref{variancecorr} but is not used in the analysis.

The number of pairs of time points of distance $r$ in a trace of length $N$ is simply $N-r$ and the combination of Eqs.~\ref{SIcremp} and Eqs.~\ref{SIinsidecr} becomes:
\begin{eqnarray}
c(r)& =&  \frac{1}{M(N-r) ( \bar{v^2} - \bar{v}^{2} )} \sum_{\alpha = 1}^{M}  \left[   \sum\limits_{i=1}^{N-r} \left\{  \left( v_{\alpha i} - \frac{1}{N} \sum\limits_{l=1}^{N} v_{\alpha l} \right) \left( v_{\alpha (i+r)} - \frac{1}{N} \sum\limits_{l=1}^{N} v_{\alpha l}  \right)  \right\} \right] \\ \nonumber
&=  &\frac{1}{M(N-r) ( \bar{v^2} - \bar{v}^{2} )}   \sum_{\alpha = 1}^{M} \Big[   \sum\limits_{i=1}^{N-r}  \Big\{  v_{\alpha i} v_{\alpha (i+r)}  - v_{\alpha i} \left( \frac{1}{N} \sum\limits_{l=1}^{N} v_{\alpha l} \right)   -  \left( \frac{1}{N} \sum\limits_{l=1}^{N} v_{\alpha l} \right) v_{\alpha (i+r)}+ \\ \nonumber
&& (N-r)  \left( \frac{1}{N} \sum\limits_{l=1}^{N} v_{\alpha l} \right)^2  \Big\} \Big]\\ \nonumber
&=&  \left\langle  \frac{1}{ ( \bar{v^2} - \bar{v}^{2} )}    \left\{ \sum\limits_{i=1}^{N-r}  \frac{v_{\alpha i} v_{\alpha (i+r)} }{N-r}  -                   \sum\limits_{i=1}^{N-r}  \frac{v_{\alpha i}}{N-r} \left( \frac{1}{N} \sum\limits_{l=1}^{N} v_{\alpha l} \right)    -  \sum\limits_{i=r+1}^{N}  \frac{v_{\alpha i}}{N-r} \left( \frac{1}{N} \sum\limits_{l=1}^{N} v_{\alpha l} \right)  + \frac{1}{N^2}  \left(  \sum\limits_{l=1}^{N} v_{\alpha l} \right)^2  \right\} \right\rangle_{\alpha},
 \end{eqnarray}
where we have explicitly written out the terms and  in the last line we introduced the average over traces $\avg{\cdot}_{\alpha}=1/M\sum_{\alpha = 1}^{M} \cdot$. In steady state due to time invariance:
\begin{equation}
\label{timeinv}
\left\langle \sum\limits_{i=N-r+1}^{N}  \frac{v_{\alpha i}}{N-r} \left( \frac{1}{N} \sum\limits_{l=1}^{N} v_{\alpha l} \right) \right\rangle_{\alpha} = \left\langle  \sum\limits_{i=1}^{r}  \frac{v_{\alpha i}}{N-r} \left( \frac{1}{N} \sum\limits_{l=1}^{N} v_{\alpha l} \right)  \right\rangle_{\alpha} 
\end{equation}
and the  theoretical (not connected) correlation between two points is  a function only of the distance between these two points:
\beq
\label{Ctilde}
\tilde{C}_{r}= \avg{v_{ i} v_{ i+r}}=1/M \sum_{\alpha = 1}^{M} v_{\alpha i} v_{\alpha i+r}.
\eeq
We have assumed that $M$ is large and a population average over the $M$ traces for points separated by $r$ on each trace approximates the $M\rightarrow \infty$ limit of the theoretical average over different realizations of the process. Using Eq.~\ref{Ctilde} we obtain:
\beq
 c(r) =   \frac{\tilde{C}_{r} }{ \bar{v^2} - \bar{v}^{2}}  +  \frac{1}{  \bar{v^2} - \bar{v}^{2} }   \left\langle  \sum\limits_{i=1}^{r}  \frac{2 v_{\alpha i}}{N-r} \left( \frac{1}{N} \sum\limits_{l=1}^{N} v_{\alpha l} \right)   +\frac{1}{N} \left( \frac{1}{N} - \frac{2}{(N-r)} \right) \left(  \sum\limits_{l=1}^{N} v_{\alpha l} \right)^2   \right\rangle_{\alpha}.
\eeq
To rewrite $\left\langle \left(  \sum\limits_{l=1}^{N} v_{\alpha l} \right)^2   \right\rangle_{\alpha}$ as a sum over $\tilde{C}_r$ we calculate the number of pairs of time points separated by a distance $k$ in the whole trace of length $N$. For $k=0$ it is equal to $N$ and for $1 \leq k \leq N-1$ it is equal to $2(N-k)$:
\begin{equation}
\left\langle \left(  \sum\limits_{l=1}^{N} v_{\alpha l} \right)^2   \right\rangle_{\alpha} = N \tilde{C}_{0} + \sum_{k=1}^{N-1} 2 (N-k) \tilde{C}_{k}.
\end{equation}
Similarly
\begin{eqnarray}
\left\langle  \sum\limits_{i=1}^{r} \sum_{l=1}^{N} v_{\alpha i} v_{\alpha l}   \right\rangle_{\alpha} & =& \left\langle  \left( \sum\limits_{i=1}^{r} v_{\alpha i} \right)^2   \right\rangle_{\alpha}  + \left\langle \sum_{i=1}^{r} \sum_{l=r+1}^{N} v_{\alpha i} v_{\alpha l}  \right\rangle_{\alpha} \\ \nonumber\\
&=&r \tilde{C}_{0} + \sum_{k=1}^{r-1} 2 (r-k) \tilde{C}_{k} + \sum_{l=r+1}^{N} \sum_{i=1}^{r} \tilde{C}_{\abs{l-i}} \\ \nonumber\\
&=& r \tilde{C}_{0} + \sum_{k=1}^{r-1} 2 (r-k) \tilde{C}_{k}+   \sum_{m=1}^{N-1} \tilde{C}_{m} [ \min(m+r,N)-\max(r,m)].
\end{eqnarray}
Collecting the empirical connected autocorrelation function in Eq.~\ref{SIcremp} is expressed in terms of the theoretical non-connected correlation function in Eq~\ref{Ctilde} as:
\begin{multline}
c(r) = \frac{1}{ \bar{v^2} - \bar{v}^{2}} \left[ \tilde{C}_{r} + \frac{1}{N} \left( \frac{1}{N} - \frac{2}{(N-r)} \right) \left(N \tilde{C}_{0}+ \sum_{k=1}^{N-1} 2 (N-k) \tilde{C}_{k} \right) \right. \\  \left. + \frac{2}{N(N-r)} \left(  r \tilde{C}_{0}+ \sum_{k=1}^{r-1} 2 (r-k) \tilde{C}_{k}+   \sum_{m=1}^{N-1} \tilde{C}_{m} [ \min(m+r,N) -\max(r,m)]  \right)   \right].
\label{corrfinal}
\end{multline}
$\tilde C_k$ is the theoretical steady state non-connected correlation function of the process (given in Eq.~6 of the main text for the two state model, Eq.~8 of the main text for the cycle model and as the Fourier transform of Eq.~12 for the $\Gamma$ model) and the average is over random realizations of the process. The mean and variance of the signal, $\bar{v}$ and $\bar{v^2}$, provide a normalization factor that is constant for all time differences $r$. We normalize the autocorrelation function setting the second term to $1$ and these terms are not needed for the inference.

\subsection{Correction to the autocorrelation function from correlations in the variance}
\label{variancecorr}

In  Section~\ref{meancorrection} we calculated the finite size correction due to short traces for the empirical connected correlation function assuming that differences between the empirical variance and the theoretical variance for infinite traces do not affect the connected autocorrelation function. This approximation is valid for our data. For completeness we now calculate the finite size correction coming from spurious correlations in the variance obtained when computing the variance trace by trace, before averaging over the traces (Eq.~\ref{SIcremp}). Analyzing the data, we normalize the autocorrelation function of each trace before taking the average over all traces because of potential nucleus-to-nucleus variability in the signal calibration. This is equivalent to dividing each autocorrelation function by its  variance, before averaging over the traces and can introduce errors. 

The empirical connected correlation function in Eqs.~\ref{SIcremp} and~\ref{SIinsidecr0} can be rewritten by adding and subtracting $1$ in the denominator as:
\begin{equation*}
c(r) =  \frac{N}{(N-r) (\bar{v^2} - \bar{v}^{2} ) } \avg{\left[  \frac{ \sum\limits_{(i,j), |i-j| = r} \left\{  \left( v_{\alpha i} - \frac{1}{N} \sum\limits_{l=1}^{N} v_{\alpha l} \right) \left( v_{\alpha j} - \frac{1}{N} \sum\limits_{l=1}^{N} v_{\alpha l}  \right)  \right\} }{ 1 + \frac{1}{\bar{v^2} - \bar{v}^{2} } \left(  \sum\limits_{j=1}^{N} \left( v_{\alpha j} - \frac{1}{N} \sum\limits_{l=1}^{N} v_{\alpha l}  \right)^2 - ( \bar{v^2} - \bar{v}^{2} )\right) } \right] }_{\alpha},
\end{equation*}
where the average $\avg{\cdot}_{\alpha}$ is over $M$ traces as defined in  Section~\ref{meancorrection}.
Assuming the true variance of the process is close to the empirical variance we linearize the denominator :
\begin{multline}
c(r) =  \frac{N}{(N-r) (\bar{v^2} - \bar{v}^{2} ) } \Big \langle  \left[   \sum\limits_{(i,j), |i-j| = r} \left\{  \left( v_{\alpha i} - \frac{1}{N} \sum\limits_{l=1}^{N} v_{\alpha l} \right) \left( v_{\alpha j} - \frac{1}{N} \sum\limits_{l=1}^{N} v_{\alpha l}  \right)  \right\} \right] \times \\ \left[2 - \frac{1}{\bar{v^2} - \bar{v}^{2} }   \sum\limits_{j=1}^{N} \left( v_{\alpha j} - \frac{1}{N} \sum\limits_{l=1}^{N} v_{\alpha l}  \right)^2  \right]  \Big \rangle_{\alpha}.
\end{multline}
The first term in the paranthesis is proportional to the connected correlation function in Eq.~\ref{corrfinal} we calculated in  Section~\ref{meancorrection} assuming constant variance. We focus on the second term:
\begin{multline}
d(r) =    \Big\langle  \left[   \sum\limits_{(i,j), |i-j| = r} \left\{  \left( v_{\alpha i} - \frac{1}{N} \sum\limits_{l=1}^{N} v_{\alpha l} \right) \left( v_{\alpha j} - \frac{1}{N} \sum\limits_{l=1}^{N} v_{\alpha l}  \right)  \right\} \right]  \cdot\left[  \sum\limits_{j=1}^{N} \left( v_{\alpha j} - \frac{1}{N} \sum\limits_{l=1}^{N} v_{\alpha l}  \right)^2  \right]  \Big\rangle_{\alpha}\\ \nonumber
=    \Big\langle   \left\{ \sum\limits_{i=1}^{N-r}  v_{\alpha i} v_{\alpha (i+r)}   -                   \sum\limits_{i=1}^{N-r}  v_{\alpha i} \left( \frac{1}{N} \sum\limits_{l=1}^{N} v_{\alpha l} \right)   \right.    \left. -  \sum\limits_{i=r+1}^{N}  v_{\alpha i} \left( \frac{1}{N} \sum\limits_{l=1}^{N} v_{\alpha l} \right)  + \frac{N-r}{N^2}  \left(  \sum\limits_{l=1}^{N} v_{\alpha l} \right)^2  \right\}  \times \\ \left[  \sum\limits_{j=1}^{N} v_{\alpha j}^2 - \frac{2}{N} \sum\limits_{j=1}^{N} \sum\limits_{l=1}^{N} v_{\alpha j} v_{\alpha l} + \frac{N}{N^2} \sum\limits_{j=1}^{N} \sum\limits_{l=1}^{N} v_{\alpha j} v_{\alpha l} \right]  \Big\rangle_{\alpha}.
\end{multline}
Using time invariance at steady state (Eq.~\ref{timeinv}) in the first factor and simplifying the algebra in the second factor:
\begin{multline}
d(r) =   \Big \langle   \left[ \sum\limits_{i=1}^{N-r}  v_{\alpha i} v_{\alpha (i+r)}  - 2 \sum\limits_{i=1}^{N-r}    v_{\alpha i} \left( \frac{1}{N} \sum\limits_{l=1}^{N} v_{\alpha l} \right) + \frac{N-r}{N^2}  \left(  \sum\limits_{l=1}^{N} v_{\alpha l} \right)^2  \right]  \cdot \left[  \sum\limits_{j=1}^{N} v_{\alpha j}^2 - \frac{1}{N} \sum\limits_{j=1}^{N} \sum\limits_{l=1}^{N} v_{\alpha j} v_{\alpha l}  \right] \Big\rangle_{\alpha} \\ \nonumber
=    \Big\langle  \sum\limits_{j=1}^{N}  \sum\limits_{i=1}^{N-r}  v_{\alpha j}^2 v_{\alpha i} v_{\alpha (i+r)} - \frac{1}{N} \sum\limits_{i=1}^{N-r}  \sum\limits_{j,l=1}^{N}v_{\alpha j} v_{\alpha l}  v_{\alpha i} v_{\alpha (i+r)} 
-\frac{2}{N}  \sum\limits_{i=1}^{N-r} \sum\limits_{j,l=1}^{N}   v_{\alpha i} v_{\alpha l} v_{\alpha j}^2 + \frac{2}{N^2} \sum\limits_{i=1}^{N-r} \sum\limits_{j,k,l=1}^{N}   v_{\alpha i}v_{\alpha j}v_{\alpha k}v_{\alpha l} \\
+ \frac{N-r}{N^2} \sum\limits_{j,k,l=1}^{N} v_{\alpha j}^2 v_{\alpha k}v_{\alpha l} -\frac{N-r}{N^3} \sum\limits_{j,k,l,m=1}^{N} v_{\alpha j}v_{\alpha k}v_{\alpha l} v_{\alpha m}
 \Big \rangle_{\alpha}.
\end{multline}
The final correction for correlation due to correlations in the variance coming from short time traces is easily evaluated {in} terms of four-points correlation function $F(s,t,u)= \overline{ v_i v_{i+s} v_{i+s+t} v_{i+s+t+u}  } $.

\begin{figure}[h]
\includegraphics[width=0.85\textwidth]{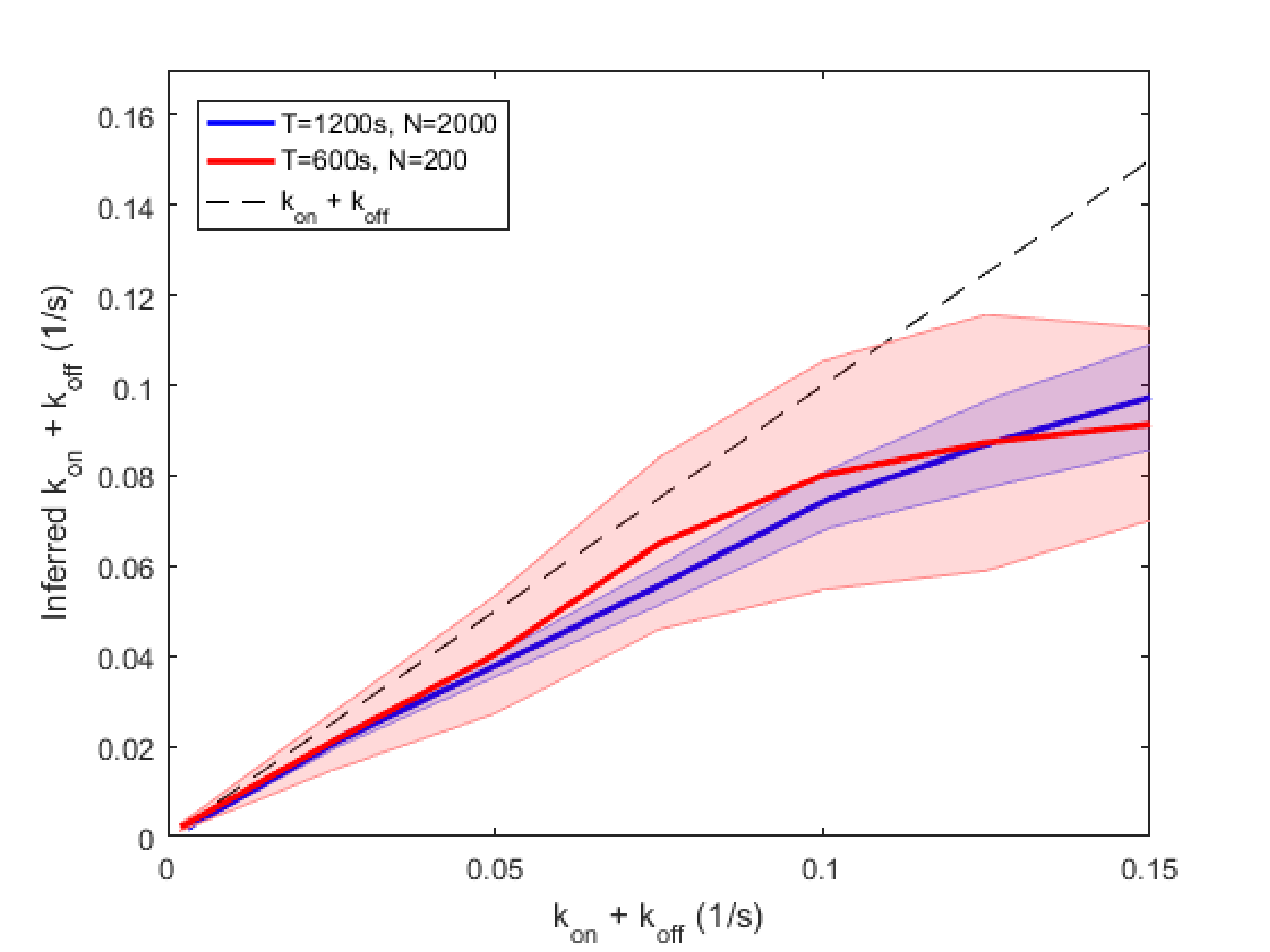} 	\caption{\AW{{\bf Inference of the two-state model from the cross-correlation function between $3'$ signals and $5'$ signals}.  The gene cassette contains two identical arrays of MS2 binding sites on the $3'$ and $5'$ ends, separated by a gene of $3$ kbp in length. The input parameters $k_{\rm on}$, $k_{\rm off}$ are varied so as to maintain the same $P_{\rm on}=0.1$.   \label{trace_example}}}
\end{figure} 

\begin{figure}[h]
\includegraphics[width=0.85\textwidth]{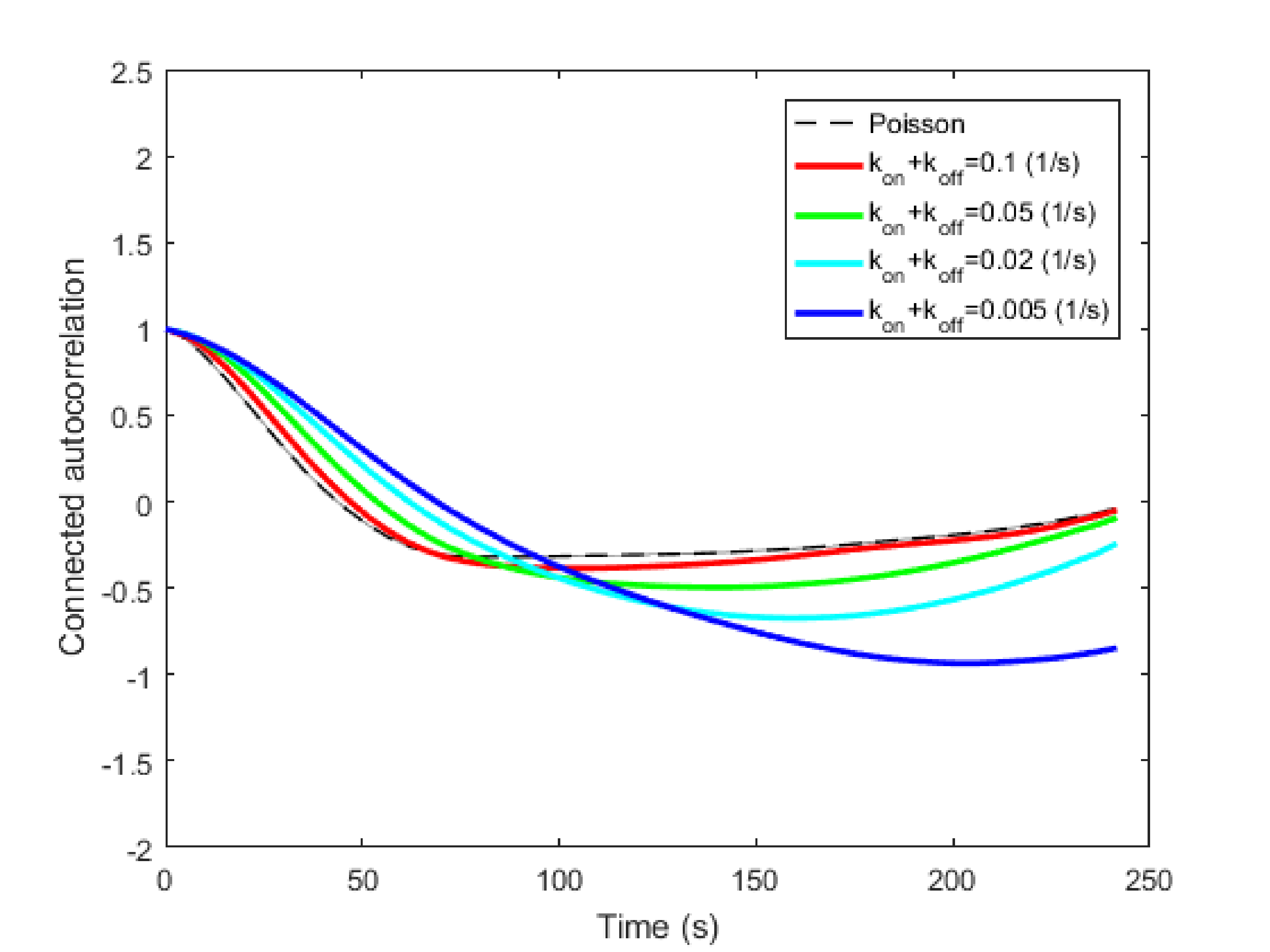} 	\caption{\AW{{\bf Comparison between the autocorrelation functions of the Poisson-like model and the two-state model}. Shown are the autocorrelation functions (calculated from $1000$ traces of $250$ s in length) of the Poisson-like model (dashed black) and the two-state model (solid) with varying $k_{\rm on}$ and $k_{\rm off}$. The model parameters are set to achieve the same effective transcription rate, $P_{\rm on}=0.1$, that we infer in the boundary region. For large $k_{on}+k_{off}$ values the shape of the autocorrelation function is dominated by the autocorrelation of the fluorescent probe and the Poisson-like and two state model autocorrelation functions look very similar. The inferred two state param{eters} are close to the green line. Since it is difficult to estimate the number of independent measurements, we cannot use standard statistical measures to compare these models with different numbers of parameters, whereas to determine the value of parameters within a given model we use a statistical measure (the mean square distance between the model prediction and data). For this reason we can differentiate between parameter values for the two state model that result in similar looking autocorrelation functions, but we cannot differentiate between two classes of models that result in similar differences in the autocorrelation functions. \label{poisson_vs_twostate}}}
\end{figure}

 \subsection{Cross-correlation}
The presented correlation analysis can also be extended to constructs with two colored promoters inserted at two differen{t} positions on the same gene. In this case, each construct can have a different loop design function $L^{\nu}_i$, where ${\nu}=1,2$, and the cross-correlation of the normalized fluorescence intensity is:
\begin{equation}
\avg{F_1(t) F_2(s)} = \sum_{i=1}^{r_1} \sum_{j=1}^{r_2} L^1_i L^2_j  <a_{i}(t) a_{j} (s) >.
\label{crossauto}
\end{equation}
The $L^{\nu}_i$ functions start at the same point (the one describing the downstream construct is $0$ for the first steps). 

After the loop design functions $L^{\nu}_i$ have been defined, the calculation of the theoretical cross-correlation function and autocorrelation rely only on calculating the correlations of the gene expression state, which is the same for both. So the results presented for the particular models are valid, after correcting for the two different loops functions. For examples, the steady state connected cross-correlation function of the two state model is:
\begin{eqnarray}
\label{crossco2}
\avg{F_1(t) F_2(t+\tau)} -\avg{F_1(t)}^2= \sum_{i=1}^r \sum_{j=1}^r  & L_i   L_j P_{\rm on}     P_{\rm off} e^{|\tau-j +i|(\delta-1)},
 \end{eqnarray}
where $P_{\rm on}$ and $\avg{F_1(t)}^2=\avg{F_2(t)}^2$ can be independently calculated from either probe, which provides an independent estimate of the experimental noise. 

The differences in the use of the cross-correlation function and autocorrelation function arise when calculating the finite size corrections from short traces,  because assumptions about the statistical time invariance of the signal in steady state are no longer valid. The non-connected theoretical correlation function (equivalent of Eq.~\ref{Ctilde}) is now defined on two signals, $ v_{i}$ and $w_{i}$: 
\beq
\tilde C_r=\avg{{ v_{\alpha, i} w_{\alpha, i+m}}},
\eeq
where $\avg{\cdot}$ define the average over random realizations of the process and in steady state is independent of $i$. 
Unlike for the autocorrelation function, $\tilde C_r$ is no longer symmetric with exchange of $v_{ i} $ and $w_{i}$. The empirical cross-correlation function is (assuming the variance is well approximated by the empirical variance):
\begin{multline}
 c(r) =  \left\langle  \frac{1}{ ( \bar{v^2} - \bar{v}^{2} )}    \left\{ \sum\limits_{i=1}^{N-r}  \frac{v_{\alpha i} w_{\alpha (i+r)} }{N-r}  -                   \sum\limits_{i=1}^{N-r}  \frac{v_{\alpha i}}{N-r} \left( \frac{1}{N} \sum\limits_{l=1}^{N} w_{\alpha l} \right)   \right. \right. \\  \left. \left. -  \sum\limits_{i=r+1}^{N}  \frac{w_{\alpha i}}{N-r} \left( \frac{1}{N} \sum\limits_{l=1}^{N} v_{\alpha l} \right)  + \frac{1}{N^2}  \left(  \sum\limits_{l=1}^{N} v_{\alpha l} \right)\left(  \sum\limits_{l=1}^{N} w_{\alpha l} \right)  \right\} \right\rangle_{\alpha},
\end{multline}
which in terms of the $\tilde C_m$ is:
\beq
 c(r) =  \frac{1}{ ( \bar{v^2} - \bar{v}^{2} )} \left\{   \tilde C_r -       \frac{1}{N(N-r)}            \sum\limits_{i=1}^{N-r}   \sum\limits_{l=1}^{N} \tilde C_{l-i} - \frac{1}{N(N-r)}    \sum\limits_{i=r+1}^{N}   \sum\limits_{l=1}^{N} \tilde C_{i-l}  + \frac{1}{N^2}   \sum\limits_{i,l=1}^{N} \tilde C_{i-l}\right\}.
\eeq
 Repeating the steps in  Section~\ref{meancorrection} we obtain the finite size correction for the cross-correlation function. 
\begin{multline}
 c(r) =  \frac{1}{ ( \bar{v^2} - \bar{v}^{2} )} \left\{  \tilde C_r -       \frac{1}{N(N-r)} \sum_{k=-N+r+1}^{N-r-1}  (N-r-|k|)\tilde C_k \right. \\ \left. -  \frac{1}{N(N-r)}            \sum\limits_{i=1}^{N-r}   \sum\limits_{l=N-r+1}^{N} \tilde C_{l-i} -  \frac{1}{N(N-r)} \sum_{k=-N+r+1}^{N-r-1}  (N-r-|k|) \tilde C_k  \right. \\  \left. - \frac{1}{N(N-r)}    \sum\limits_{i=r+1}^{N}   \sum\limits_{l=1}^{r} \tilde C_{i-l}  + \frac{1}{N^2} \sum_{k=-N+1}^{N-1}  (N-|k|) \tilde C_k
 \right\} \\ \nonumber
= \frac{1}{ ( \bar{v^2} - \bar{v}^{2} )} \left\{   \tilde C_r -       \frac{2}{N(N-r)} \sum_{k=-N+r+1}^{N-r-1}  (N-r-|k|) \tilde C_k  -  \frac{1}{N(N-r)}  \sum_{m=1}^{N-1} \tilde C_m [ \min (m+N-r, N )- \max ( N-r,  m)] \right. \\  \left. - \frac{1}{N(N-r)} \sum_{m=1}^{N-1} \tilde C_m [ \min (m+r, N )- \max ( r,  m)] + \frac{1}{N^2} \sum_{k=-N+1}^{N-1}  (N-|k|) \tilde C_k
 \right\}.
\end{multline}

\begin{figure}
\begin{center}
\noindent\includegraphics[width=\linewidth]{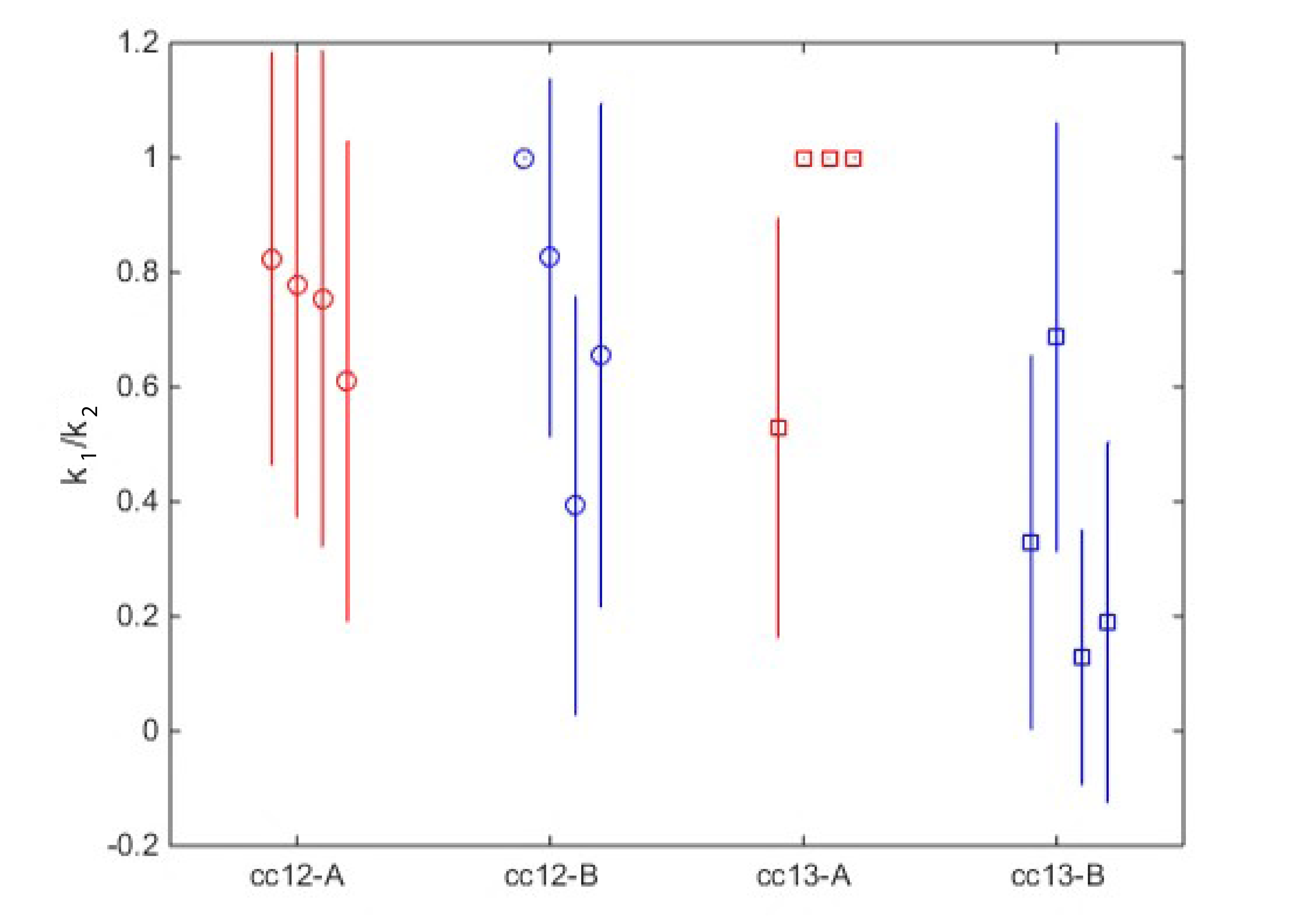}
\caption{{\bf{The fit of the three state cycle model to  the data.}} The fit of the ratio of the two rates for leaving the two OFF states, $k_1/k_2$, to the steady state traces from four embryos in the anterior and boundary region of cell cycle 12 and 13. Each point is data from one embryo. The error bars represent the standard deviation of the inferred value. The fit is for a randomized $60\%$ of the data. The sum of the switching rates  $k_{\rm on}+k_{ 1}+k_2$ is shown in Fig.~5B of the main text.}
\label{3state} 
\end{center}
\end{figure}

\subsection{Precision of the translational process}
\label{appprecision}

The precision of the total  mRNA produced during a cell cycle presented in the main text is proportional to the activity of the gene and requires a careful calculation of the variability of the probability of the gene to be ON in different nuclei at the same position. 
The total activity of a nucleus, defined as the integral of the normalized fluoresce $\sum_i^K F_i$, where $i<K$ are the sampling times in steady state window of the cycle, in steady state is proportional to the probability of the gene to be ON in a given trace,  $P_{\rm on}^{\alpha}$. To keep our analysis independent of normalization, we will calculate the relative error defined as the variance over the mean of $P_{\rm on}^{\alpha}$, ${\rm var}({P_{\rm on}^{\alpha}})/\avg{ P_{\rm on}^{\alpha}}_{\alpha}$, where the averages are taken over traces. 

First, we can calculate the relative error of the probability of the gene to be ON $P_{\rm on}^{\alpha}$ directly from the traces. We compute the mean and standard deviation of the distribution of $P_{\rm on}^{\alpha}$ in a given window along the AP axis. $P_{\rm on}^{\alpha}$ for each trace is calculated from Eq.~\ref{SImeanF}.

We can compare the results of the empirically estimated relative error to predictions of the steady state models. 
We know that the expected average over traces $\sum_{\alpha=1}^M P_{\rm on}^{\alpha}$ is $P_{\rm on}$. 
Within the assumption of our model presented in  Section~\ref{apptwo}, the expectation value of the square of the $P_{\rm on}^{\alpha}$ is expressed in terms of the expression states of the gene, $X(t)$: 
\beq
\langle  P_{\rm on}^{\alpha, 2} \rangle_{\alpha} =\left\langle  \frac{1}{T^2}  \int_0^T dt \int_0^T ds X(t) X(s) \right\rangle_{\alpha} ,
\eeq
where the average is over $M$ traces and $T$ is the total duration of the trace in real time.
In terms of the probability that the gene is ON at time $\tau$ given that it was ON at time $0$, $A(\tau)$ defined in Eq.~\ref{Adef}, we obtain
\beq
 \langle P_{\rm on}^{\alpha, 2} \rangle_{\alpha} = { \frac{1}{T^2}  \int_0^T dt \int_0^T ds P_{\rm on} A(t-s)} ,
\label{genprec}
\eeq
where $A(\tau)$ has units of seconds. 
The relative error is obtained by replacing $A(\tau)$ by the appropriate function for each model. 
For the two state model:
\beq
\langle  P_{\rm on}^{\alpha, 2} \rangle_{\alpha} = \frac{P_{\rm on}}{T^2}  \int_0^T dt \int_0^T ds (P_{\rm on} + P_{\rm off} e^{- |t-s| (k_{\rm on} + k_{\rm off})} ).
\eeq  
Integrating and substracting the mean squared we obtain the relative error:
\beq
\frac{\delta P_{\rm on}}{P_{\rm on}} = \frac{1}{T}\sqrt{ 2 \frac{   k_{\rm off} }{k_{\rm on}(k_{\rm on}+k_{\rm off})} (T - \frac{1-e^{-T(k_{\rm on}+k_{\rm off})}}{k_{\rm on}+k_{\rm off}}) } .
\eeq
The probability of the gene to be on is proportional to the total mRNA produced and for large $T$ we reproduce the result in Eq.~4 in the main text:
\beq
\frac{\delta {\rm mRNA}}{{\rm mRNA}} = \sqrt{ \frac{2}{T} \frac{   k_{\rm off} }{k_{\rm on}(k_{\rm on}+k_{\rm off})}  }=\sqrt{ 2\frac{\tau_i (1-P_{\rm on})} { T P_{\rm on}}}.
\eeq
For the three state cycle model the same calculation is valid until Eq.~\ref{genprec} and is then carried out numerically. 

\AW{In the Poisson-like firing model  the accuracy of {\it hunchback} mRNA production over one cell-cycle the average number of events within a cell-cycle of duration $T$ is $n=T/T_{\rm eff} = rT/(6r+1)$. The amount of mRNA produced during the cell-cycle is proportional to the number of polymerase arrival events. 
Polymerase arrivals are Poisson-distributed and followed by a deterministic delay of $6s$ when the polymerase binding site is still occupied by the previous polymerase, so we find that the squared relative error is proportional to the inverse of the number of events times the error on the effective arrival rate of polymerases:
\begin{equation}
\frac{\langle mRNA^2 \rangle - \langle mRNA \rangle^2}{\langle mRNA \rangle^2} = \frac{1}{n} \frac{\langle T_{\rm eff}^2 \rangle - \langle T_{\rm eff} \rangle^2}{T_{\rm eff}^2} = \frac{6r+1}{rT} \frac{1/r}{(1/r+6)^2} = \frac{1}{T(6r+1)}.
\end{equation}
Replacing with the expression of Eq.~\ref{poisspon}
\begin{equation}
\frac{\langle mRNA^2 \rangle - \langle mRNA \rangle^2}{\langle mRNA \rangle^2} = \frac{1}{T} \frac{1}{ 1+P_{\rm on}/(1-P_{\rm on})  }  = \frac{6(1-P_{\rm on})}{T},
\end{equation}
and the relative error in the produced mRNA is
\begin{equation}
\frac{\delta {\rm mRNA}}{{\rm mRNA}}  = \sqrt{\frac{6(1-P_{\rm on})}{T}}.
\end{equation}
The predicted accuracy for both the boundary and anterior regions in the embryo is much higher than the experimentally observed accuracy.}

Precision from static (Fluorescent In Situ Hybridization -- FISH) images is calculated as the {standard deviation} over the mean of the distribution of a binary variable, which for each nucleus is $1$ if the gene is on in the static image and $0$ if it off \cite{Gregor2007, Porcher2010, Little2013}. The signal in FISH datasets in an average over an unknown timeframe. To compare our analysis of the time dependent signal to these previous measurements, we use a binary variable, which is $1$ for each nucleus that was ON during the steady state interphase and $0$ for each nucleus that was always OFF. The results of the relative error as a function of position obtained using this empirical analysis in Fig.~\ref{precissionSI} show agreement with previous reports \cite{Little2013}: for most traces the relative error in the anterior is zero -- all nuclei in a given AP axis window express, and it increases to $\sim 50\%$ at the boundary. 

\AW{
\begin{figure}
\begin{center}
\noindent\includegraphics[width=\linewidth]{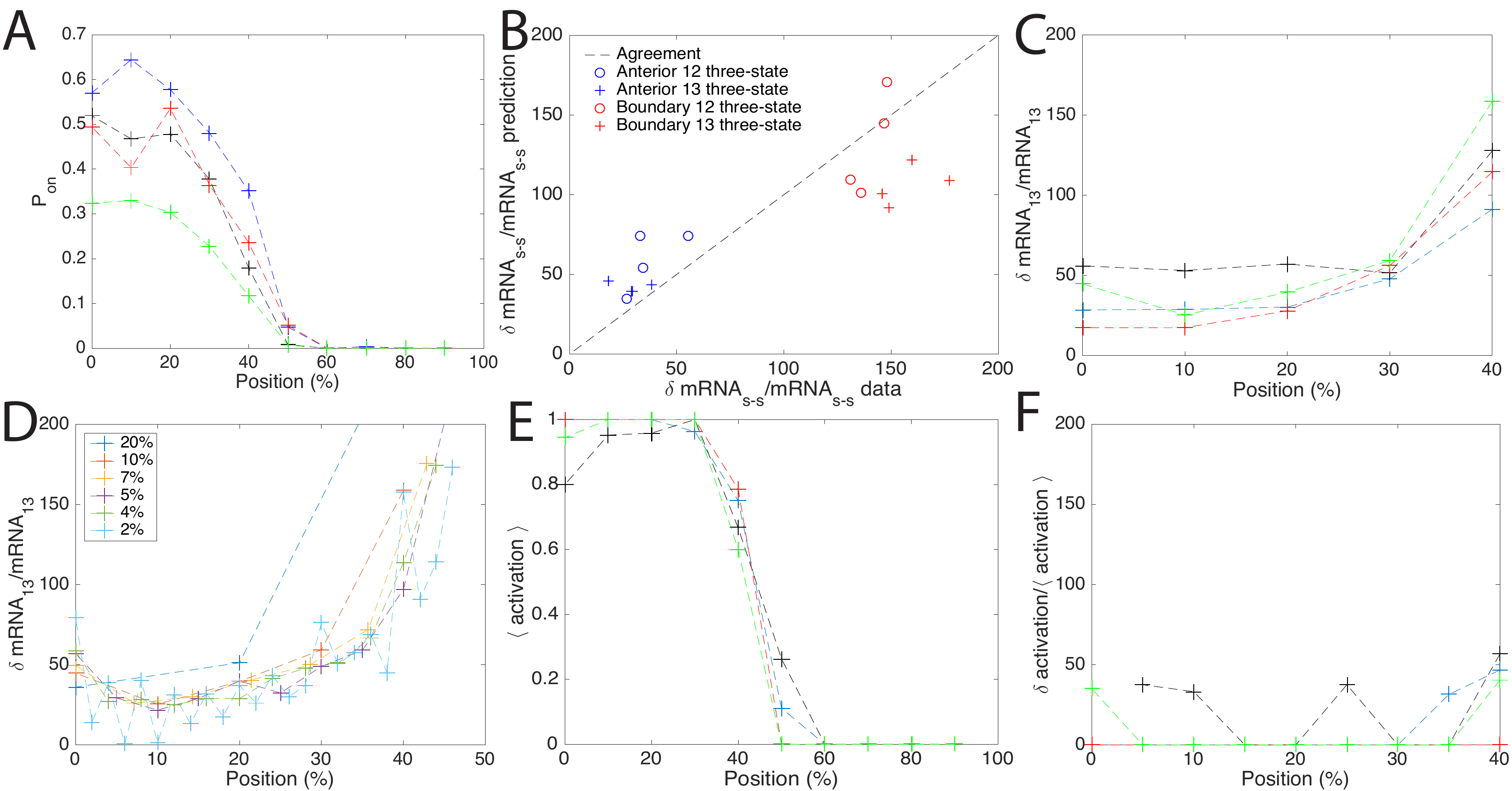}
\caption{\AW{{\bf{The relative error of gene expression.}} A. The mean probability of the gene to be ON at any time during the cell cycle as a function of the embryo length (binary approximation). B. Comparison of the relative error in the mRNA produced during the steady state of the interphase estimated empirically from data (abscissa) and from theoretical arguments in Eq.~\ref{genprec} using the inferred parameters from the autocorrelation function (ordinate), in the anterior (blue) and the boundary (red) regions, show very good agreement. C. The conclusions about precision do not depend on the embryo. The relative error of the total mRNA produced in cell cycle 13 as a function of position for windows equal to $10\%$ of the embryo length. Each colored line represents one embryo. The same data plotted as an average over embryos with the variance as error bars is shown in Fig.~7 of the main text. D. The conclusions about precision do not depend on the window size. The total mRNA produced in cell cycle 13 as a function of position for different window sizes. Except for very large scales ($20\%$) and very small scales comparable to one nuclear width ($2\%$, the relative error as a function of position is reproducible.  E.The mean probability for the gene to be ON averaged over the cell cycle. F. The relative error of the discrete variable that describes the probability of the gene to be ON at any time during the cell cycle as function of position. The relative error is much lower in the anterior compared to the error in the total produced mRNA, but remains high at the boundary.  In A, C, E and F each colored lines describe different embryos. }
\label{precissionSI} }
\end{center}
\end{figure}

\begin{figure}[h]
\includegraphics[width=0.85\textwidth]{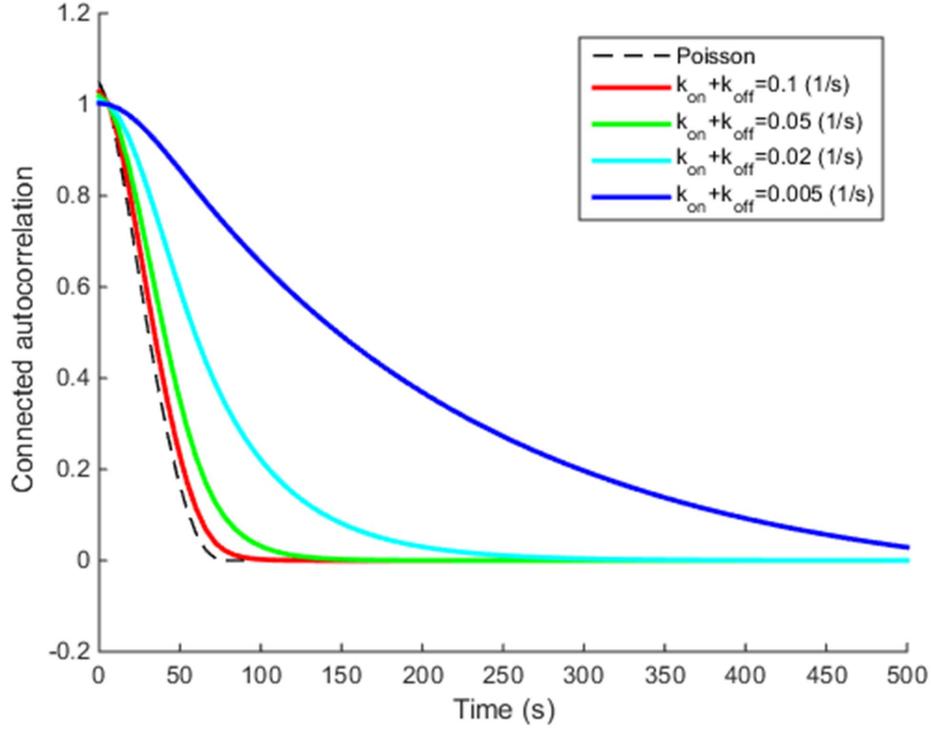} 	\caption{\AW{{\bf The autocorrelation function for Poisson-like model and the two-state model for infinitely-long time traces}. Autocorrelation functions of the Poisson-like model (dashed black) and the two-state models (solid) with $P_{\rm on}=0.1$ (similar to the inferred value in the boundary region) and varying $k_{on}+k_{off}$. In the inferred parameter regime  (approximately green line), longer time traces do not help distinguish the two models based on the autocorrelation function. For large $k_{on}+k_{off}$ values the shape of the autocorrelation function is dominated by the autocorrelation of the fluorescent probe and the Poisson-like and two state model autocorrelation functions look very similar, even for long traces.) \label{poisson_vs_twostatelong}}}
\end{figure}
}

\begin{figure}
\begin{center}
\noindent\includegraphics[width=\linewidth]{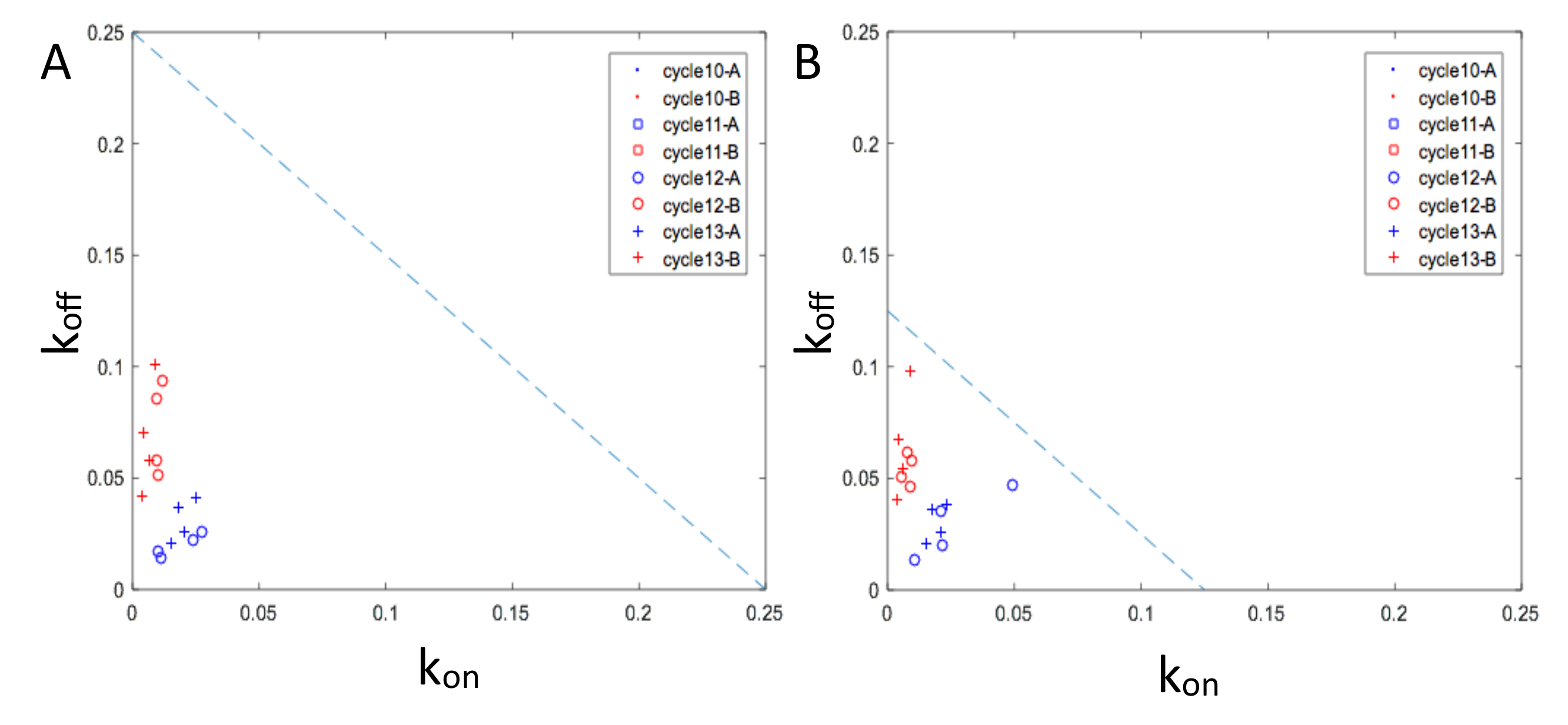}
\caption{{\bf{The dependence of the data fit on polymerase buffering time.}} Assuming different buffering times for the polymerase does not strongly affect the fit of the switching rates: a fit with $\tau_{\rm buffering}=4 s$ (A) and $\tau_{\rm buffering}=8 s$. $\tau_{\rm buffering}=6 s$ is used in the main text in Fig.~5D.}
\label{3state} 
\end{center}
\end{figure}

\end{document}